\documentclass[twocolumn]{aastex62}
\usepackage{longtable, multirow}

\newcommand{\kms}{km\,s$^{-1}$}
\newcommand{\cmsq}{cm$^{-2}$}
\newcommand{\degree}{\ensuremath{^\circ}}

\newcommand{\mum}{\mbox{$\mu{\rm m}$}}

\newcommand{\cross}{$\mathrm{\times}$}

\newcommand{\hi}{\mbox{H\,{\sc i}}}
\newcommand{\mgii}{\mbox{Mg\,{\sc ii}}}

\newcommand{\ciii}{\mbox{C\,{\sc iii}}}

\newcommand{\civ}{\mbox{C\,{\sc iv}}}
\newcommand{\heii}{\mbox{He\,{\sc ii}}}

\def\h2{$\rm H_2$}
\def\Nh2{$N$(H${_2}$)}

\def\lya{\ensuremath{{\rm Ly}\alpha}}

\def\kms{km\,s$^{-1}$}

\def\zem{$z_{\rm em}$}

\def\21{21-cm}

\def\t0{T$_{0}$}

\def\c21{$C_{21}$}

\def\oii{[O\,{\sc ii}]}
\def\oiii{[O\,{\sc iii}]}

\submitjournal{ApJ}
\shorttitle{SALT-NOT survey of MIR-selected AGN} 
\shortauthors{Gupta et al.}
\begin{document}

\title{MALS SALT-NOT survey of MIR-selected powerful radio-bright AGN at $0<z<3.5$} 

\correspondingauthor{N. Gupta}
\email{ngupta@iucaa.in}

\author{N. Gupta}  
\affil{Inter-University Centre for Astronomy and Astrophysics, Post Bag 4, Ganeshkhind, Pune 411 007, India}

\author{G. Shukla} 
\affil{Inter-University Centre for Astronomy and Astrophysics, Post Bag 4, Ganeshkhind, Pune 411 007, India}

\author{R. Srianand} 
\affil{Inter-University Centre for Astronomy and Astrophysics, Post Bag 4, Ganeshkhind, Pune 411 007, India}

\author{J.-K. Krogager}  
\affil{Institut d'Astrophysique de Paris, UMR 7095, CNRS-SU, 98bis bd Arago, 75014  Paris, France}

\author{P. Noterdaeme}  
\affil{Institut d'Astrophysique de Paris, UMR 7095, CNRS-SU, 98bis bd Arago, 75014  Paris, France}

\author{A. J. Baker}  
\affil{Department of Physics and Astronomy, Rutgers, the State University of New Jersey, 136 Frelinghuysen Road, Piscataway, NJ 08854-8019, USA}

\author{F. Combes}  
\affil{ Observatoire de Paris, Coll\`ege de France, PSL University, Sorbonne University, CNRS, LERMA, Paris, France}

\author{J. P. U. Fynbo}  
\affil{Cosmic Dawn Center (DAWN), University of Copenhagen, Jagtvej 128, DK-2200, Copenhagen N, Denmark}
\affil{Niels Bohr Institute, University of Copenhagen, Jagtvej 128, DK-2200, Copenhagen N, Denmark}

\author{E. Momjian}  
\affil{National Radio Astronomy Observatory, P.O. Box O, Socorro, NM 87801, USA}

\author{M. Hilton}  
\affil{Astrophysics Research Centre and School of Mathematics, Statistics and Computer Science, University of KwaZulu-Natal, Durban 4041, South Africa }

\author{T. Hussain} 
\affil{Inter-University Centre for Astronomy and Astrophysics, Post Bag 4, Ganeshkhind, Pune 411 007, India}

\author{K. Moodley}  
\affil{Astrophysics Research Centre and School of Mathematics, Statistics and Computer Science, University of KwaZulu-Natal, Durban 4041, South Africa }

\author{P. Petitjean}  
\affil{Institut d'Astrophysique de Paris, UMR 7095, CNRS-SU, 98bis bd Arago, 75014  Paris, France}

\author{H.-W. Chen}  
\affil{Department of Astronomy \& Astrophysics, The University of Chicago, 5640 S Ellis Ave., Chicago, IL 60637, USA}

\author{P. Deka}
\affil{Inter-University Centre for Astronomy and Astrophysics, Post Bag 4, Ganeshkhind, Pune 411 007, India}

\author{R. Dutta}  
\affil{Dipartimento di Fisica G. Occhialini, Universit\`a degli Studi di Milano Bicocca, Piazza della Scienza 3, 20126 Milano, Italy}

\author{J. Jose}  
\affil{ThoughtWorks Technologies India Private Limited, Yerawada, Pune 411 006, India}

\author{G. I. G. J\'ozsa}  
\affil{South African Radio Astronomy Observatory, 2 Fir Street, Black River Park, Observatory 7925, South Africa}
\affil{Department of Physics and Electronics, Rhodes University, P.O. Box 94, Makhanda, 6140, South Africa}
\affil{Argelander-Institut f\"ur Astronomie, Auf dem H\"ugel 71, D-53121 Bonn, Germany}

\author{C. Kaski}  
\affil{ThoughtWorks Technologies India Private Limited, Yerawada, Pune 411 006, India}

\author{H.-R. Kl\"ockner}  
\affil{Max-Planck-Institut f\"ur Radioastronomie, Auf dem H\"ugel 69, D-53121 Bonn, Germany}

\author{K. Knowles}  
\affil{Astrophysics Research Centre and School of Mathematics, Statistics and Computer Science, University of KwaZulu-Natal, Durban 4041, South Africa}

\author{S. Sikhosana}  
\affil{Astrophysics Research Centre and School of Mathematics, Statistics and Computer Science, University of KwaZulu-Natal, Durban 4041, South Africa}

\author{J. Wagenveld}  
\affil{Max-Planck-Institut f\"ur Radioastronomie, Auf dem H\"ugel 69, D-53121 Bonn, Germany}

\begin{abstract}

We present results of an optical spectroscopic survey using SALT and NOT to build a WISE mid-infrared color-based, dust-unbiased sample of powerful radio-bright ($>$200\,mJy at 1.4\,GHz) AGN for the MeerKAT Absorption Line Survey (MALS).  Our sample has 250 AGN (median $z=1.8$) showing emission lines, 26 with no emission lines, and 27 without optical counterparts.    Overall, our sample is fainter ($\Delta i$=0.6\,mag) and redder ($\Delta(g-i)$=0.2\,mag) than radio-selected quasars, and representative of fainter quasar population detected in optical surveys.  About 20\% of the sources are narrow line AGN (NLAGN) -- $65\%$ of these, at $z < 0.5$ are galaxies without strong nuclear emission, and 10\% at $z>1.9$, have emission line ratios similar to radio galaxies.   The farthest NLAGN in our sample is M1513$-$2524 ($z_{em}=3.132$), and the largest radio source (size$\sim$330\,kpc) is M0909$-$3133 ($z_{em}=0.884$). We discuss in detail 110 AGN at $1.9 < z < 3.5$. Despite representing the radio loudest quasars (median $R$=3685), their Eddington ratios are similar to the SDSS quasars having lower $R$.  We detect 4 C~{\sc iv} BALQSOs, all among AGN with least $R$, and highest black hole masses and Eddington ratios. The BAL detection rate ($4^{+3}_{-2}$\%) is consistent with that seen in extremely powerful ($L_{1.4GHz}>10^{25}$\,W\,Hz$^{-1}$) quasars. Using optical light-curves, radio polarization  and $\gamma$-ray detections, we  identify  7  high-probability BL Lacs. We also summarize the full MALS footprint to search for \hi\ 21-cm and OH 18-cm lines at $z<2$.
\end{abstract}


\section{Introduction} 
\label{sec:intro}  

Active Galactic Nuclei (AGN) radiate across the entire electromagnetic spectrum but in their early phases of evolution the thermal emission from the accretion disk especially at UV to optical wavelengths can be obscured by the dusty material fueling the central engine.  Subsequently, as the radiative and mechanical feedback clears the obscuring material the central AGN can become observable as optically selected AGN \citep[e.g.,][]{Sanders96}. The direct view of the accretion disk may also be blocked by a dusty `torus' which has been postulated to explain the appearances of different types of AGN through orientation-based unification schemes \citep[e.g.,][]{Urry95}.  In this paradigm, type\,{\sc i} and type\,{\sc ii} AGN appear different merely due to varying amounts of dust along the observer's line of sight.  The redder AGN are then more obscured simply because they are being viewed at inclinations closer to the equatorial plane of the torus. 
%

In principle, AGN can also be obscured by dust from high-column density \hi\ absorbers such as damped \lya\ systems (DLAs), which are defined to have $N$(\hi)$> 2\times 10^{20}$\cmsq\ \citep[][]{Wolfe00} and associated with intervening galaxies intercepting the line of sight.  Despite the availability of large samples of intervening DLAs \citep[e.g.,][]{Noterdaeme09dla, Parks18} and the fact that they trace the bulk of neutral hydrogen in the Universe \citep[][]{Wolfe05}, the detected DLAs show little dust \citep[e.g.][]{Murphy04} and only a small fraction exhibit the presence of molecular hydrogen  \citep[][]{Petitjean00, Ledoux03, Noterdaeme08, Srianand12dla, Noterdaeme15, Muzahid15, Balashev18, Zahedy20, Boettcher21}.  

Interestingly, none of the DLAs detected to date have properties similar to dense molecular clouds in the Galaxy. In fact, PKS\,1830-211 which exhibits extreme reddening \citep[visual extinction, $A_V > 5.8$;][]{Mathur97} due to an intervening absorber at $z=0.89$ with properties similar to dense molecular clouds, was identified on the basis of its peculiar radio spectrum and morphology \citep{Rao1988,Jauncey1991}.  The redshift $z=0.89$ of the lensing galaxy responsible for the intervening absorption was discovered through a spectroscopically blind search of molecular absorption at millemeter wavelengths \citep{wiklind1996}.  A dust-unbiased census of DLAs is required to correctly estimate the key observables such as \hi\ and metal mass densities of the Universe, and the extent of dust-obscured AGN missed in UV-optical colour based surveys \citep[e.g.,][]{Pei91, Srianand97, Ellison01, Pontzen09, Frank10, Krogager19}.    

Regardless of the origin of obscuration, it is desirable to build dust-unbiased samples of AGN to distinguish between competing paradigms based on evolution or orientation to understand the AGN population itself and its impact on galaxy evolution via feedback \citep[e.g.,][]{Fabian12, Heckman14}. 
The most effective techniques to achieve this are naturally based on photometric selection at hard X-ray, mid-infrared (MIR) or radio wavelengths.  In recent years the AllWISE catalog from the Wide-field Infrared Survey Explorer \citep[WISE;][]{Wright10, Cutri14} in its four bands, namely $W1$ (3.4\,$\mu$m), $W2$ (4.6\,$\mu$m), $W3$ (12\,$\mu$m), and $W4$ (22\,$\mu$m), has emerged as an excellent resource to identify obscured AGN across the full sky. The 5$\sigma$ point-source sensitivities of the WISE All-Sky release in the four bands in Vega-based magnitudes are 16.83, 15.60, 11.32 and 8.0, respectively. 
%

The MIR selection of AGN works exactly in the same manner as the traditional optical color selection except that it is less susceptible to dust extinction.  The near-IR spectrum of an AGN is simply an extension of the UV-optical power law. The MIR and longer wavelength IR emission comes from the hot dust in  the `torus' which absorbs UV-optical emission and reprocesses it \citep[][]{Netzer07}. 
All these give rise to a power law spectrum which dominates at wavelengths longer than $\sim$1\,$\mu m$ and is easily distinguishable from a declining Rayleigh-Jeans spectrum of galaxies and stars in MIR color space up to $z\sim3$. Thus, the MIR color selection can potentially detect obscured AGN  missed in UV-optical and even X-ray selections over a wide redshift range  \citep[e.g.,][]{Lacy04, Stern05, Mateos12, Stern12}.

Taking advantage of the segregation of various classes of objects i.e., stars, early-type and spiral galaxies, brown dwarfs and AGN in the WISE color space \citep[see Fig. 12 of][]{Wright10}, several AGN identification techniques based on the ALLWISE data release have been proposed in the literature.
The majority of these are based on the $W1$ and $W2$ i.e., the two most sensitive bands of WISE \citep[see][for a large all sky catalog of AGN]{Assef18}, and  have been shown to yield AGN space density much higher than those from the optical and X-ray surveys \citep[][]{Stern05, Assef10}. 
The MIR colors corresponding to {\it Fermi} detected AGN have been used to build large samples ($\sim$15,000) of blazars \citep[][]{Dabrusco19}. 
The MIR color diagnostic has also been applied to identify dual AGN candidates which represent a crucial and rare stage in the galaxy evolution \citep[e.g.,][]{Satyapal17}.

In this paper, we present a MIR-color scheme based on $W1$, $W2$ and $W3$ that efficiently selects quasars at $z > 1.4$. 
We have also carried out a large spectroscopic campaign using the Nordic Optical Telescope (NOT) and the Southern African Large Telescope (SALT) to measure the redshifts and confirm the nature of a subset of AGN candidates identified using our MIR-color scheme.  The NOT component of the survey has already been presented by \citet[][]{Krogager18}.
Here we present the complete SALT--NOT catalog of the 303 AGN at  $0<z<3.5$ that constitutes radio sources brighter than 200\,mJy at $\sim$1.4\,GHz in the NRAO VLA Sky Survey \citep[NVSS;][]{Condon98} and at declination, $\delta < +20^\circ$.

The SALT-NOT sample represents a small minority of AGN that are extremely powerful in radio with 1.4\,GHz spectral luminosity, $L_{1.4GHz}$ $>10^{24}$\,W\,Hz$^{-1}$.  In past many studies have attempted to isolate the physical mechanisms responsible for the quasar radio-loudness dichotomy.  The massive black holes (BHs) in radio loud AGN are found to be systematically a few times heavier than those in their radio quiet counterparts \citep[e.g.,][]{Laor00}. A further interesting trend that has emerged from these studies is the anti-correlation between the Eddington ratio and radio loudness \citep[e.g.,][]{Sikora07}.  
The physical parameters such as BH mass, spin and accretion rate along with the in situ magnetic field are believed to determine the extent of radio loudness and overall appearance of the AGN \citep[e.g.,][]{Sikora13}.
The SALT-NOT sample offers an opportunity to revisit these issues for extremely radio loud quasars selected through MIR colors.

This paper is organized as follows.  In Section~\ref{sec:samp}, we present 
details of radio-infrared-optical cross-matching to select candidates for the SALT-NOT survey and the sample selection criteria.   The latter is based on the requirements of the MeerKAT Absorption Line Survey \citep[MALS; see][for key science objectives]{Gupta17mals}, an ongoing large survey at the South African precursor \citep[][]{Jonas16} of the upcoming Square Kilometer Array (SKA).
In Section~\ref{sec:obs}, we present the details of SALT observations and data analysis.  These details for the NOT component of the survey are available in  \citet[][]{Krogager18}. 
The details of emission line identification, redshift measurement and AGN classifications based on the SALT-NOT spectra are provided in Section~\ref{sec:spec}.
Section~\ref{sec:res} presents selected results based on radio continuum and optical emission line properties.  For AGN at $1.9<z<3.5$, we also estimate BH masses, accretion rates and the occurrence of broad absorption lines, and revisit the issue of their dependence on radio loudness. We also discuss emission line less AGN with highest optical photometric variability and the targets with no optical counterparts.  
Finally, in Section~\ref{sec:summ} we summarize the results and complement the SALT-NOT catalog with AGN from the literature to assemble a target list of $\sim$650 AGN defining the MALS footprint to search for cold atomic and molecular gas at $z<2$. 

Throughout this paper we use the $\Lambda$CDM cosmology with $\Omega_m$=0.315, $\Omega_\Lambda$=0.685 and H$_{\rm 0}$=67.4\,\kms\,Mpc$^{-1}$ \citep[][]{Planck20}.

\section{Sample Selection}      
\label{sec:samp}   

The SALT-NOT sample definition is based on the requirements of MALS. MALS  is using MeerKAT's L and UHF bands covering 900--1670 and 580--1015\,MHz to carry out a sensitive search for \hi\ 21-cm (OH 18-cm) absorbers at  $0 \le z \lesssim 0.6$ ($0 \le z \lesssim 0.9$) and $0.4 \lesssim z \lesssim 1.4$ ($0.6 \lesssim z \lesssim 1.9$), respectively \citep[see][for key science objectives]{Gupta17mals}. 
In order to ensure reasonable observability with the MeerKAT telescope located in South Africa, the majority of MALS pointings will be at  $\delta < +20^\circ$.
The survey is well underway and the first L- and UHF-band spectra based on the science verification data are presented in \citet[][]{Gupta21} and \citet[][]{Combes21}, respectively.  

Each MALS pointing is centered at a radio source brighter than 200\,mJy at $\sim$1\,GHz.
For L-band we require these bright AGN to be at $z>0.6$ whereas for the UHF-band all must be at $z>0.4$ and a significant fraction at $z>1.4$, i.e., the \hi\ 21-cm line redshift corresponding to the lower frequency edge of the band.

Since dust bias can be very significant towards sight lines with cold atomic and molecular gas, it is also desirable that the central targets for MALS are selected without any bias towards dust extinction. 
Thus, to efficiently identify quasars that can be used for both the L- and UHF-bands we have developed a selection scheme optimized to identify $z>1.4$ quasars based primarily on MIR colors from the AllWISE catalog.   
The WISE color space of radio sources identified as quasars in the SDSS spectroscopic catalog is presented in Fig.~2 of \citet[][]{Krogager18}.   
This figure shows that low- and high-$z$ quasars are separated in $W1$, $W2$ and $W3$ color space, and the following color cuts:
\begin{equation}
	\begin{array}{c}
W1 - W2 < 1.3\times(W2 - W3) - 3.04; \\
\\
W1 - W2 > 0.6,
\\
\end{array}
\label{eqwise}
\end{equation}
can be used to efficiently select high-$z$ quasars. 

To summarize, SALT--NOT sample is based on the following three criteria: {\it (i)} flux density $>$ 200\,mJy at $\sim$1\,GHz,  {\it (ii)}  $\delta < +20^\circ$, and {\it (iii)} MIR colors consistent with Equation~\ref{eqwise}.  The last condition requires that the object is detected in the first three WISE bands. Note that only $\sim$10\% of 747 million objects are detected in all three bands.
We revisit the efficacy of the MIR-color selection scheme in subsequent sections. 

\subsection{Cross-matching and reference sample}      
\label{sec:cross}   
%
\begin{figure} 
\centerline{\vbox{
\centerline{\hbox{ 
\includegraphics[trim = {0cm 2cm 0cm 2cm}, width=0.5\textwidth,angle=0]{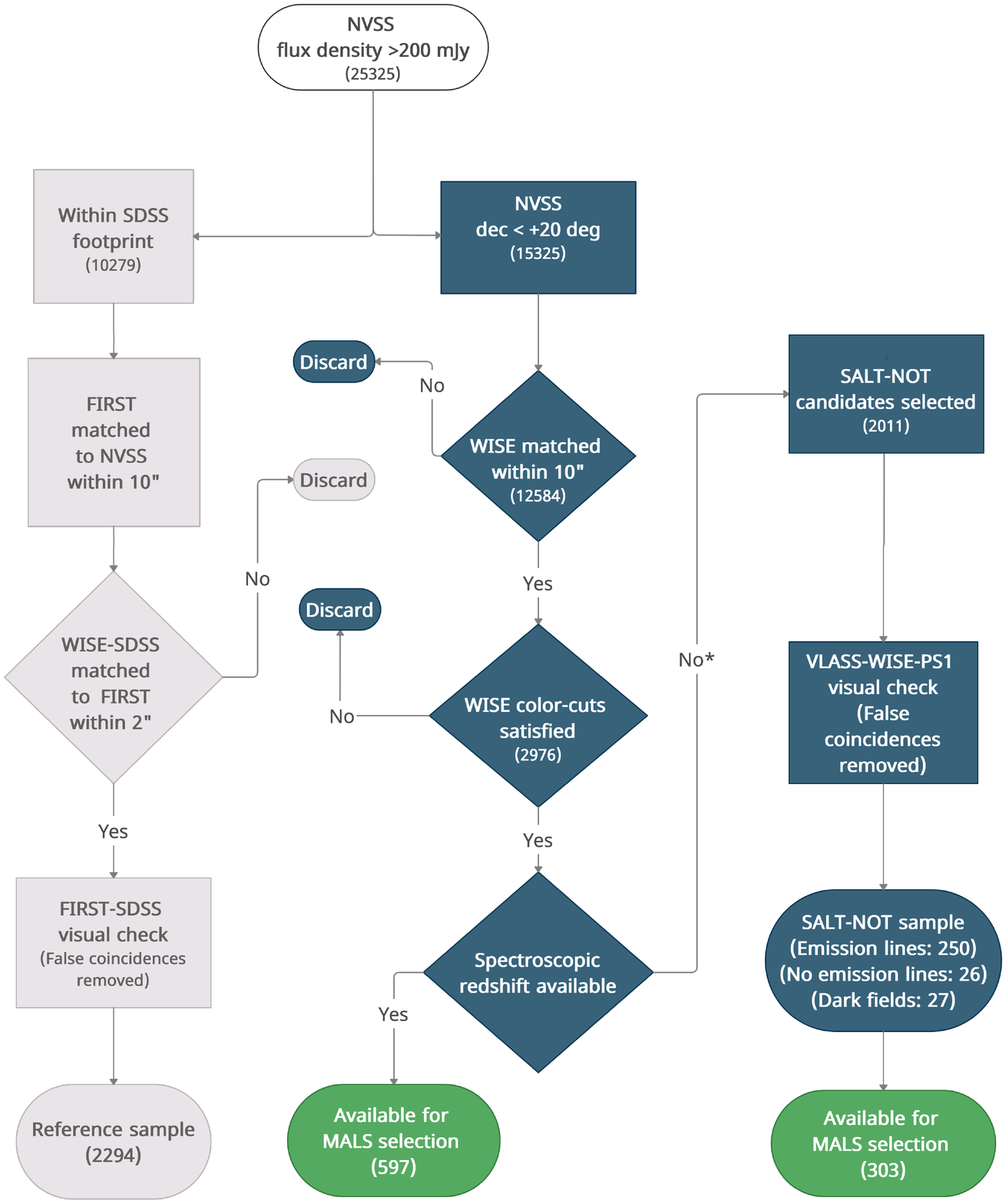}  
}} 
}}  
\vskip+0.0cm  
\caption{
	Schematic illustration of the selection process for the quasar {\it reference} sample and SALT--NOT targets. The numbers refer to sources at each step. 
	The `$\star$' indicates that a handful of targets with spectra in SDSS were included in the SALT--NOT survey to verify our selection and observational strategy. See Sect.~\ref{sec:spec} for more details.
} 
\label{fig:xmatch}   
\end{figure} 

Cross-matching multi-wavelength datasets poses multiple challenges.  Firstly, not all objects may be detected in all the datasets.  Secondly, due to different spatial resolutions of datasets, an object in one dataset may have multiple counterparts in the other.  This is particularly relevant for radio sources which often display extended emission made up of components such as core, jets and lobes. Depending on the frequency the strongest radio component may be far away from the optical or IR counterpart.  For reliable cross-matching it is desirable to have radio data with a spatial resolution of few arcseconds or better.  
%
\begin{figure*} 
\centerline{\vbox{
\centerline{\hbox{ 
\includegraphics[trim = {0cm 0cm 0cm 0cm}, width=1.0\textwidth,angle=0]{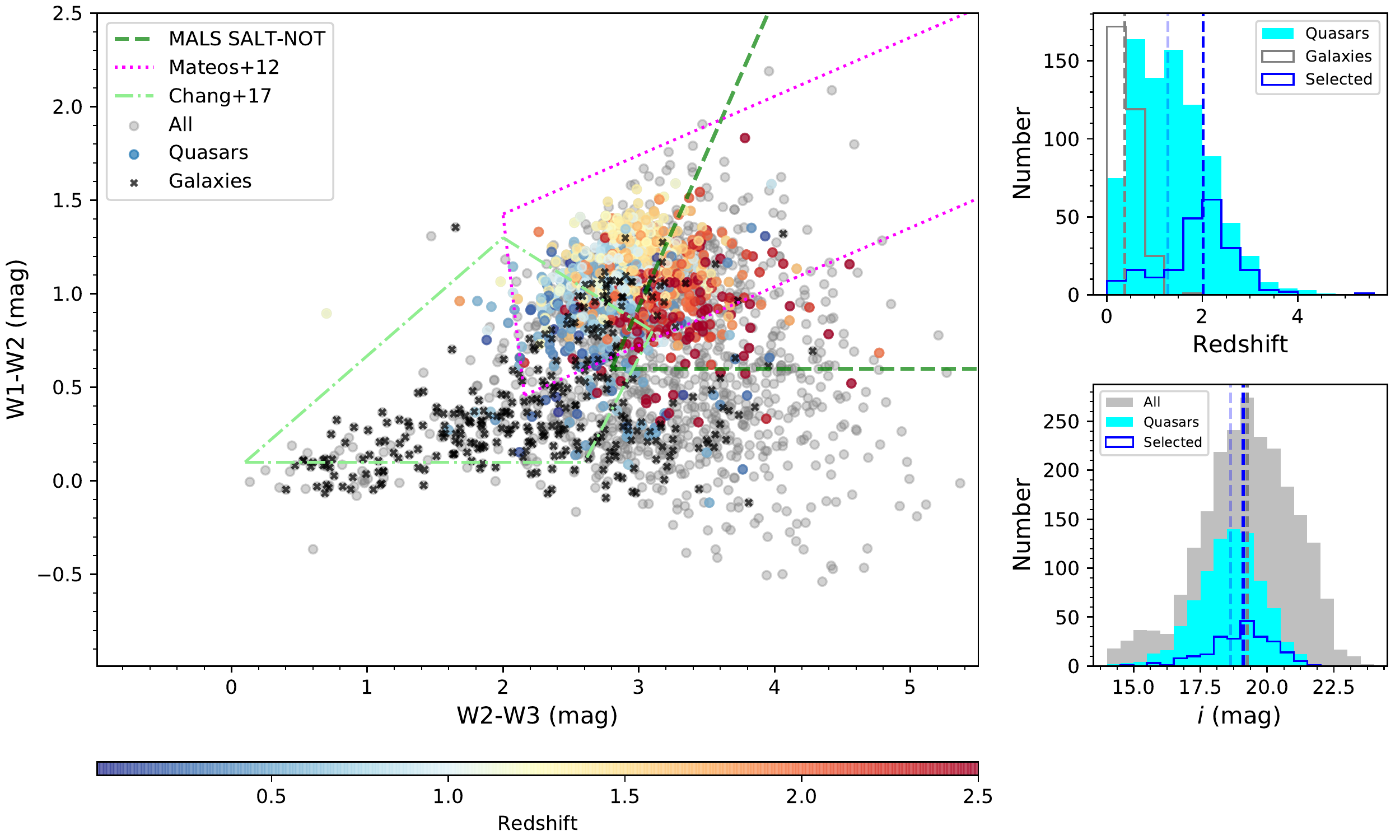}  
}} 
}}  
\vskip+0.0cm  
\caption{
Left: WISE color-color plot in Vega magnitudes for radio bright sources in the quasar {\it reference} sample.  Sources with known spectroscopic redshifts are color-coded according to redshift. The AGN wedges of \citet[][]{Mateos12}, \citet[][]{Chang17} and our SALT--NOT criteria are shown using dotted, dash-dotted and dashed lines, respectively.  	
Right, top panel:  Redshift distributions of quasars, galaxies and WISE-selected quasars (i.e., quasars satisfying SALT--NOT WISE color cuts). Right, bottom panel: Distributions of SDSS $i$-band AB magnitude for all sources, quasars and WISE-selected quasars are shown.  The vertical dashed lines indicate the median for each distribution. Note that SDSS 5$\sigma$ depth for the $i$-band is 22.2\,mag.
} 
\label{fig:refsamp}   
\end{figure*} 

Another major challenge is that each selection method has its own systematics.  For quasars, it is quite possible that selections using different methods may lead to objects with different characteristics such as different Eddington ratios, clustering properties and host galaxy masses.  The selected quasars may then be intrinsically different,  therefore  comparison with samples from other selection approaches ought to be done with caution. 

Concerning our sample definition, only\footnote{The Rapid ASKAP Continuum Survey \citep[][]{Mcconnell20} was unavailable at the time of planning this project.} the NRAO VLA Sky Survey \citep[NVSS;][]{Condon98} and the Sydney University Molonglo Sky Survey \citep[SUMSS;][]{Mauch03} at 1.4\,GHz and 0.843\,GHz, respectively, have the sensitivity, positional accuracy and the sky coverage required to perform cross-matching with the AllWISE catalog \citep[][]{Cutri14} at MIR wavelengths. But the spatial resolution offered by NVSS is typically $45$\arcsec.
In comparison, the angular resolutions in the four WISE bands are $6.1$\arcsec, $6.4$\arcsec, $6.5$\arcsec, and $12.0$\arcsec, respectively. 
The coarse resolution at radio wavelength might then lead to a bias against extended lobe-dominated quasars: either they will be assigned a wrong WISE counterpart or completely missed.
To quantify the impact of this and the other above-mentioned challenges in our selection process, we start with constructing a {\it reference} sample.\\  

The process of building the {\it reference} sample is shown in Fig.~\ref{fig:xmatch}.  We start with NVSS which has 25,325 radio sources brighter than 200\,mJy at 1.4\,GHz and $\delta > -40^{\circ}$.  We use optical data from SDSS and higher spatial resolution (i.e $\sim$5") radio data from the Faint Images of the Radio Sky at Twenty-Centimeters \citep[FIRST;][]{Becker95} to examine our selection process.  The FIRST survey has observed $\sim$10,000 square degrees of the North and South Galactic Caps at 1.4\,GHz, which overlaps very well with the SDSS sky coverage.  These two surveys provide the best available radio-optical data covering a large area of the sky.  Since the bulk of FIRST-SDSS coverage is at positive declination, we do not apply any declination cuts to the {\it reference} sample.  
%

We cross-match 10,279 NVSS sources which are within the SDSS footprint with FIRST.  We adopt a search radius of $30$\arcsec.  Since FIRST has a spatial resolution of $5$\arcsec\ and positional accuracy of $0.1$\arcsec, it will allow us to identify the resolved radio source components associated with the NVSS sources.
Next, we cross-match FIRST radio source components with SDSS (DR14 photometry) and AllWISE catalogs to identify the closest match. Note that we require a valid WISE match to have detections in all three bands, i.e., $W1$, $W2$ and $W3$. For this we adopt a search radius of $2$\arcsec, which has been shown to provide a good compromise between completeness and random association for FIRST--SDSS quasars \citep[e.g.,][]{Lu07}.  
We visually examined all 2327 FIRST--SDSS associations using the FIRST and SDSS $i$-band images, and excluded: ({\it i}) 8 unreal radio source components: these are due to an artefact in the FIRST image; and ({\it ii}) 12 radio sources with complex morphology: for these it is not possible to reliably assign an optical--MIR counterpart to the radio source. 
Spectra of 1164 sources are available in SDSS.  We inspect these and reject 13 cases with false spectral line identifications.
Finally, we have a {\it reference} sample of 2294 sources from NVSS, with WISE detections in $W1$, $W2$ and $W3$ bands, and cross-matching further refined by FIRST and SDSS.

The WISE colors of the {\it reference} sample are shown in the {\it left} panel of Fig.~\ref{fig:refsamp}.  Among 1151 reference sample sources with SDSS spectra, 834 are identified as quasars and 317 as galaxies. 
Clearly, as expected the majority of radio bright quasars have red $W1 -W2$ colors.  Only beyond $z\sim3$, the rest frame 1$\,\mu$m minimum would shift into the $W2$ band causing the color to turn blue \citep[see ][]{Stern12}. But for high-$z$ quasars this can be compensated by even small amounts of dust, allowing the method based only on $W1 - W2$ color to perform rather well even at high redshifts.
However, significant host galaxy contribution can also turn $W1 - W2$ blue.
In the figure, we show the color wedge of \citet[][]{Mateos12} defined based on X-ray selected AGN.  The additional constraint based on $W2 - W3$ allows the wedge of  \citet[][also see their Fig.~5]{Mateos12} to efficiently identify AGN as long as the host galaxy contribution is less than 20\%. Indeed the wedge efficiently identifies the bulk of powerful {\it reference} quasars.

In Fig.~\ref{fig:refsamp}, we also demarcate the region corresponding to Equation~\ref{eqwise}. The redshift and $i$-band magnitude distributions of the reference sample are shown in the {\it right} panels of Fig.~\ref{fig:refsamp}.  
The median redshift and $i$-band magnitude are 0.9 and 18.6\,mag, respectively.
If we consider only quasars, these values are 1.3 and 18.6\,mag, respectively.  For  quasars within our MIR-wedge these values are 2.0 and 19.1\,mag, respectively (see vertical dashed lines in Fig.~\ref{fig:refsamp}).
Also, within the wedge the fraction of AGN classified as quasars increases to 94$\pm$6\%  (214/228) whereas for the full sample it is 73$\pm$3\% (834/1151). 
The fraction of $z>1.4$ quasars also increases from 30\% (full sample) to 75\% (within the wedge).   
Thus, our MIR-wedge preferentially selects more powerful and higher redshift quasars.

\subsection{SALT-NOT candidate sample}      
\label{sec:cands}   

The process of identifying candidates for SALT--NOT survey is shown in Fig.~\ref{fig:xmatch}.  We start with 25,325 sources brighter than 200\,mJy in NVSS but first restrict ourselves to 15,325 at $\delta < +20^\circ$ and then to 12,584 with WISE matches within $10$\arcsec.  Since we use NVSS all the candidates are at $\delta > -40^\circ$.
Next we consider only 2976 sources that satisfy our WISE-color cuts.  Of these, 597 sources ($\sim$340 at $z>1.4$) have spectroscopic redshifts from the literature\footnote{Nasa/IPAC Extragalctic Database (NED) and the Million Quasars (Milliquas) Catalog v4.0}, and 
will be used later for MALS target selection (see Section~\ref{sec:summ}).
For the SALT-NOT survey, we exclude these and consider the remaining 2011 radio sources which have a high probability of being a quasar at $z>1.4$.

Next, targets were drawn from the pool of 2011 candidates to fill the allocated and observable local sidereal time (LST) ranges at the NOT and SALT locations.  The final sample of 303 candidates which could be observed in the allocated time is presented in Table~\ref{tab:wisesamp}.
The sample has been subjected to visual checks using the quick look radio images at 2--4\,GHz from the Very Large Array Sky Survey \citep[VLASS; ][]{Lacy20}.  These quick look images have a spatial resolution of $\sim2.5$\arcsec\ but are not the final data products meeting the overall survey requirements.  The positional accuracy is limited to $\sim0.5$\arcsec.  Therefore, we used these images only to verify the coincidence between the MIR source and the radio peak.  In the cases where a VLASS image was unavailable or not of reasonable quality, the corresponding candidate was assigned a lower priority. Following the process described in Sect.~\ref{sec:cross}, we have rejected dubious coincidences which primarily meant rejecting sources with complex radio morphologies.  It was straightforward to verify and accept the compact radio sources (i.e., single component in VLASS), and indeed our overall selection process preferentially picks quasars associated with compact radio emission.  But wherever possible we also included targets with clear core-jet or compact symmetric object (CSO) type morphology.  In both these cases it is possible to identify the location of MIR/optical AGN. The details of radio structure of AGN observed with SALT-NOT are presented in Sect.~\ref{sec:res}.

\begin{figure*} 
\centerline{\vbox{
\centerline{\hbox{ 
\includegraphics[trim = {0cm 0cm 0cm 0cm}, width=0.85\textwidth,angle=0]{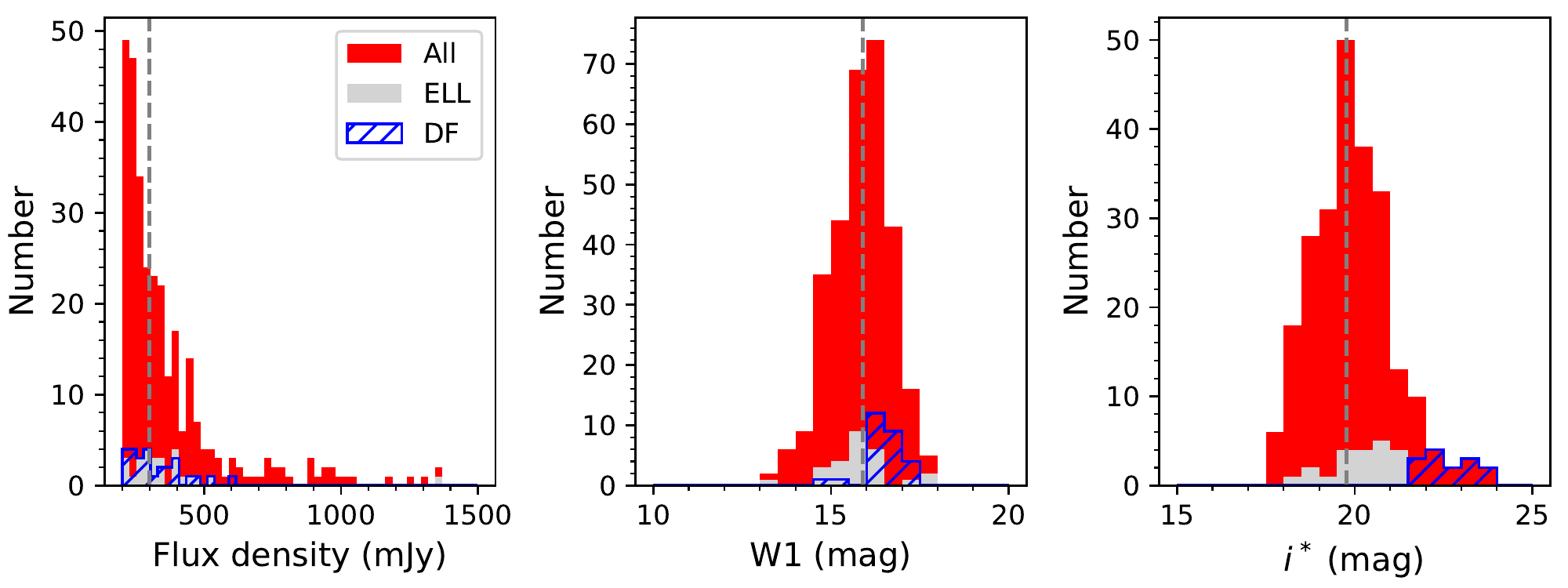}  
}} 
}}  
\vskip+0.0cm  
\caption{
Distribution of flux density at 1.4\,GHz, $W1$ and $i$-mag for our MIR selected sample. The gray and blue-hashed distributions show objects with no clear emission lines (emission line less, ELL) and objects with no detected spectrum (dark field, DF), respectively.  The vertical dashed lines mark the median for all the sources in each panel. In the first panel, four targets brighter than 1500\,mJy are out of frame. Note that for DFs, the $i$-band magnitudes are from the more sensitive PS1 stack catalog \citep[][]{Waters20}. 
} 
\label{fig:fluxw1mag}   
\end{figure*} 

For candidates at $\delta > -30^\circ$, we have also examined the $i$-band images from  the Panoramic Survey Telescope and Rapid Response System \citep[PanSTARRS;][]{Chambers16} by considering the nearest match within 2\arcsec.  Ideally, in the pursuit of dust-obscured quasars one does not include any criteria based on optical wavelengths. 
Therefore, we limited the majority of our targets to those that could be successfully observed with SALT and NOT in a modest amount of observing time but included a small ($\sim10$\%) fraction with very faint ($i > 22$\,mag) optical counterparts.  
Note that in Table~\ref{tab:wisesamp} we have excluded 7 targets (list provided in the notes to Table~\ref{tab:wisesamp}) from the NOT sample presented in table~1 of \citet[][]{Krogager18}. The VLASS--WISE--PS1 visual check revealed that the lack of spectroscopic outcome in these cases is due to the misidentification of the optical/IR source associated with the radio source.  Such misidentifactions have already been removed from the sample for the SALT component.

The sample presented in Table~\ref{tab:wisesamp} is split into three categories: ({\it i}) with emission lines in the optical spectrum (250 objects), ({\it ii}) with no emission lines in the optical spectrum i.e., emission line less (ELLs; 26), and ({\it iii}) dark fields (DFs; 27) i.e., neither emission line nor a continuum source in optical spectra and images, respectively despite being detected in WISE images. 
The distributions of 1.4\,GHz flux density, $W1$ and $i$-band magnitudes are shown in Fig.~\ref{fig:fluxw1mag}.  
Further details and comparison with the {\it reference} sample are provided in subsequent sections.
%

\subsection{Efficacy of selection process}      
\label{sec:effic}   

\begin{figure} 
\centerline{\vbox{
\centerline{\hbox{ 
\includegraphics[trim = {0cm 0cm 0cm 0cm}, width=0.5\textwidth,angle=0]{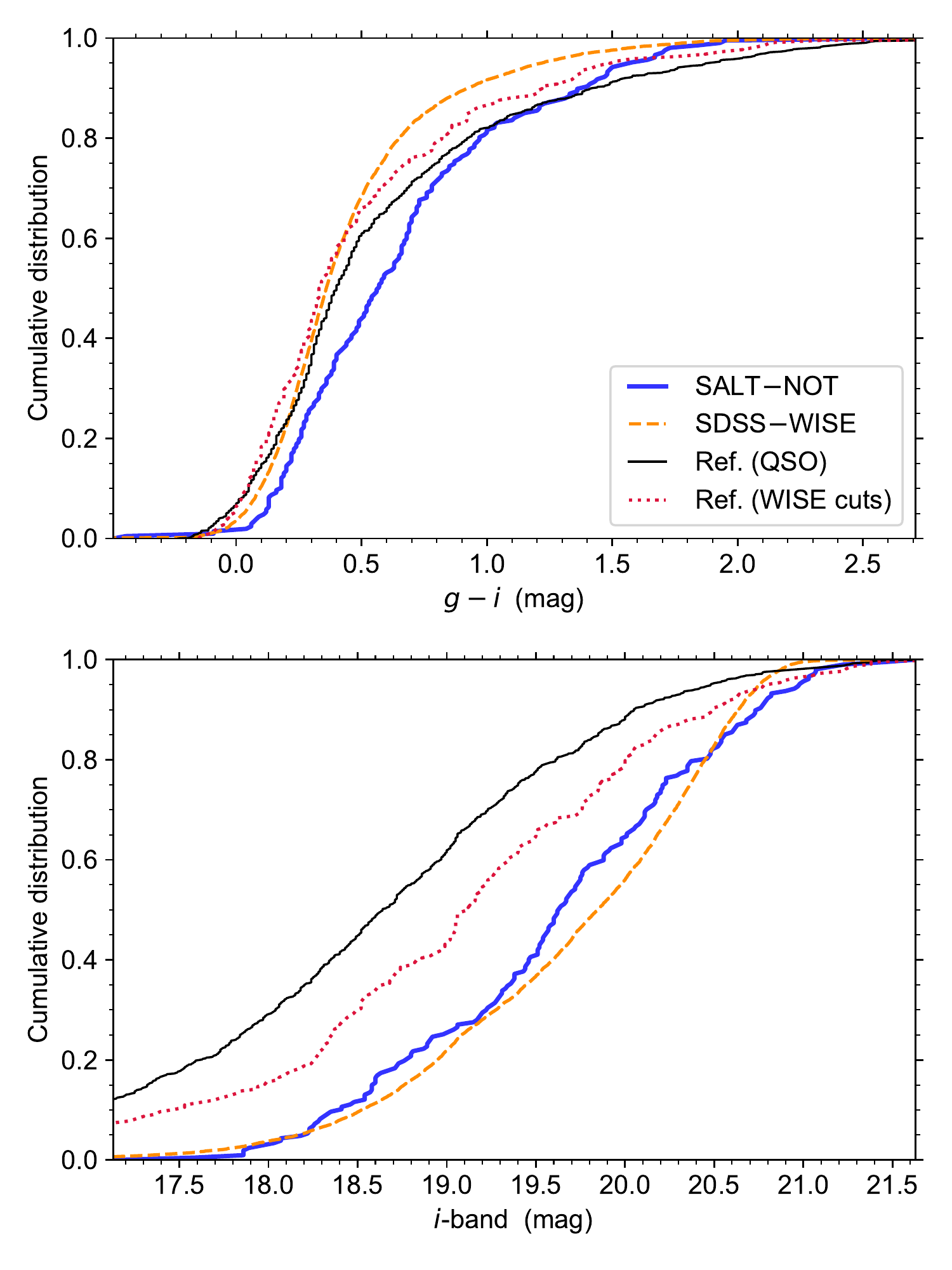}
}} 
}}  
\vskip+0.0cm  
\caption{
	Comparison of the MALS SALT--NOT sample (blue, thick line) with the full quasar {\it reference} sample (black, thin line), the quasar {\it reference} sample that fulfils Eq.~(\ref{eqwise}; red, dotted line), and the SDSS+WISE quasar comparison sample (orange, dashed line). 
} 
\label{fig:compar}
\end{figure} 

As previously mentioned, our target selection process is optimized for identifying powerful quasars at $z>1.4$. We have therefore included a set of criteria on the WISE photometry (see Eq.~\ref{eqwise}). The mere fact that we require detections in all three WISE bands might introduce a bias in the resulting quasar sample, and the subsequent color criteria might further affect their optical properties. In this section, we aim to quantify any such selection effects.

For this purpose, we compile a highly complete spectroscopic sample of 10,498 quasars from the SDSS (DR16) in the so-called `Stripe 82' region where the spectroscopic completeness is the highest in SDSS. We then cross-match this sample with the AllWISE catalog to produce our optical comparison sample.  We note that although no radio flux density limits have been applied here, only 347/10,498 are detected in FIRST, and only 7 of these are brighter than 200\,mJy. We will ignore this distinction in our statistical analysis. 

We first check the effect of requiring detections in all three WISE bands used in our criteria. We find that out of the 10,498 quasars, only 56\% have WISE photometry in all three bands. 
Overall, the quasars {\it with} WISE detections are brighter and have slightly redder colors compared to the quasars without WISE detections in all three bands. The median $i$-band brightness is 19.9 and 20.4~mag for quasars with and without WISE detections, respectively, and the median $g-i$ color differs by 0.034~mag (Fig.~\ref{fig:compar}). When restricting the color comparison to quasars with the same brightness ($i<20$~mag), the color difference diminishes to 0.023~mag; i.e., at the order of the photometric uncertainty.

The average redshift of quasars with WISE detections is lower ($<z>=1.48$) than quasars without WISE detections ($<z>=1.86$), which is in line with the brighter $i$-band magnitudes for quasars with WISE detections. Yet, there are no significant color differences as a function of redshift.

The next step is to quantify the effect of our WISE color criteria. We use the same optical comparison sample as defined above to check the optical brightness and colors of those quasars that meet the criteria in Eq.~(\ref{eqwise}) and those that do not. Our WISE criteria have been designed to reduce the number of $z<1.4$ quasars, and as shown in Fig.~\ref{fig:refsamp}, our sample effectively reduces the number of low-redshift quasars. Due to the largely different redshift-distributions, it is not surprising to find that the quasars that follow Eq.~(\ref{eqwise}) are 0.4~mag fainter on average. Considering only quasars at $z>2$, the quasars that follow our WISE criteria Eq.~(\ref{eqwise}) are on average only 0.2~mag fainter. We find no significant difference in optical color between quasars that meet our criteria and those that do not.

Lastly, we compare the SALT--NOT sample with the {\it reference} sample of quasars as well as the optically-selected SDSS quasar sample. In Fig.~\ref{fig:compar}, we show the cumulative distributions of $g-i$ color and $i$-band brightness. We find that the SALT--NOT sample has on average redder colors than the other samples, with the optically-selected sample showing the bluest colors. The median color difference is 0.2\,mag.
On the other hand, the SALT--NOT sample is fainter than the {\it reference} quasar sample and shows a distribution of optical brightness similar to the SDSS quasar sample.

In conclusion, we find that our imposed WISE criteria result in a sample of radio-bright quasars that is fainter and slightly redder than  radio-selected ({\it reference}) quasars. Part of this difference in color and brightness can be attributed to the difference in the respective redshift distributions. Nonetheless, our sample is representative in terms of redder optical colors commonly observed in radio-selected samples while probing fainter quasars more representative of deep, optically-selected samples.

\section{Observations and data analysis}      
\label{sec:obs}   

We observed 255 and 99 targets with SALT and NOT, respectively.  Of these, 40 targets, labelled as SUGAR (NOT-SALT) in column\,7 of Table~\ref{tab:wisesamp}, are common to both the sub-surveys. 
In this section, we present details of SALT observations and data analysis. The corresponding details for NOT observations carried out in 2016 August ({\tt P53-012}) and 2017 February ({\tt P54-005}; PI: Krogager) are presented by \citet[][]{Krogager18}.  

We used the Robert Stobie Spectrograph \citep[RSS; ][]{burgh2003,kobul2003} on SALT to obtain optical spectra of 255 targets.
These service mode observations were carried out between November 2014 and April 2017 through the programmes: {\tt 2014-2-SCI-027}, {\tt 2015-1-SCI-012}, {\tt 2015-2-SCI-030}, {\tt 2016-1-SCI-022} and {\tt 2016-2-SCI-17} (PI: N. Gupta). Each observing block typically contained two $\sim$10 minutes science exposures, in the long-slit mode having a slit width of 2\arcsec. We note that the calibration lamp and the flat field images required for the wavelength calibration and flat fielding were obtained separately, as part of {\tt 2015-1-SCI-012}. The detector in RSS consists of a mosaic of three CCDs with a total size of 3171$\times$2052 pixels.  The pixel size is 15\mum\ which corresponds to a spatial sampling of $0.1267$\arcsec\ per pixel. We opted for 2 \cross\ 2 pixel binning to improve the signal-to-noise ratio (SNR). We used RSS PG0900 grating with a grating tilt angle of 15.875\degree. This provides a wavelength coverage of 4486--7533~\AA, but with gaps at 5497--5551 and 6542--6589~\AA. The resulting spectral resolution is $R=1064$ at 6041~\AA. The observations were performed on clear nights with a median seeing of $\sim1.5$\arcsec.  The fainter targets were assigned higher priority for observations on dark and gray nights. 

The preliminary data reduction, i.e., gain correction, overscan bias subtraction, cross-talk correction and amplifier mosaicing were performed by the SALT observatory staff using the semi-automated pipeline PySALT\footnote{\href{http://pysalt.salt.ac.za}{http://pysalt.salt.ac.za}}, the Python/PyRAF software package for SALT data reduction \citep[][]{Crawford2010}. Using IRAF\footnote{IRAF is distributed by the National Optical Astronomy Observatories, which are operated by the Association of Universities for Research in Astronomy, Inc., under cooperative agreement with the NSF.} packages, we further removed cosmic rays and applied the flat-field correction. The wavelength calibration was performed using a Xenon arc lamp and the spectrophotometric standard star L377 was used for flux calibration.
We extracted 1D spectra from individual exposures after subtracting the background.  We also corrected the 1D spectra for atmospheric extinction and applied the air to vacuum correction to the wavelengths. The final 1D spectra of the quasars used for analysis is the combination of all the available exposures (typically $\sim$20\,min on-source).

\section{Spectral analysis}      
\label{sec:spec}   

\begin{figure*} 
\centerline{\hbox{ 
\includegraphics[viewport=23 162 565 690, clip=true,width=0.98\textwidth,angle=0]{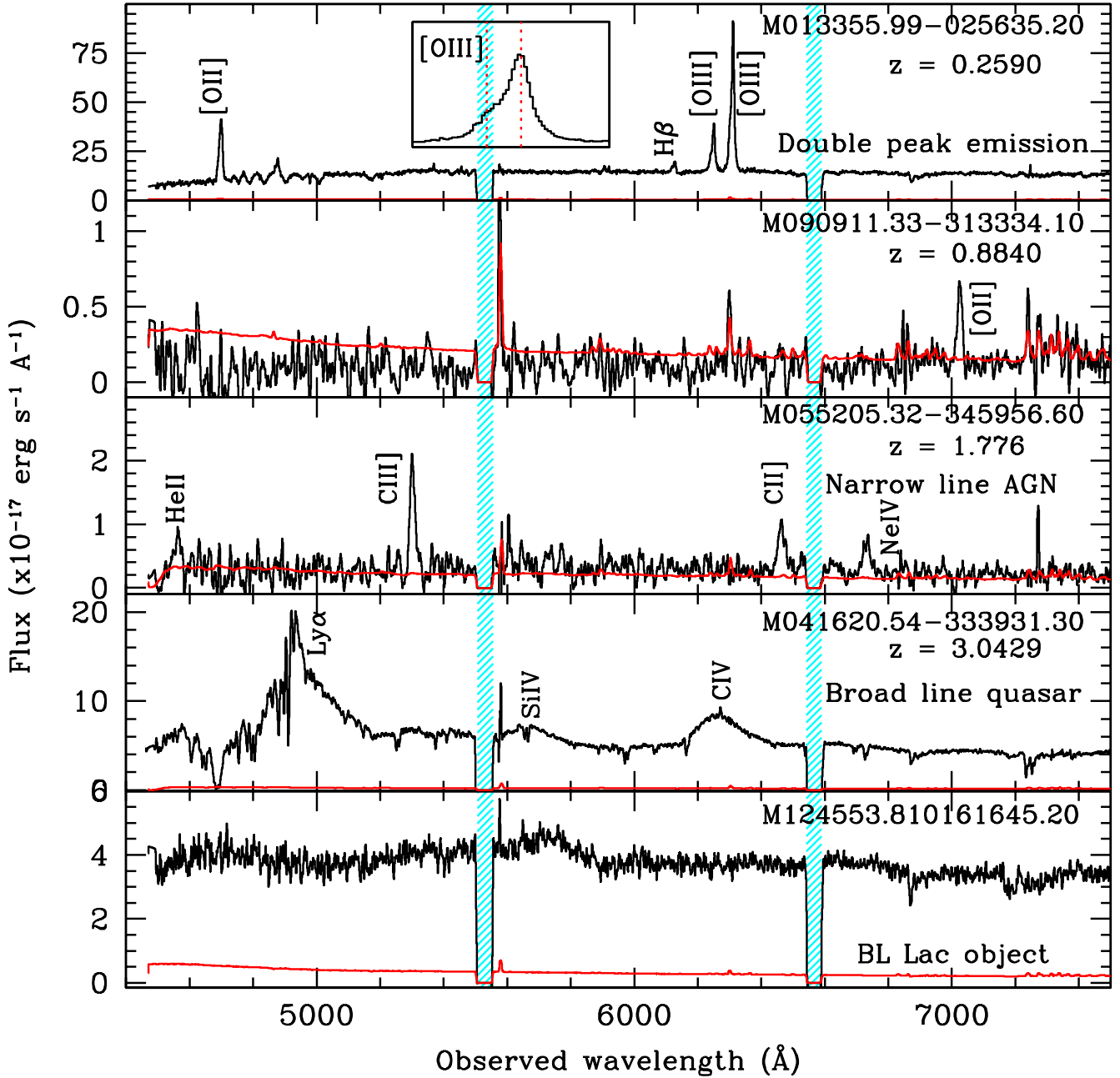}
}}  
\vskip+0.0cm  
\caption{
Spectra of five representative AGN from our sample at increasing redshifts from top to bottom. 
Different emission lines based on which the redshift was determined are identified.
All the spectra are smoothed by 3 to 7 pixels for visual clarity. The error spectra are shown in red in each panel. The hashed regions mark the spectral range falling in the CCD gaps. Inset in the top most panel shows the double peak [O~{\sc iii}]$\lambda$5007 profile. In the bottom panel an example of AGN without detectable emission lines is shown. We use photometric variability to identify such objects as  BL Lacs (see Section~\ref{sec:emless}).
} 
\label{fig:emission}   
\end{figure*} 

We visually examined all the spectra for the presence of emission lines. In Fig.~\ref{fig:emission}, we show 5 representative SALT spectra.  
These represent: low-$z$ ($z<1.0$; top two panels), intermediate-$z$ ($1.0<z<1.9$; middle panel), high-$z$ ($z>1.9$; second panel from the bottom) and the emission line less (ELL; bottom panel) AGN.  
Flux scale for each spectrum is corrected for slit losses using 
the PS1 photometry. Here, we have ignored the possibility of continuum variability between our spectra and the PS1 observations.

\begin{figure} 
\centerline{\vbox{
\centerline{\hbox{ 
\includegraphics[trim = {0cm 0cm 0cm 0.0cm}, width=0.4\textwidth,angle=0]{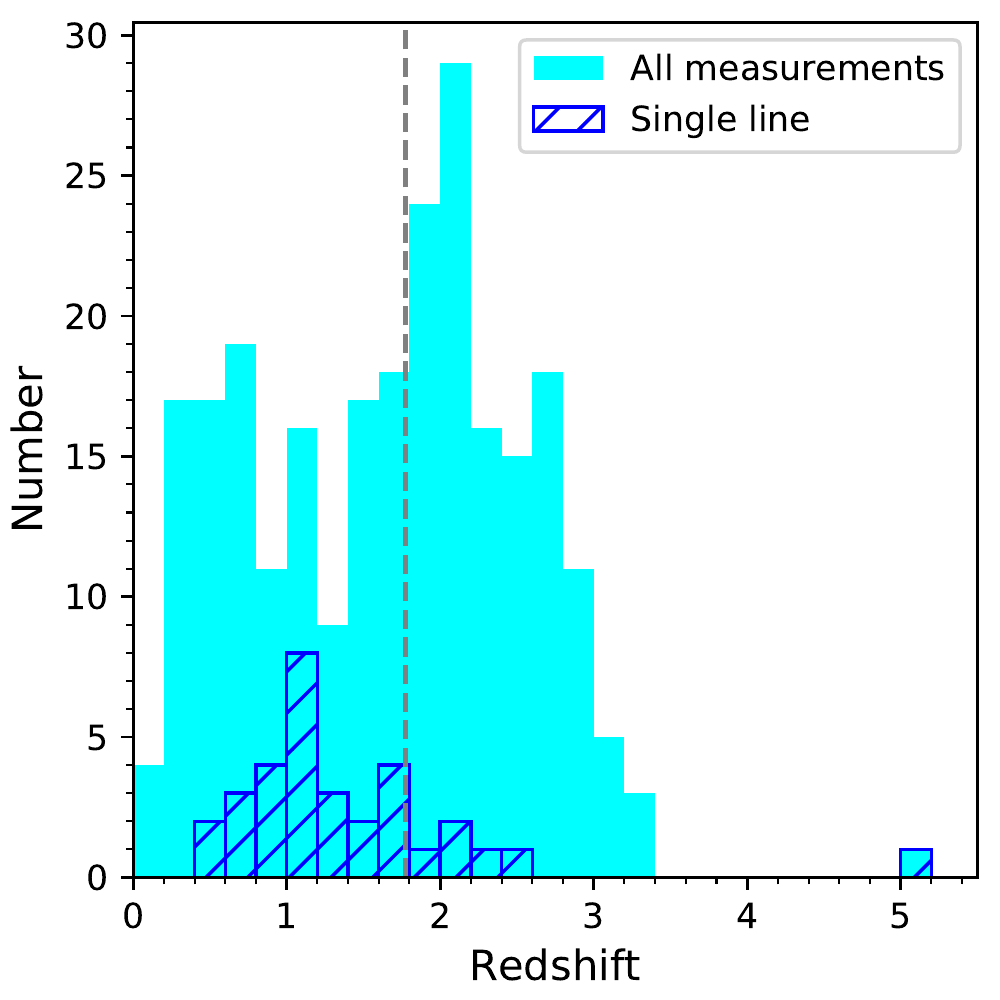}
}} 
}}  
\vskip+0.0cm  
\caption{
Redshift distribution for all the SALT--NOT targets with spectroscopic redshifts.
	The vertical dashed lines mark the median value. The distribution of redshifts based on single emission line (blue-hashed histogram) is also shown.  The redshifts based on a single emission line should be treated with caution. The correct redshift of AGN marked at $z\sim5$ is $z=0.977$ (see Section~\ref{sec:z5} for details). 
} 
\label{fig:zdist}   
\end{figure} 

For the spectra exhibiting two or more emission lines, the line identifications are robust, and hence the redshifts estimated are also correct within the wavelength uncertainty. However, the line identification is more challenging in the cases with very faint trace/emission lines or with only single emission line detection (see second panel from top in Fig~\ref{fig:emission} for example). 
For such sources the identification is based on line widths, expected relative strengths and the absence of other lines within the observed wavelength range. 
We also used various AGN (QSO, Seyfert1 and Seyfert2) templates from the literature\footnote{\href{https://archive.stsci.edu/hlsps/reference-atlases/cdbs/grid/agn}{https://archive.stsci.edu/hlsps/reference-atlases/cdbs/grid/agn}} to aid the process of line identification.

After line identification, we used peaks of prominent lines, i.e., \lya\  for $z>2.66$, \civ\ for $1.87<z<3.86$, \mgii\ for $0.58<z<1.69$, \ciii\ for $1.328<z<2.938$, [O~{\sc ii}] for $0.19<z<1.01$ and, [O~{\sc iii}] and H$\beta$ for $z<0.50$, to measure the redshifts.  The rest wavelengths are taken from the SDSS emission line list\footnote{\href{http://classic.sdss.org/dr6/algorithms/linestable.html}{http://classic.sdss.org/dr6/algorithms/linestable.html}}. 
In the cases where the peak of the prominent emission line is affected by absorption or it falls in the CCD gap, we fit the emission line avoiding these regions and take the centroid of the fit to estimate the redshift. 
The overall uncertainty in redshift ($\Delta z$) that is introduced due to wavelength calibration issues and uncertainty in measuring the peak is roughly  $\Delta z$ $\approx 0.005$.

The redshift distribution of all the AGN (250) with emission line identifications is shown in Fig.~\ref{fig:zdist}.  In 32 cases the redshift is based on a single emission line assumed to be \civ, \ciii, \mgii, \oii\ or \lya.  In Table~\ref{tab:wisesamp}, these are marked with *a, *b, *c, *d and *f, respectively.
In 26 cases, no emission lines (i.e., ELLs) are detected in the spectra.  In 27 cases, no continuum or emission lines are detected (i.e., DFs).

\section{Results}      
\label{sec:res}   

The properties of the SALT--NOT sample consisting of 250 quasars with emission lines, 26 ELLs and 27 DFs are provided in Table~\ref{tab:wisesamp}. 
In this section, we present selected results based on continuum and emission lines detected in our optical spectra. It is well beyond the scope of this paper to discuss all the aspects in detail.  Hence, we present overall properties of the observed sample and comparison with the {\it reference} sample in Sect.~\ref{sec:overall}.  This is followed by short overviews of AGN detected at $z<3.5$ split across Sections~\ref{sec:zl0p5} to \ref{sec:zg2} on the basis of different groups of observable emission lines. For AGN at $z<1.9$, we discuss only the fraction of narrow emission line objects (type II quasars or radio galaxies) found in different redshift intervals and the morphology of associated radio emission.  For quasars at $1.9 < z <3.5$, we also present the distribution of BH masses and the fraction of broad absorption line (BAL) quasars.
The farthest narrow line AGN in our sample is M1513$-$2524 ($z_{em}=3.132$), and the one with largest projected linear size of radio emission (size$\sim$330\,kpc) is M0909$-$3133 ($z_{em}=0.884$). 
For some individual objects the details, such as the morphology of associated radio emission, host galaxy and emission line properties, relevant to the discussion presented in these sections are provided in Appendix~\ref{sec:sampnotes}.  The properties of M1312-2026, the potential highest redshift AGN in the sample, are discussed in Sect.~\ref{sec:z5}. In Sections~\ref{sec:emless} and \ref{sec:dark}, we discuss AGN with no optical emission lines and dark fields, respectively.

\subsection{Overall properties}      
\label{sec:overall}   

\begin{figure*} 
\centerline{\vbox{
\centerline{\hbox{ 
\includegraphics[trim = {0cm 0cm 0cm 0.0cm}, width=0.33\textwidth,angle=0]{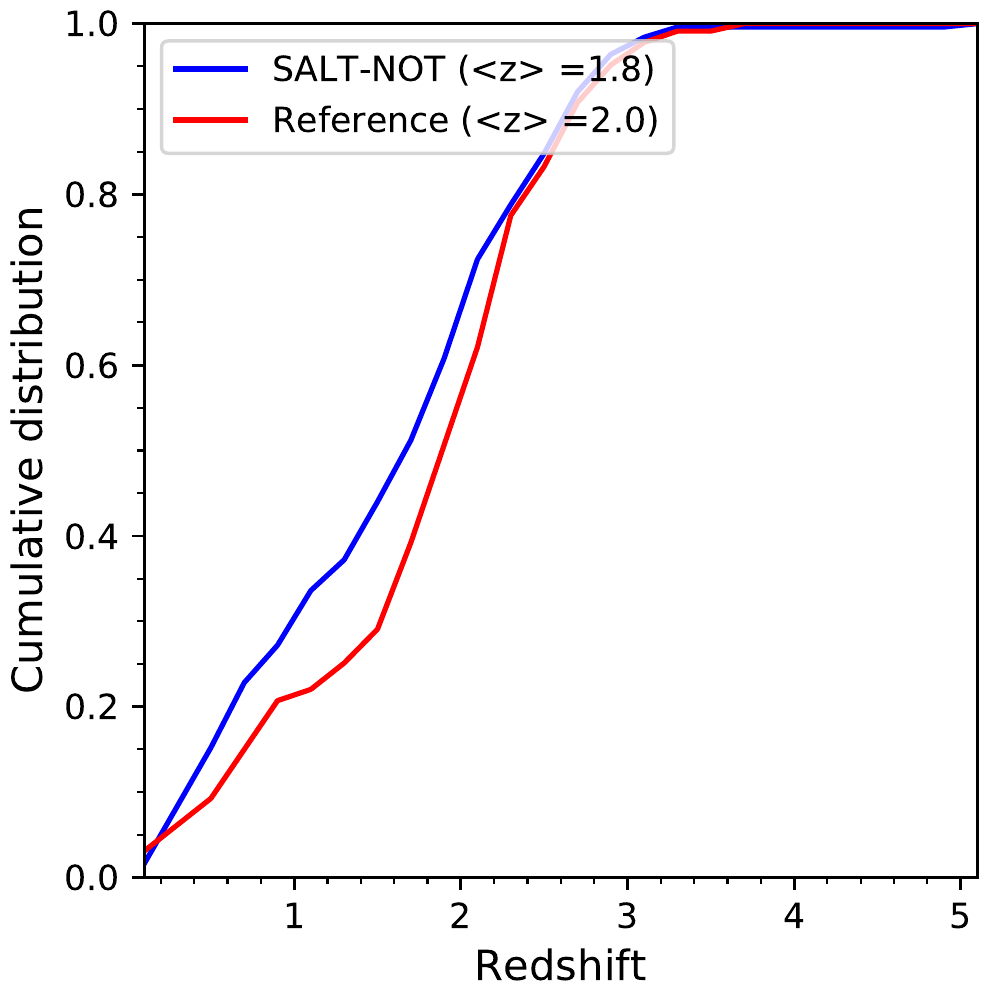}
\includegraphics[trim = {0cm 0cm 0cm 0.0cm}, width=0.33\textwidth,angle=0]{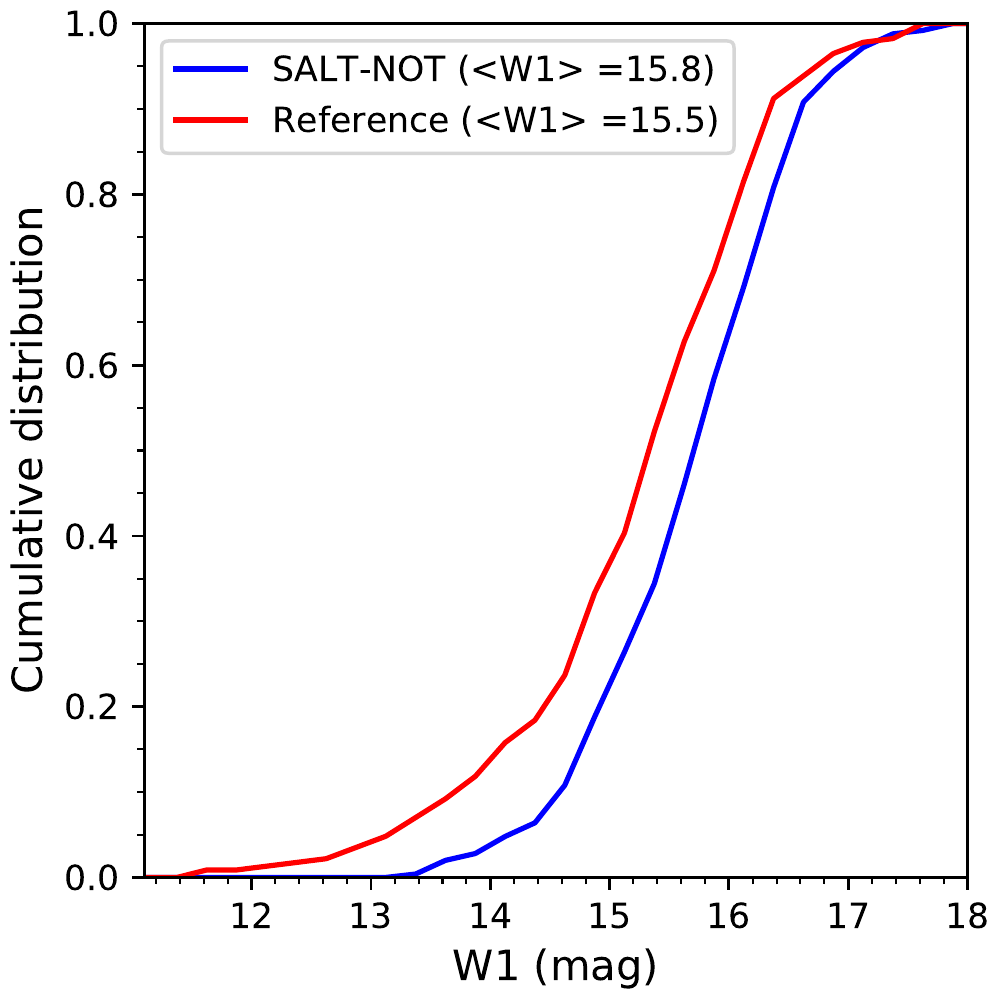}
\includegraphics[trim = {0cm 0cm 0cm 0.0cm}, width=0.33\textwidth,angle=0]{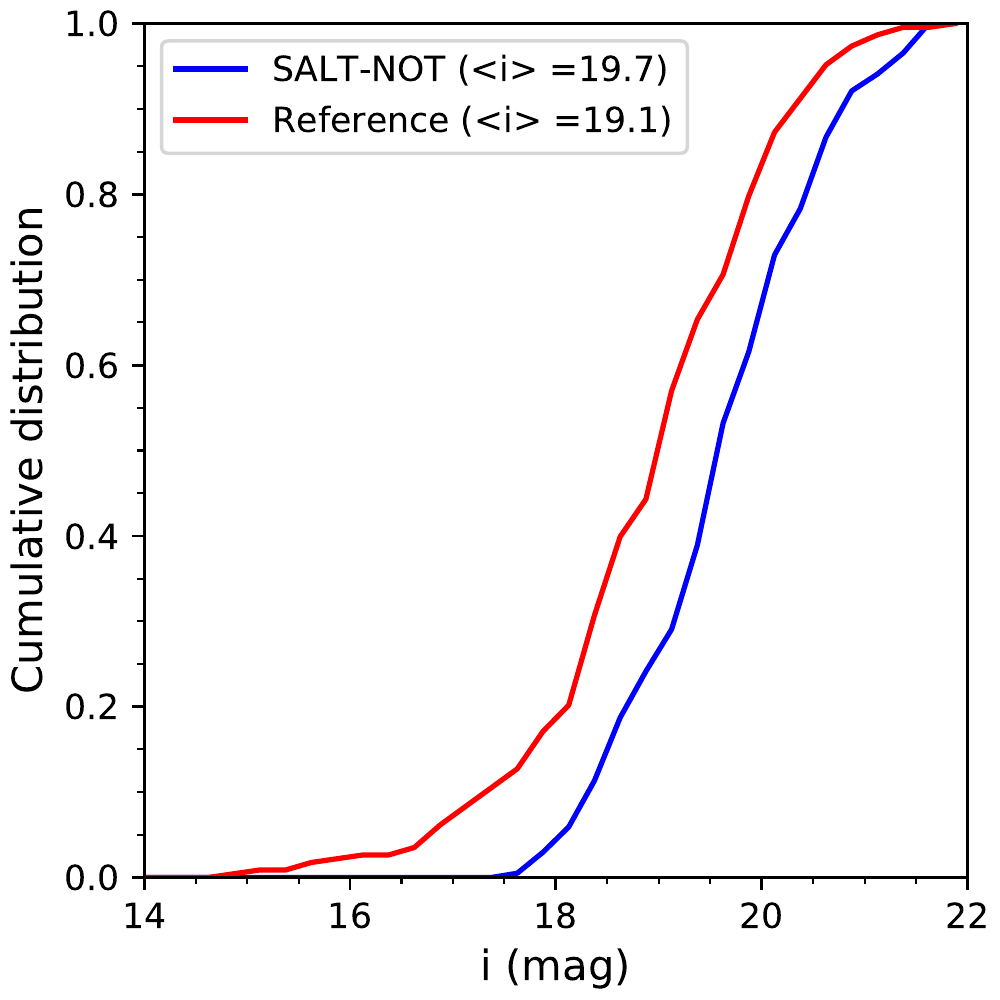}
}} 
}}  
\vskip+0.0cm  
\caption{
Comparison of our SALT--NOT sample and the subset of the {\it reference} sample satisfying our WISE criteria.  The median values are also provided.
} 
\label{fig:zcumul}   
\end{figure*} 

The basic properties, i.e., redshift, $W1$ and $i$-band magnitudes of the SALT--NOT sample are summarized in Fig.~\ref{fig:zcumul}.  We have also used emission lines detected in our spectra to classify objects as narrow-line (NL) or broad-line (BL) AGN. In Table~\ref{tab:wisesamp}, these are labeled as `G' and `Q', respectively. In total, 51/250 objects are NLAGN and 33 of these are at $z<0.5$.  As expected the fraction of NLAGN in our sample falls off at higher redshifts.

Fig.~\ref{fig:zcumul} also provides a comparison with the {\it reference} sample of quasars satisfying our MIR cut (see Section~\ref{sec:cross}).  The median redshift of the SALT--NOT sample is only slightly lower but a deficit of lower redshift AGN in the {\it reference} sample is apparent. A two-sample Kolmogorov--Smirnoff (KS) test shows that there is a less than 1 percent chance (p-value = 0.008) that the two samples are different purely by chance.
Related to this, we also note that in the SALT--NOT sample, $20\pm3$\% objects are NLAGN whereas based on the {\it reference} sample only $6^{+2.1}_{-1.6}$ \% are expected.

In Fig.~\ref{fig:zcumul}, we also compare $W1$ and $i$-band magnitudes of the two samples. Clearly, the SALT--NOT sample corresponds to a population of fainter AGN. We note that the probabilities of two samples being drawn from the sample populations are $1.7\times10^{-4}$ for $W1$ brightness and $1.6\times10^{-8}$ for $i$-band brightness. This further confirms the assertion made in Sect.~\ref{sec:effic} that the inclusion of $W3$, the less sensitive WISE band, in our target selection has not biased our sample towards brighter objects.  
In particular, the $i$-band magnitudes of AGN in our sample at $z<0.5$ and $z>0.5$ are similar.
Therefore, the excess of NLAGN at $z<0.5$ in our sample is intriguing. It is likely a direct consequence of the selection of optically fainter targets through MIR color selection.  Excluding these $z<0.5$ NLAGN renders the $z$-distributions of the SALT--NOT sample and the {\it reference} sample statistically identical.

\subsection{AGN at $z<0.5$}      
\label{sec:zl0p5}   

For $z<0.5$ AGN, our SALT spectra cover \oii, \oiii\ and H$\beta$ emission lines. The NOT spectra additionally cover the Mg~{\sc ii} emission line as well. Overall, the SALT--NOT sample has 33/303, i.e., 11\% AGN in our sample are at $z<0.5$.  We find the spectra in all the 33 cases to be mostly dominated by narrow permitted emission lines (with FWHM$<$2000 \kms). The available optical images are also consistent with their hosts being galaxies with weak or no nuclear emission.
Compared to this, in the {\it reference} sample (SDSS), only $\sim$32\% (6/19) of the $z<0.5$ AGN satisfying our MIR wedge are classified as galaxies. As mentioned above, this difference with respect to the {\it reference} sample is most likely due to the optical faintness of the SALT--NOT sample and the specific goal to reject low-redshift AGN \citep[][]{Barrows21}. 

The radio emission associated with these NLAGN is often extended, the largest projected linear size being $\sim$250\,kpc (details in Appendix~\ref{sec:sampnotes}).  An examination of optical emission lines led to identification of 2 dual AGN showing double peaked spectra (an example is shown in top panel of Fig.~\ref{fig:emission}; more details in Appendix~\ref{sec:sampnotes}).  Our detection rate of $6^{+8}_{-4}$\% is marginally higher, but only at 1$\sigma$, than the detection rate (1\%) of dual AGN in SDSS \citep[][]{Liu10}.

\subsection{AGN at $0.5 < z < 1.0$}      
\label{sec:zl1}   

For AGN at $0.5 \le z\le 1.0$ (35/303), the SALT--NOT spectra cover both Mg~{\sc ii} and [O~{\sc ii}] emission lines. 
About $32^{+13}_{-10}$\% (12/35) of these show emission lines of [O~{\sc ii}], Ne~{\sc iv} and Ne~{\sc v} without (or narrow) Mg~{\sc ii} broad emission lines.  The fraction of NLAGN (8/28 i.e., $29^{+14}_{-9}$\%) is consistent with the reference sample.

In VLASS 3\,GHz images, 8/12 ($75^{+25}_{-23}$\%) of the NLAGN in the SALT--NOT sample exhibit extended radio emission with lobes, the largest projected linear size being 330\,kpc.  Among BLAGN, as expected, a much smaller fraction ($\sim$40\%) is associated with extended radio emission \citep[][]{Padovani17, Hickox18}. The largest extent is only 180\,kpc (see Appendix~\ref{sec:sampnotes} for details).
We also examined the radio morphology and optical emission line profiles for evidences of interaction between radio jet and the host galaxy emission \citep[see e.g.,][]{Mullaney13}.  In three cases, the radio morphology is distorted. But only in one of these (details in Appendix~\ref{sec:sampnotes}), the \oii\ line is asymmetric, which may be an indication of radio jet-ISM interaction \citep[e.g.,][]{Gupta05}.

\subsection{AGN at $1.0 < z < 1.9$}      
\label{sec:zl2}   

At $1.0\le z\le 1.9$, our sample has 71 AGN.  Only two of these  are possibly NLAGN and the radio emission in both the cases has a double lobed morphology  (details in Appendix~\ref{sec:sampnotes}). This low fraction (3\%; 2/71) of NLAGN is consistent with the {\it reference} sample, which has 0/53 such AGN in this redshift range.

\subsection{AGN at $1.9 < z < 3.5$}      
\label{sec:zg2}   

As previously mentioned, for these redshifts the SALT-NOT spectra of 110 AGN in our sample allow detection of \civ\ emission lines. Based on the width of \civ\ emission line (i.e FWHM $<$ 2000 \kms) we identify 4 of these objects to be the NLAGN. In comparison, the {\it reference} sample has no NLAGN (0/128) at $z>1.9$, which is consistent at 2$\sigma$ considering one-sided Poissonian statistics. The details of NLAGN at $1.9 < z < 3.5$ in our sample are provided in Appendix~\ref{sec:sampnotes}).  The emission line ratios are typical of what is observed in high-$z$ radio galaxies. Two of these i.e., M1315$-$2745 and M1513$-$2524 are among the largest radio galaxies at these redshifts \citep[see figure 13 of][for the latest compilation]{Shukla2021}.  
In the rest of this section, we discuss BH masses, Eddington ratios and BALS associated with the BLAGN.

\subsubsection{BH masses and Eddington ratios}      
\label{sec:lum}   

\begin{figure*} 
\centerline{\vbox{
\centerline{\hbox{ 
\includegraphics[trim = {0cm 0cm 0cm 0.0cm}, width=0.95\textwidth,angle=0]{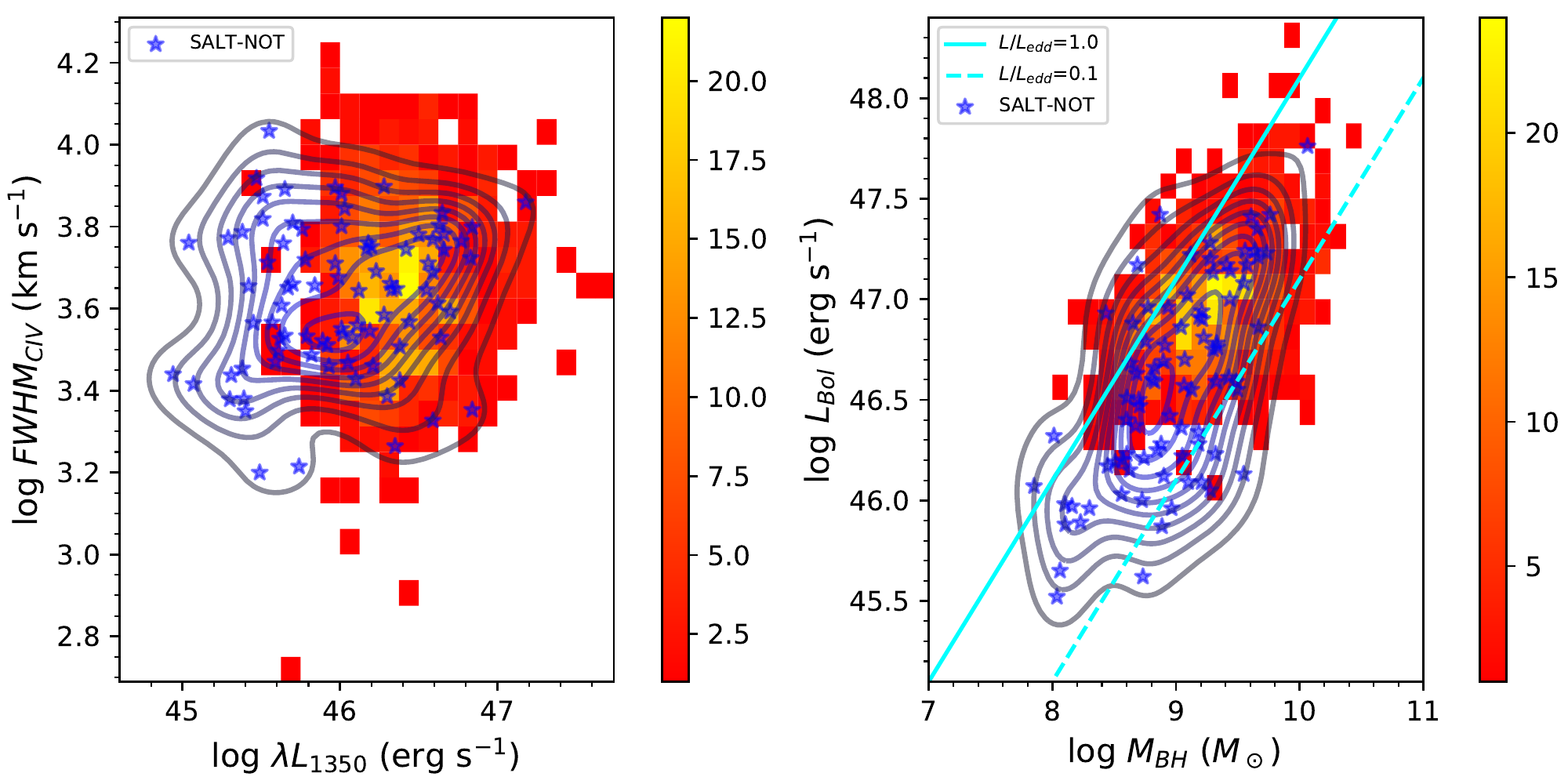}
}} 
}}  
\vskip+0.0cm  
\caption{Comparison of luminosity $\lambda L_{1350}$ vs.\ full \civ\ FWHM ({\it left}), and $L_{Bol}$ vs.\ $M_{\rm BH}$ ({\it right}).  In both the panels, the core-dominated quasars from from \citep[][]{Shen11} are shown as 2D histogram, and SALT--NOT quasars are $\star$ and contours. 
} 
\label{fig:mbh}   
\end{figure*} 

Under the assumption that broad line region (BLR) clouds are in virial equilibrium, the mass of the black hole, $M_{\rm BH}$, is given by 
%
\begin{equation}
    M_{\rm BH} = f \ \frac{R_{\rm BLR}\ {\rm FWHM}^2}{G}
\label{eq:mbh2}    
\end{equation}
where $G$ is the gravitational constant and $f$ is the virial coffecient that depends on the geometry and kinematics of the BLR.  The radius of the BLR, $R_{\rm BLR}$, is generally estimated using empirical relations between continuum or line luminosity \citep[e.g.,][]{Kaspi00}.  The value of $f$ is also empirically determined \citep[e.g.,][]{Onken04}. 
Following \citet[][]{Shen11}, we estimate virial BH masses for quasars with the \civ\ emission line using a relation of the form 
\begin{equation}
  \log \left( \frac{M_{\rm BH}}{M_\odot}  \right) = a + b\,{\rm log}(\lambda L_\lambda) + 2\,\log({\rm FWHM}) 
\label{eq:mbh1}
\end{equation}
where $a$ and $b$ are calibrated from reverberation mapping for a particular emission line \citep[e.g.,][]{Denney10}. The broad line FWHM is in \kms\ and the continuum luminosity, $\lambda L_\lambda$, which is a proxy for $R_{BLR},$ is in units of $10^{44}$\,erg\,s$^{-1}$. 
We adopt $a$ = 0.660 and $b$ = 0.53 for equation~\ref{eq:mbh1} based on the empirical relationship provided by \citet[][]{Vestergaard06} to estimate BH mass for our sample.  

Further, to place our SALT--NOT sample in the broader context of the overall quasar population, we compare our measurements to the  compilation of 105,783 SDSS DR7 quasars provided by \citet[][]{Shen11}.  For direct comparability of the measurements, we largely follow the method of \citet[][]{Shen11} to measure \civ\ FWHM and  $\lambda L_\lambda$ at 1350\AA.  
In short, we fit a power law to the continuum in the rest wavelength ranges of 1445 -- 1465 and 1700 -- 1705\,\AA.  Prior to continuum fitting we mask any unwanted spikes stronger than 5$\sigma$.  Then, we fit the continuum subtracted emission line in the wavelength range 1500 -- 1600\,\AA\ using three Gaussians.  The FWHM of the emission line, referred to as the {\it full} line FWHM, is then estimated from the fit after rejecting any individual components with strength $<$5\%.  
The continuum luminosity at 1350\AA\ ($L_{1350}$) is estimated using the same power law used to fit the continuum underlying the \civ\ emission line.  We are able to reliably estimate {\it full} \civ\ line FWHM, continuum luminosity at 1350\AA\ and BH mass of 86/110 quasars ($z>1.9$) from our sample. We have excluded 24 AGN with low SNR profiles or lines affected by the CCD edge. 
Similar to \citet[][]{Shen11}, the bolometric luminosity ($L_{\rm Bol}$) have been estimated by applying the bolometric correction factor of 3.81 to $L_{1350}$ \citep[][]{Richards06}. The Eddington ratio ($\lambda_{\rm Edd}$) is estimated as $L_{\rm Bol}$/$L_{\rm Edd}$, where $L_{\rm Edd}$ is the Eddington luminosity.

The compilation of \citet[][]{Shen11} consists of 11,840 non-BAL quasars at $1.9 < z < 3.5$ (median $z$ = 2.29), the redshift range covered by our sample (median $z$ = 2.34). We have restricted the sample of \citet[][]{Shen11} to the subset that has been identified by a uniform target selection algorithm and flux limited to $i$ = 19.1 at $z<2.9$ or $i$ = 20.2 at $z>2.9$ \citep[][]{Richards02}.
A distinguishing feature of the SALT--NOT sample is the high radio brightness of quasars. In particular, the $z>1.9$ quasars in SALT--NOT sample have predominantly flat radio spectra. The median spectral index\footnote{Spectral index $\alpha$ is defined by the power law, $S_\nu \propto \nu^\alpha$}, $\alpha^{1.4}_{0.4}$, derived using the NVSS 1420\,MHz and the uGMRT 420\,MHz total flux densities is $\sim-0.38$ \citep[][]{Gupta21hz}.  
In the uGMRT images, 79/88 of these are represented by a single Gaussian component.
Therefore, we further restrict the sample of \citet[][]{Shen11} to 786 core-dominated quasars, hereafter {\it Shen11-core} in short, detected in FIRST.

\begin{figure*} 
\centerline{\vbox{
\centerline{\hbox{ 
\includegraphics[trim = {0cm 0cm 0cm 0.0cm}, width=0.98\textwidth,angle=0]{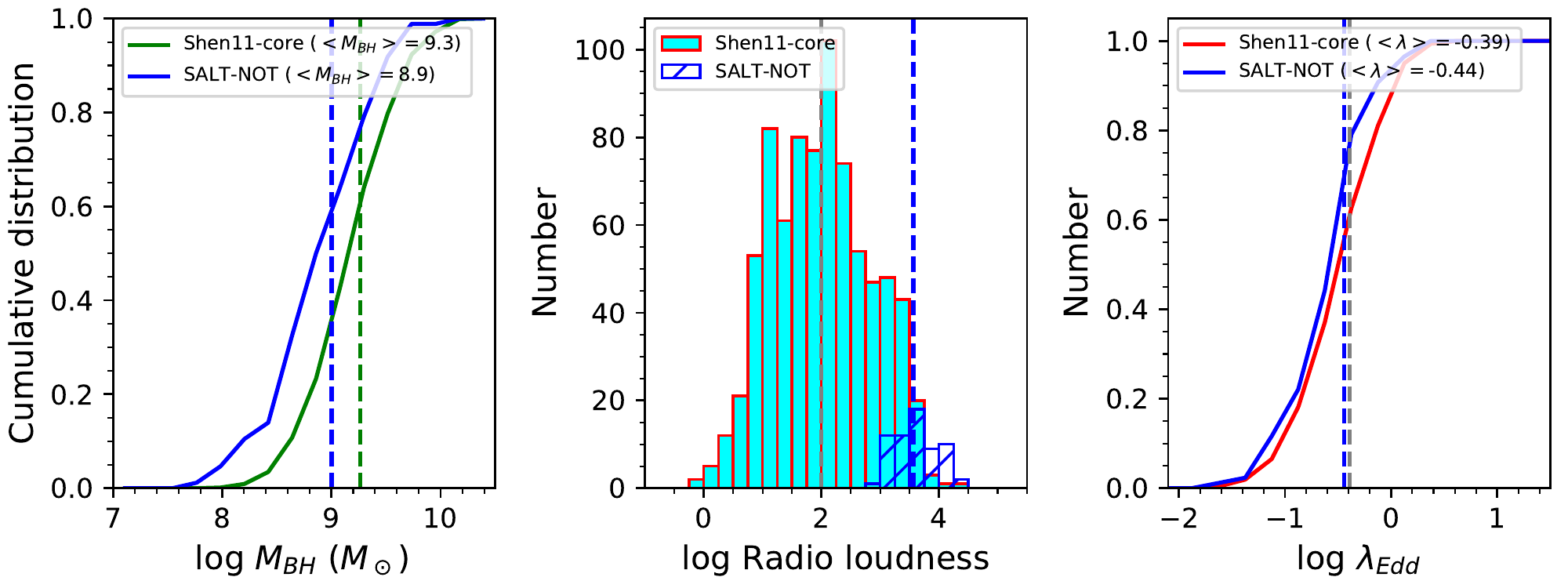}
}} 
}}  
\vskip+0.0cm  
\caption{
Comparison of BH mass (left), Radio loudness (middle) and Eddington ratios (right) for quasars from \citep[][]{Shen11} and SALT-NOT sample.  The vertical lines represent median values. 
} 
\label{fig:accr}   
\end{figure*}

In left panel of Fig.~\ref{fig:mbh}, we show the distributions of \emph{full} \civ\ FWHM and 1350~\AA\ luminosity for SALT--NOT and {\it Shen11-core} quasars. The SALT--NOT sample extends about 0.5\,dex fainter in 1350~\AA\ luminosity.  The two-sample KS test probability of these two samples being drawn from the same parent sample is extremely small (p-value = $4\times10^{-13}$). The \civ\ FWHM of the SALT--NOT sample (median FWHM = 4445\,\kms) is marginally smaller than the {\it Shen11-core} sample (median FWHM = 4660\,\kms), and this difference is statistically insignificant. We note that if the \civ\ profiles are affected by non-virial motions such as outflows \citep[][]{Coatman16}, the influence ought to be similar in both the samples.
Due to the luminosity dependence of $\lambda L_\lambda^{0.66}$ in the virial mass estimator, the overall BH masses estimated using equation~\ref{eq:mbh1} are lower for the SALT--NOT sample (see right panel of Fig.~\ref{fig:mbh}).  The KS test probability of two $M_{BH}$ distributions being drawn from the same parent sample is extremely small (p-value = $3\times10^{-5}$; see left panel of Fig.~\ref{fig:accr}).  
%

In the middle panel of Fig.~\ref{fig:accr}, we show the radio-loudness parameter ($R = f_{\rm 5GHz}/f_{2500}$), defined as the ratio of the rest-frame flux densities at 5\,GHz ($f_{5GHz}$) and at 2500\,\AA\ ($f_{2500}$), for the two samples.  For SALT-NOT quasars, the values of $f_{5GHz}$ have been estimated using the flux density at 1.4\,GHz from NVSS using the spectral slopes $\alpha^{1.4}_{0.4}$ from \citet[][]{Gupta21hz}. In cases where $\alpha^{1.4}_{0.4}$ is unavailable, we have used the median $\alpha^{1.4}_{0.4}$ of $-0.38$ instead.  The values of $f_{2500}$ have been estimated by interpolating the PS1 photometry.
Clearly, the SALT--NOT sample represents extremely radio-loud quasars (median $R = 3685$).
The corresponding Eddington ratio is only slight smaller (median log\,$\lambda_{\rm Edd}$ = $-0.44$ vs. $-0.39$) as compared to the {\it Shen11-core} sample and the difference is statistically insignificant ($p$-value = 0.02). Thus, over four orders of magnitude covered by the radio-loudness in the SALT--NOT and Shen11-core quasars, we see no dependence of radio-loudness on the accretion rate.  This confirms the saturation of radio-loudness observed at low Eddington ratios \citep[see Fig.~3 of][]{Sikora07}.

\subsubsection{BAL occurrence}      
\label{sec:bal}   

\begin{figure*} 
\centerline{\vbox{
\centerline{\hbox{ 
\includegraphics[trim = {1.0cm 1.0cm 1.0cm 1.0cm}, width=0.7\textwidth,angle=270]{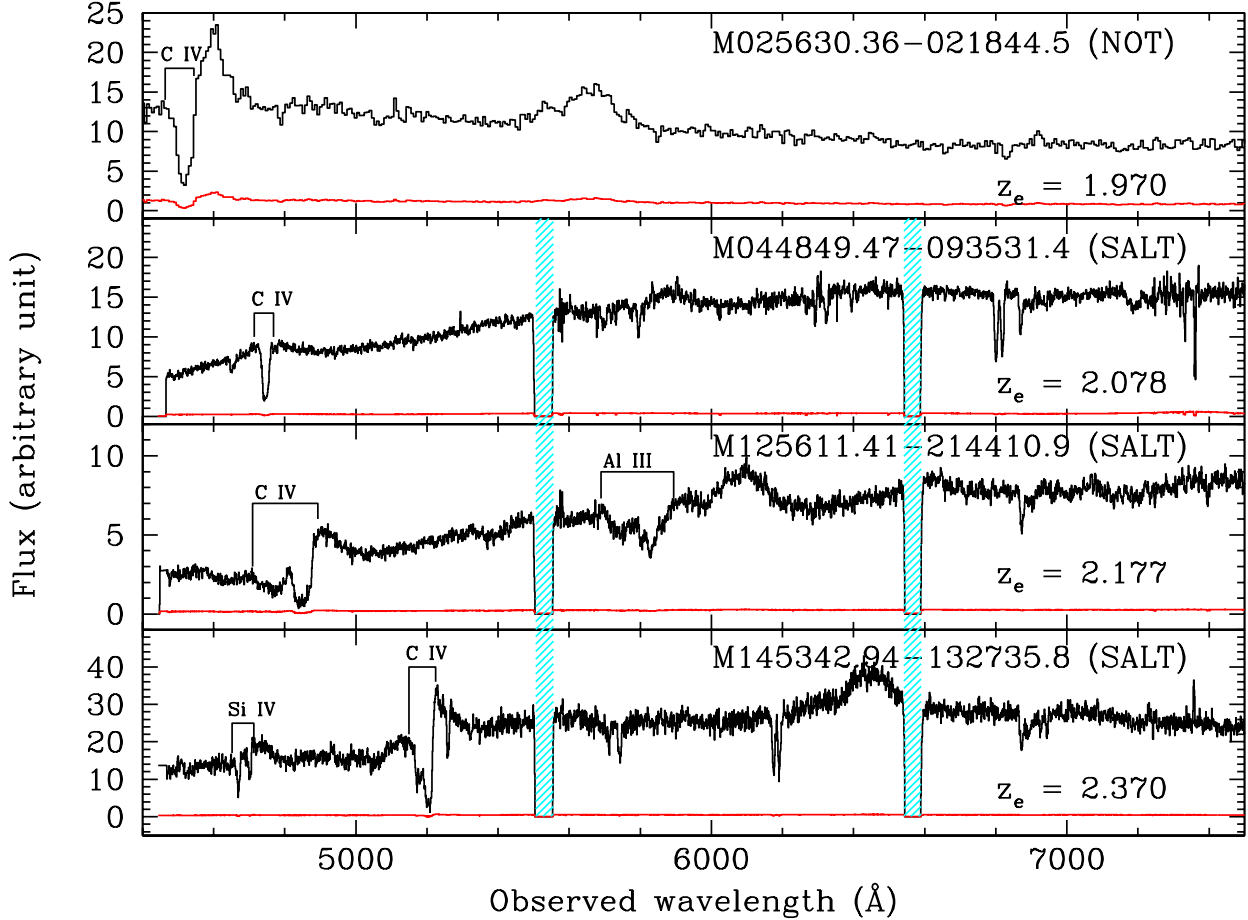}
}} 
}}  
\vskip+0.0cm  
\caption{
Spectra of four BAL QSOs found in our sample and associated error spectra (in red) are shown. Different BAL absorption are marked in each panel. Hashed regions are the CCD gaps in our SALT observations. 
} 
\label{fig:bal}   
\end{figure*} 

We focus now on the 105 SALT--NOT AGN showing \civ\ emission at $z_e \ge 1.9$ to search for broad absorption lines (BALs) representing outflowing gas. We exclude 7 cases at $z\sim2.6$ where the blue wing of \civ\ emission line falls in the CCD gap.  In the remaining, we detect 4 \civ\ BAL quasars (see Fig.~\ref{fig:bal}), i.e., an overall BAL detection rate of 4/98 = $4^{+3}_{-2}$\%. 
We find a consistent BAL fraction in our {\it reference} sample, without any MIR cut, where 9 out of 175 quasars at $z\ge2.0$ show \civ\ BALs (5$\pm$2\%).

Typically, 10-20\% of radio sources are detected as BALQSOs.  \citet{Becker00} studied the properties of radio selected BAL quasars from First Bright Quasar Survey sample. They also noticed that while the BAL detection rate is high among radio loud quasars their dependence on the radio luminosity is complex. In particular, the authors find a deficiency of BALQSOs among very radio bright quasars. Specifically, the BAL fraction drops from  $20.5^{+7.3}_{-5.9}$\% at  $L_{1.4\,GHz}$ $\sim10^{25}$\,W\,Hz$^{-1}$ to $<8\%$ at $L_{1.4\,GHz}$ $\sim3\times 10^{26}$\,W\,Hz$^{-1}$ \citep[][]{Shankar08}.  The low BAL detection rate in the SALT--NOT sample which has $L_{1.4\,GHz} >$ $\sim10^{27}$\,W\,Hz$^{-1}$ is consistent with this trend in luminosity.

The radio emission associated with M1453$-$1327 is extended (largest angular size, LAS$\sim17.1^{\prime\prime}$) with a projected linear size of 140\,kpc. In the other three cases, the radio emission is dominated by a compact component (projected linear size $<$5\,kpc). This is consistent with the previous results that the radio emission associated with BALQSOs is compact \citep[][]{Becker00}.
Notably, the 4 BALQSOs have $R_{\rm 5GHz/2500\AA}$, $\lambda_{\rm Edd}$ and $M_{\rm BH}$ in the range of $10^{2.9-3.2}$, $10^{0.2-0.5}$ and $10^{9.1-9.8}$~M$_\odot$, respectively, which correspond to objects with least radio loudness, and highest BH mass and Eddington ratios.

\subsection{ M1312--2026: a powerful AGN at $z\sim5$?  }      
\label{sec:z5}

\begin{figure*} 
\centerline{\vbox{
\centerline{\hbox{ 
\includegraphics[viewport=35 95 740 520,clip=true,width=0.90\textwidth,angle=0]{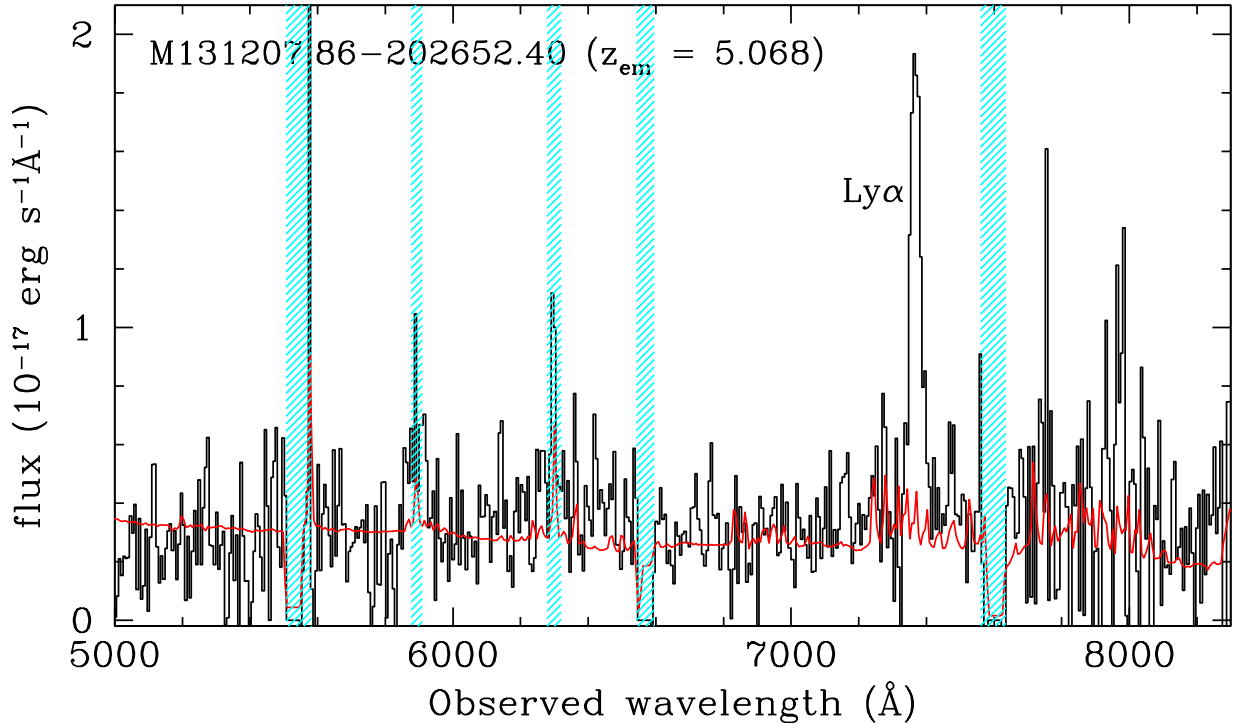}  
}} 
}}  
\vskip+0.2cm  
\caption{
The SALT spectrum of M1312-2026.  The overall SALT spectrum and the single emission line assumed to be Ly$\alpha$ at $z = 5.068$ have been consistently reproduced in four individual exposures, two of which were taken using a different spectral set-up. The error spectrum is shown in red. The shaded regions (cyan) mark the spectral gap and spectral range affected by sky subtraction residuals.  The VLT X-shooter spectrum shows that the actual \zem = 0.977 (see text for details). 
} 
\label{fig:saltz5}   
\end{figure*} 

\begin{table}
\caption{Flux densities of M1312--2026.}
\begin{tabular}{ccccccccc}
\hline

Frequency & Flux    & Deconvolved Size  &   Ref.  \\
  (GHz)   &    (mJy)       &            &              \\ 

\hline
   0.23   &  2577          &  7.1\arcsec$\times$3.3\arcsec, 0.0$^{\circ}$     &  (1)                  \\
   1.39   &   677          &  0.8\arcsec$\times$0.2\arcsec, 5.0$^{\circ}$     &  (2)                  \\
   1.40   &   778          &  $<$17.9\arcsec                                            &  (3)                       \\
   4.86   &   233          &  $<$2\arcsec                                               &  (4)       \\
\hline 
\end{tabular}
\\
{\flushleft
{\bf References:} (1) \citet[][]{Gupta21hz}; (2) This work. Observed on 2018 July 14 for 10\,mins (on-source time) using uGMRT with 200\,MHz wideband covering 1260-1460\,MHz.); (3) NVSS; (4) \citet[][]{Kapahi98}.}
\label{z5flux}
\end{table}

The bright radio sources at high-$z$ are extremely rare. This is partly due to the interaction between the emitting electrons and the cosmic microwave background (CMB) which makes AGN less luminous in radio \citep[][]{Ghisellini14}.  In the SALT spectrum of M1312-2026 (Fig.~\ref{fig:saltz5}), we identify only a single narrow emission line over the wavelength range covered.  If this emission line is \lya, then the corresponding redshift will be $z = 5.068$. At 4.86\,GHz, the radio source has a flux density of 233\,mJy and a size of $<2^{\prime\prime}$ (Table~\ref{z5flux}), and is unambiguously identified with a WISE and PS1 object with colors that are consistent with it being a high-$z$ radio loud quasar. 

In order to confirm the redshift of M1312-2026, we reobserved the object with X-shooter of the Very Large Telescope on 2022 February 01.  In the X-shooter spectrum, we do not detect expected Mg~{\sc ii}, C~{\sc iii}] and C~{\sc iv} lines corresponding to the presumed \lya\ at z=5.068.  Instead, the detection of [O~{\sc iii}], H$\alpha$ and H$\beta$ emission lines
confirm the above identified line to be [O~{\sc ii}] and the systemic redshift of the quasar to be $z=0.977$.  Further details of the X-shooter spectrum of this highly reddened object will be presented in a future paper.

\subsection{Objects without emission lines}      
\label{sec:emless}   

In recent times, the locus of MIR colors of known {\it Fermi} detected AGN have been used to build large samples of blazars \citep[e.g.,][]{Chang17, Dabrusco19}. 
These MIR color-selection techniques use the first three or all four bands of WISE, and typically extract $\sim$0.5\% ($<0.001$\%) of the overall radio\footnote{Based on the NVSS sensitivity.} (WISE) source population as blazars.  
They include both classes of blazars: {\it (i)} BL Lacs with featureless optical spectra, i.e., no emission lines  \citep[equivalent width $<$ 5\,\AA; ][]{Stickel91}, and {\it (ii)} flat spectrum, radio-loud quasars (FSRQs) with quasar-like emission lines.  In the recent literature, these two classes are also referred to as BZBs, i.e., blazars of BL Lac type and BZQs, i.e., blazars of quasar type, respectively \citep[][]{Massaro09}.

In the following, we focus on the 26 objects without emission lines (emission line less; ELL) in our sample that can be considered candidate BZBs. 
The median $q_{12}$ parameter which is defined as the logarithm of the ratio of flux densities measured in $W3$-band and at 20cm of ELL targets is $-2.8$.  This is well outside the range of (-1.85, -1) for BZB candidates based on the locus of {\it Fermi} detected BL Lacs \citep[][see also Fig.~\ref{fig:refsamp}]{Dabrusco19}. 
The SALT--NOT ELLs thus represent a population of candidate BL Lacs that is under-represented in the MIR selected population.

The most crucial and somewhat controversial aspect of establishing the BL Lac nature of AGN is to confirm its featureless optical spectra.  Our SALT/NOT spectra are adequate to detect the broad emission lines from AGN but weak lines may be missed due to the poor spectral resolution and noisy data.  This is illustrated by the case of two ELL candidates, M0051+1747 and M0105+1845, for which no redshift determination could be established based on the NOT spectra alone. However, these targets have later been observed as part of SDSS and have been identified as \zem = 1.536 and 1.074, respectively. We note that these redshifts are consistent with weak and narrow emission lines in the NOT spectra, which were originally marked tentative.  Thus, the objective of the analysis presented here is to identify BL Lac candidates based on the general properties of blazars: large time variability at all wavelengths and timescales, polarized emission, and a flat or inverted radio spectrum \citep[e.g.,][]{Stocke01}.

We evaluate the variability of ELLs using the data from the Zwicky Transient Facility \citep[ZTF;][]{Masci19}.  The $r$-band light curves are available for 15 AGN (median PS1 $r$ = 19.9 mag; see Fig.~\ref{fig:ztf} in the Appendix~\ref{sec:21opt}).  The remaining targets are either not covered (only 3) in ZTF or are fainter (median PS1 $r$ = 20.9 mag) and have time series with less than 10 data points. From the light curves, we estimate the variability amplitude ($\sigma_{rms}^2$) and error ($S_D^2$) on it using the normalized excess variance method described in \citet[][]{Nandra97} and  \citet[][]{Vaughan03}. The variability amplitude strength, VAS = $\sigma_{rms}^2$/$S_D^2$, is provided in Fig.~\ref{fig:ztf}.  

We also estimate the observed $\chi^2_{obs}$ using $\chi^2_{obs} = \sum_{i=1}^{N}(m_i - \overline{m})^2)/\sigma_i^2$, where $m_i$ are $N$ measured fluxes with individual error $\sigma_i$ and mean $\overline{m}$. The hypothesis of the AGN being intrinsically non-variable can be rejected at 99\% confidence level for 11/15 targets. The remaining 4 targets are among the faintest (PS1 $r$-band: 20.3 -- 21.7\,mag) and the conclusion is likely influenced by the low SNR of the light curve. Rejecting these and considering objects with VAS$>$3.0, we obtain a sample of 4 promising (M0022+0608, M0029$-$1740, M1245$-$1616 and M2142$-$2444) and two marginal (M0010$-$2157 and M0824$-$1029) BL Lac candidates (see Fig.~\ref{fig:ztf}). Optical spectrum of M1245$-$1616 obtained with SALT is shown in Fig.~\ref{fig:emission}.

\begin{figure} 
\centerline{\vbox{
\centerline{\hbox{ 
\includegraphics[trim = {0cm 0cm 0cm 0.0cm}, width=0.4\textwidth,angle=0]{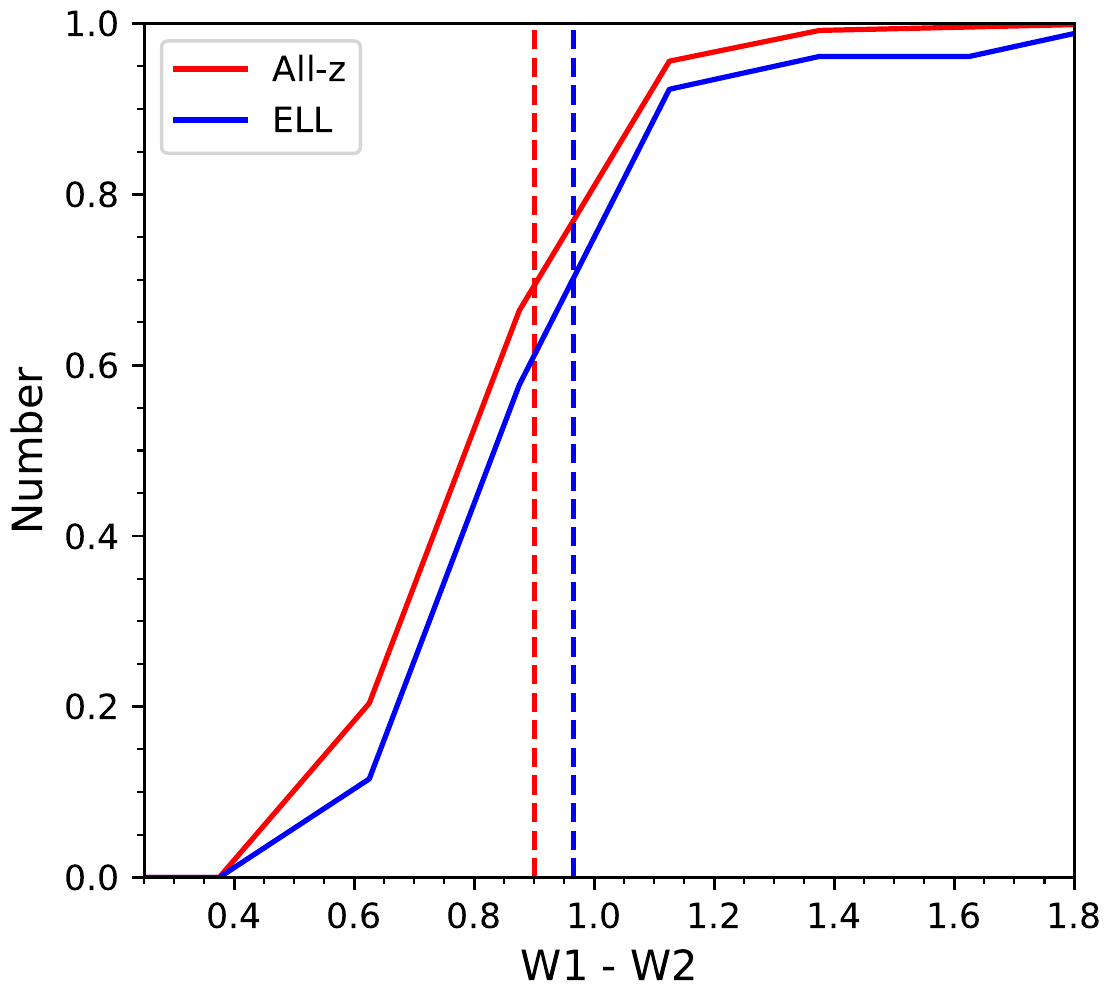}
}} 
}}  
\vskip+0.0cm  
\caption{
$W1 - W2$ color distribution for SALT--NOT targets with and without redshift determinations.  The median values are marked using dashed and dotted lines, respectively. 
} 
\label{fig:w1mw2}   
\end{figure} 

Interestingly, the candidate with the highest VAS (M0022+0608; see Fig.~\ref{fig:ztf}) is in fact a bonafide $\gamma$-ray emitting BL Lac \citep[][]{Massaro09}.  Further, the candidates M0010$-$2157, M0029$-$1740, M1245$-$1616 and M2142$-$2444 are also identified in the blazar catalog of \citet[][]{Dabrusco19} but as BZQ candidates on the basis of their position in the three dimensional ($W1$-$W2$-$W3$-$W4$) WISE  color-space based on the locus of confirmed {\it Fermi} blazars \citep[][]{Massaro09, Acero15}.
Further, 4/6 candidates are also polarized at 2-4\% level in NVSS \citep[][]{Taylor09}, and except one (M0029$-$1740) all are core-dominated in VLASS 3\,GHz images.  Note that it is not unusual for BL Lacs to exhibit extended radio emission \citep[][]{Rector01}.

If the BL Lacs are similarly represented in the 11 remaining ELLs not analysed due to the lack of ZTF light curves, then we expect another 4 candidates.  In this context, it is also useful to cross-correlate all 26 ELLs with the latest {\it Fermi} data release \citep[4FGL DR2; ][]{Ballet20}.  We recover previously noted M0022+0608, and find two new matches corresponding to M0010$-$2157 and M0553$-$0840. 

In summary, we have 7 high-probability BL Lac candidates; 6 identified on the basis of variability analysis and 1 through cross-matching with $\gamma$-ray sources, i.e., a detection rate of 7/303 ($2\pm1$\%).   
\citet[][]{Stickel91} used 1\,Jy sample to estimate the surface density (0.005 for $>$1\,Jy at 5\,GHz) of radio selected BL Lacs. The fraction of BL Lacs in the 1\,Jy sample of radio sources brighter than 20\,mag ($r$-band) is 15\%.  Therefore, we expect that better quality data through a targeted optical follow-up will confirm more of these ELLs as BL Lacs. 

\subsection{Dark fields}      
\label{sec:dark}   

The 27 targets for which there are no firm optical identifications, i.e., dark fields (DFs), are also presented in Table~\ref{tab:wisesamp}. For 14 of these, the stacked PS1 catalog implies optical counterparts with a median $i$-band magnitude of 22.5\,mag \citep[][]{Waters20}.
Clearly, these represent the faintest targets in our sample (see last panel of Fig.~\ref{fig:fluxw1mag}).  
On average they are also fainter in WISE (see last panel of Fig.~\ref{fig:fluxw1mag}).  But the $W1$ magnitudes are still well within the range over which we have successfully obtained optical spectra (middle panel of Fig.~\ref{fig:fluxw1mag}).  Thus, misidentifications due to errors in radio--MIR cross-matching is unlikely to significantly contribute to DFs. 
This is also confirmed by the VLASS--PS1 overlays of 24 DFs covered in PS1 (see Fig.~\ref{fig:maps}).  In 20/24 cases, the radio source is compact and coincides ($<1^{\prime\prime}$) with the MIR position marked by the cross. In the remaining 4 cases, the radio emission is extended, exhibits a double-lobe morphology and the MIR source is located between them, i.e., at the expected location of the AGN.

The above implies that DFs are truly fainter than the implicit magnitude limit of SALT--NOT observations and barely detectable in PS1\footnote{Note that the 5$\sigma$ depth of PS1 $i$-band stack images is 23.1\,mag.}. Thus, it is reasonable to assume that these have the same redshift distribution (median $z$ = 1.8) as the other SALT--NOT targets.
The basic statistics of MIR-wedge implies that 5\% of these are low-redshift radio galaxies. Indeed, upon careful inspection of the PS1 images, we identify diffuse, low-surface brightness emission associated with two targets: M0959$-$0236 and M1144$-$1455. These two targets could plausibly be radio galaxies at $z<1$.

The optical faintness of the remaining DFs, presumably quasars, could be due to {\it (i)} an intrinsically low luminosity; {\it (ii)} very large amounts of dust in the foreground (intrinsic or intervening) or {\it (iii)} a very high redshift ($z > 5$) such that the detectable continuum and emission lines are redshifted into the near-infrared; or {\it (iv)} a non-quasar origin of the radio/MIR emission.
There is already a tentative indication of a slight excess of intervening and associated DLAs towards optically faint quasars \citep[][]{Ellison08, Gupta21hz}.  Thus, the identified DFs in our survey may be hiding a small but important population of dusty \hi\ absorbers. The MALS observations of these central DF targets and the other optically faint AGN in the MeerKAT field-of-view will reveal any such absorbers at $z<2$.

\section{Summary and outlook}    
\label{sec:summ}  

\begin{figure*} 
\centerline{\vbox{
\centerline{\hbox{ 
\includegraphics[trim = {0.0cm 0.0cm 0.0cm 0.0cm}, width=1.0\textwidth,angle=0]{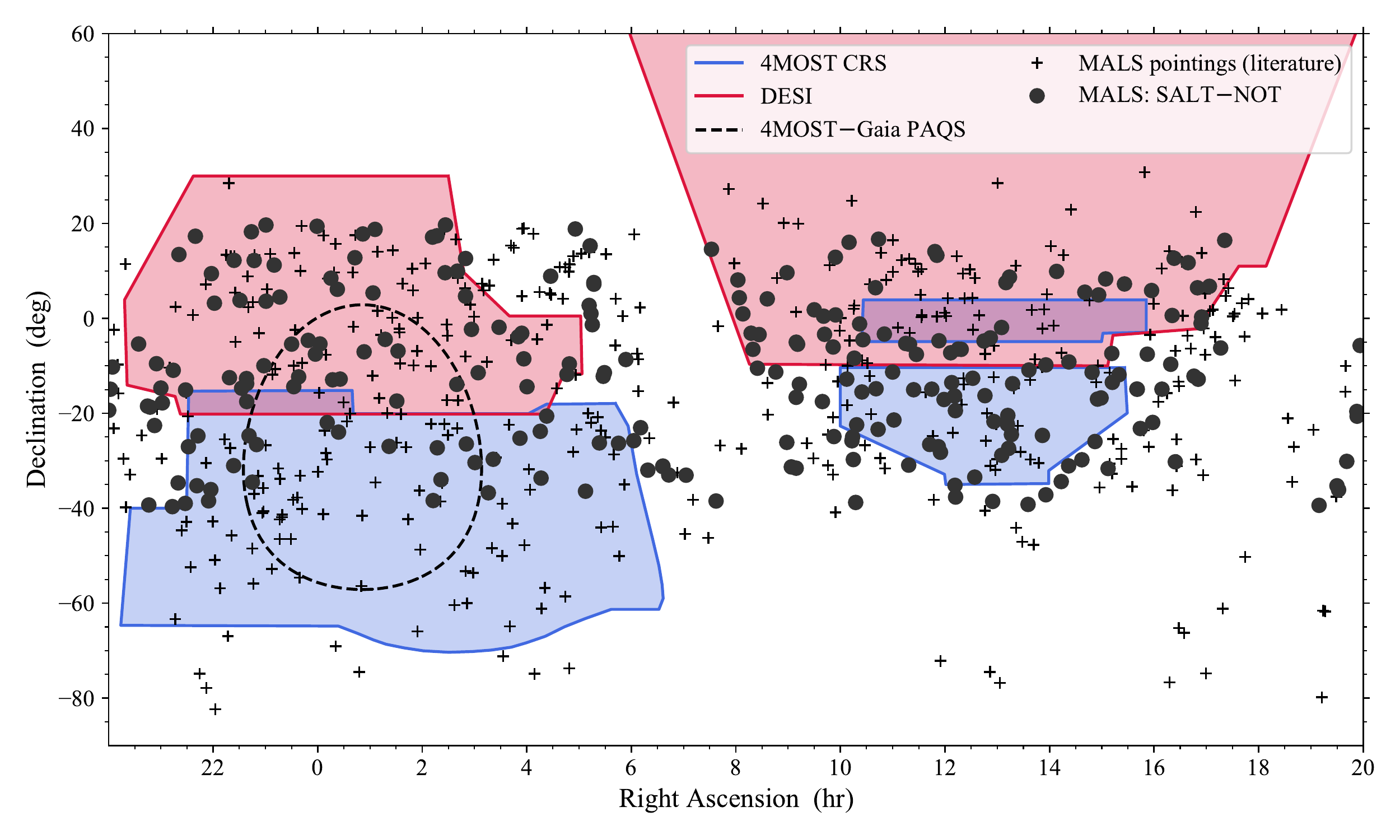}
}} 
}}  
\vskip+0.0cm  
\caption{
Compilation of MALS pointings from the literature (black crosses) and from our dedicated SALT and NOT observations (solid circles). Overlaid are the footprints of a few upcoming spectroscopic surveys: DESI (red), 4MOST Cosmology Redshift Survey (CRS; blue) and the 4MOST--Gaia Purely Astrometric Quasar Survey (PAQS; black dashed line). Note that ESO's 4MOST–Gaia PAQS survey (PI: Krogager) will observe 120,000 targets over 5 years.
} 
\label{fig:footprint}   
\end{figure*} 

\begin{deluxetable}{lcc}
\tabletypesize{\small}
\tablecaption{Summary of SALT-NOT and MALS (central AGN) sample. }
\tablehead{
\colhead{} &  \colhead{SALT-NOT} & \colhead{MALS$^a$}     \\
}
\startdata
Number of AGN                &       303           &    $650^b$         \\
Median flux density (mJy)     &       300           &    $340^{c}$       \\
Median $W1$ (mag)             &       15.9          &    15.6            \\
Median $i$ (mag)              &       19.8          &    19.4$^d$        \\
Inside MIR wedge (equation~\ref{eqwise}) &  100\%   &    70\%                \\ 
With spectroscopic redshift   &       83\%          &    79\%            \\
Median spectroscopic-$z$      &       1.8           &    1.7             \\
\label{tab:malstarget}
\enddata
\tablecomments{$a$: Includes SALT-NOT sample. $b$: 70 at $\delta < -40^\circ$ selected using SUMSS. The remaining at $\delta > -40^\circ$ are based on NVSS. $c$: Estimated at 1.4\,GHz. SUMSS 0.850\,MHz flux densities computed at 1.4\,GHz assuming $\alpha = -0.8$.   $d$: Only $\sim550$ targets overlapping with PS1 considered.  Note slight increase in the $i$-band brightness due to the inclusion of redshifts from the literature. 
}
\end{deluxetable}

In this paper, we describe a large spectroscopic survey of radio bright AGN selected based on the following criteria:
{\it (i)} radio flux density at 1.4\,GHz $>$ 200\,mJy;
{\it (ii)} $\delta < +20^\circ$; and
({\it iii}) WISE MIR-colors: $W1 - W2$ $< 1.3\times(W2 - W3) - 3.04$ and $W1 - W2 > 0.6$.
These criteria are optimized to select powerful quasars at $z>1.4$.
Starting from 25,325 sources brighter than 200\,mJy in NVSS we identify 2011 candidate quasars, of which 303 were observed with SALT (180\,hrs) and NOT \citep[6 nights;][]{Krogager18}.  

The SALT--NOT sample of 303 objects presented in this paper consists of {\it (i)} AGN with emission lines in the  optical spectrum (250); {\it (ii)} objects with no  emission lines in the optical spectrum, i.e., emission line less (ELLs; 26); and {\it (iii)} dark fields (DFs; 27), i.e., neither emission lines nor a continuum source in the optical spectra or photometry (also see Table~\ref{tab:malstarget}).  

We compare the SALT--NOT sample with the highly complete spectroscopic sample of 10,498 quasars from the SDSS (DR16) in the so-called `Stripe 82' region, and with a {\it reference} sample of 2294 AGN from SDSS selected by applying the same constraints on the radio flux density and MIR colors as the SALT--NOT sample.    
We show that our imposed WISE criteria has lead to a sample of radio-bright quasars which are fainter ($\Delta i = $0.6\,mag) and redder ($\Delta (g - i)$ = 0.2\,mag) than radio-selected quasars in the {\it reference} sample, and representative of a fainter quasar population detected in optical surveys.

About 20\% (51/303) of the AGN in our sample only have narrow emission lines. The majority of these (33/51) at $z<0.5$, are galaxies without strong nuclear emission.
In two cases the emission line is double peaked and likely corresponds to rare dual AGN.
The highest redshift NLAGN (M1513$-$2524) is at $z = 3.132$. The details of this optically faint ($r>23$\,mag) AGN associated with a large ($\sim$90\,kpc) and luminous \lya\ nebula are presented by \citet[][]{Shukla2021}.  The double-lobed radio emission associated with M1513$-$2524 has an extent of 184\,kpc.
In general, the NLAGN in our sample exhibit extended radio emission with lobes and in some cases cores -- the largest being M0909-3133 ($z_{em}$= 0.884) with a projected linear size of 330\,kpc.

Based on the {\it reference} sample, we expect only 5\% of objects in the SALT--NOT sample to be NLAGN.  The discrepancy is primarily due to the $z<0.5$ contaminants.  Excluding these would render the $z$-distribution and broad- vs.\ narrow-line AGN composition of the SALT--NOT sample identical to the {\it reference} sample.
Overall, the application of our WISE-color criteria improves the efficiency of identifying $z>1.4$ AGN by a factor of $\sim$1.7 (i.e., from $\sim$30\% to $\sim$50\%).

We discuss in detail the properties of the 110 SALT--NOT AGN at $1.9 < z < 3.5$.
The four NLAGN in this sub-sample show emission line ratios similar to high-$z$ radio galaxies.  The radio emission associated with two of these (M1315$-$2745 and M1513$-$2524) are among the largest at these redshifts (280 and 184~kpc, respectively).
We estimate Bolometric luminosities, radio-loudness parameters ($R = f_{\rm 5GHz}/f_{2500\AA}$), black hole masses ($M_{\rm BH}$) and Eddington ratios of the remaining 106 AGN in this sub-sample.  In comparison to the optically selected core-dominated SDSS quasars from \citet[][]{Shen11}, the SALT--NOT quasars are fainter ($\sim$0.5\,dex lower $L_{1350}$) and, consequently, have lower ($\sim$0.4\,dex) $M_{\rm BH}$. It is intriguing to note that despite representing the most radio-loud quasars (median $R$ = 3685), the Eddington ratios of our objects are similar to the less radio-loud (median $R$ = 100) SDSS quasars.  In fact, we do not find any dependence between radio loudness and Eddintgton ratio in the SALT--NOT and SDSS quasar samples spanning four orders of magnitude in $R$.  This confirms the saturation of radio loudness observed at low Eddington ratios \citep[][]{Sikora07}, and the significance of physical parameters such as black hole spin and the in-situ magnetic field in influencing the overall appearance of AGN.

Among $1.9 < z < 3.5$ AGN, we also detect 4 \civ\ BAL quasars.  All are associated with objects of least radio loudness, and highest BH masses and Eddington ratios. The low BAL quasar detection rate ($4^{+3}_{-2}$\%) is consistent with that seen in extremely powerful ($L_{1.4GHz} > 10^{25}$\, W\,Hz$^{-1}$) radio loud quasars \citep[][]{Shankar08}.
The systematic \hi\ 21-cm absorption line search in these quasars suggests deficiency of cold atomic gas at host galaxy scales \citep[][]{Gupta21hz}. 
In this context, it will be interesting to confirm the slight excess of proximate DLAs observed in our sample \citep[][]{Gupta21hz}. 
The details of gaseous environment of these quasars at larger ($>10$\,kpc) scales are presented in a separate paper \citep[][]{Shukla21samp}.

Based on detailed optical variability analysis, radio polarization and $\gamma$-ray properties we identify 7 high-probability BL Lacs among 26 ELLs in our sample.  Better quality optical spectra will likely confirm a larger fraction of these ELLs as bonafide BL Lacs.  Finally, we discuss 27 DFs in our sample which could be due to {\it (i)} an intrinsically low luminosity; {\it (ii)} very large amounts of dust in the foreground (intrinsic or intervening) or {\it (iii)} a very high redshift ($z > 5$) such that the continuum is strongly absorbed by the intervening \lya\ forest and emission lines are redshifted into the near-infrared; or {\it (iv)} a non-quasar origin of the radio/MIR emission.

The selection criteria of SALT-NOT sample are based on the requirements of MALS.
The entire MALS footprint of $\sim$500 pointings each at L and UHF bands is based on the SALT-NOT sample presented here and additional $\sim$350 bright radio sources selected from the literature.  
The overall properties of this larger pool of 650 AGN are presented in Table~\ref{tab:malstarget} (see also Fig.~\ref{fig:footprint} for the survey footprint and overlap with various optical surveys). 
%

The complete details of the MALS sample will be provided in future papers. But, in short, these additional sources are AGN: {\it (i)} with known spectroscopic redshift and $\delta > -40^\circ$ from the literature (see Fig.~\ref{fig:xmatch} i.e., the green box with 597 AGN), and {\it (ii)} brighter than 200\,mJy at 843\,MHz and $\delta < -40^\circ$ in the SUMSS \citep[][]{Mauch03}.  Further, priority has been given to AGN having excellent multi-wavelength data or close to nearby ($z<0.1$) galaxies.  
This extended sky coverage through a larger pool of 650 sources offers flexibility in selecting pointings for the observations, and maximising the scientific output corresponding to various galaxies and AGN detected within the MALS field-of-view \citep[see][for some of the science cases]{Gupta17mals}.  

Currently, $\sim$400 pointings mostly at L band have already been observed (Fig.~\ref{fig:footprint}). Looking forward, MALS will continue to observe with the MeerKAT through 2021-22 and afterwards primarily in the UHF band. We plan to release all radio continuum images and spectra along with the catalogs of detected sources and spectral lines through our survey website: \textcolor{blue}{https://mals.iucaa.in}.  This will include multi-wavelength data such as the SALT-NOT spectra presented here.  We expect all these to be an excellent resource for the astronomical community for a broad range of galactic and extragalactic applications.

\acknowledgments

We thank the anonymous referee for useful comments and suggestions.
PPJ was supported in part by the French National Research Agency (ANR) under contracts ANR-16-CE31–0021. PPJ thanks Camille N\'ous (Laboratoire Cogitamus) for inappreciable and often unnoticed discussions, advice, and support.
This  work  is  based  on  observations  made  with  SALT and NOT.
We thank NOT and SALT staff for their support during the observations.  
The National Radio Astronomy Observatory is a facility of the National Science Foundation operated under cooperative agreement by Associated Universities, Inc.
Based on observations obtained with the Samuel Oschin Telescope 48-inch and the 60-inch Telescope at the Palomar Observatory as part of the Zwicky Transient Facility project. ZTF is supported by the National Science Foundation under Grant No. AST-2034437 and a collaboration including Caltech, IPAC, the Weizmann Institute for Science, the Oskar Klein Center at Stockholm University, the University of Maryland, Deutsches Elektronen-Synchrotron and Humboldt University, the TANGO Consortium of Taiwan, the University of Wisconsin at Milwaukee, Trinity College Dublin, Lawrence Livermore National Laboratories, and IN2P3, France. Operations are conducted by COO, IPAC, and UW.

\appendix
\counterwithin{figure}{section}

\section{Properties of MALS SALT-NOT sample}
\label{sec:21opt}

The properties of AGNs in the SALT-NOT sample are provided here.  The ZTF light curves and radio-optical overlays for ELLs and DFs, respectively, are also provided.

\begin{figure*} 
\centerline{\vbox{
\centerline{\hbox{ 
\includegraphics[trim = {0cm 0cm 0cm 0cm}, width=0.50\textwidth,angle=0]{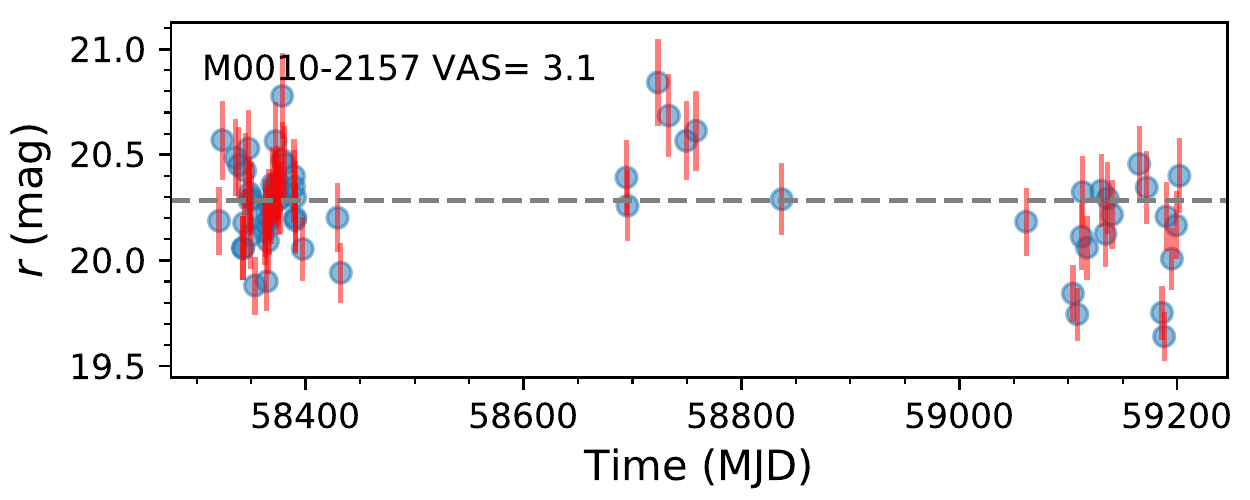} 
\includegraphics[trim = {0cm 0cm 0cm 0cm}, width=0.50\textwidth,angle=0]{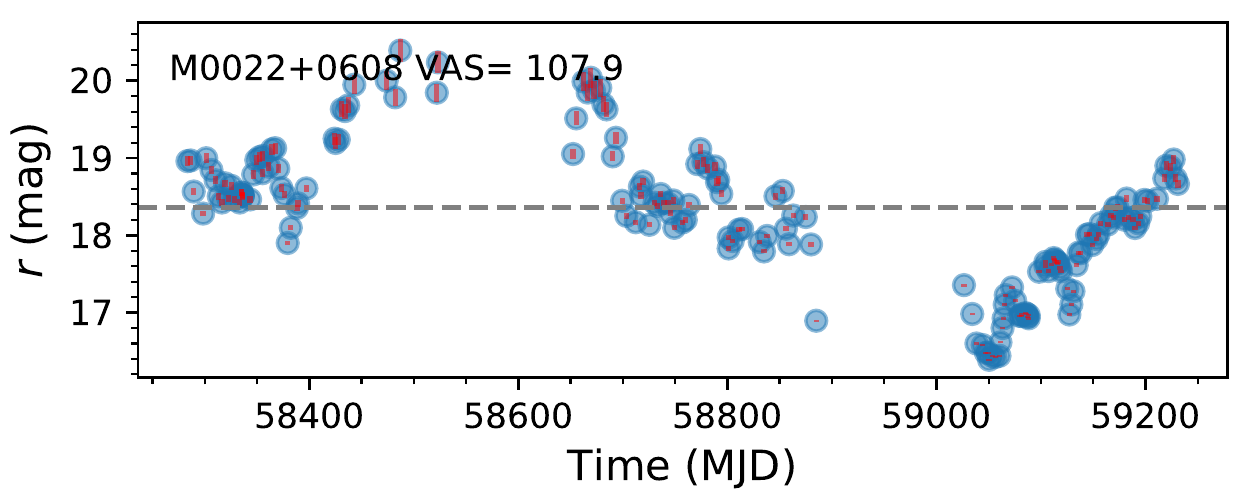} 
}}
\centerline{\hbox{ 
\includegraphics[trim = {0cm 0cm 0cm 0cm}, width=0.50\textwidth,angle=0]{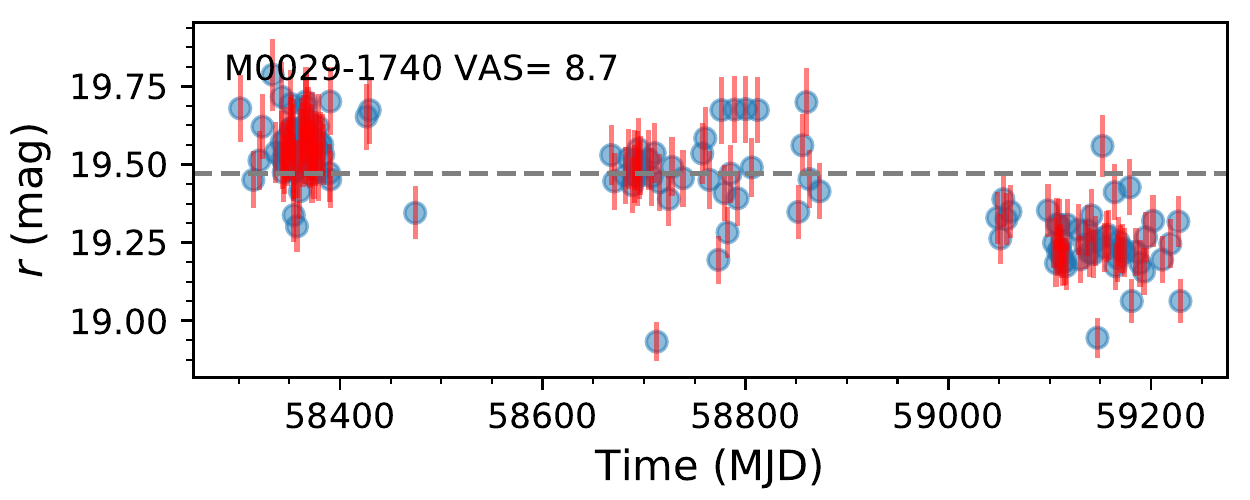} 
\includegraphics[trim = {0cm 0cm 0cm 0cm}, width=0.50\textwidth,angle=0]{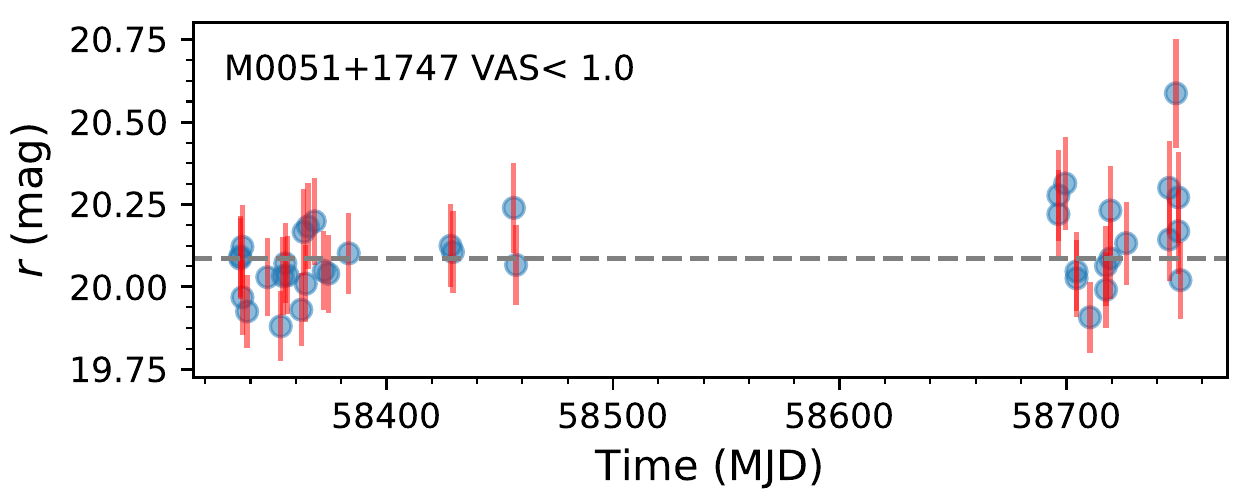} 
}}
\centerline{\hbox{ 
\includegraphics[trim = {0cm 0cm 0cm 0cm}, width=0.50\textwidth,angle=0]{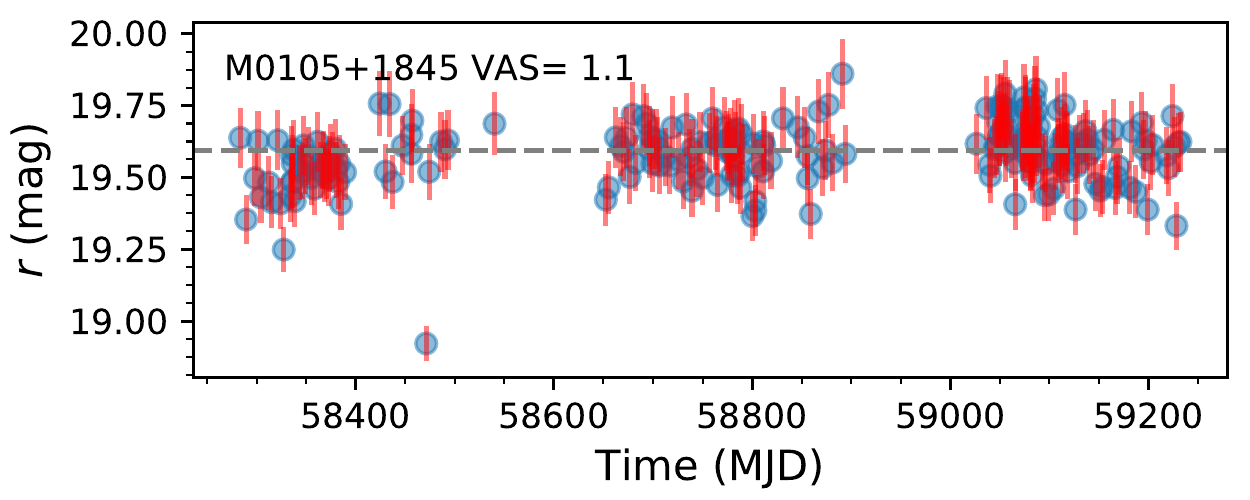} 
\includegraphics[trim = {0cm 0cm 0cm 0cm}, width=0.50\textwidth,angle=0]{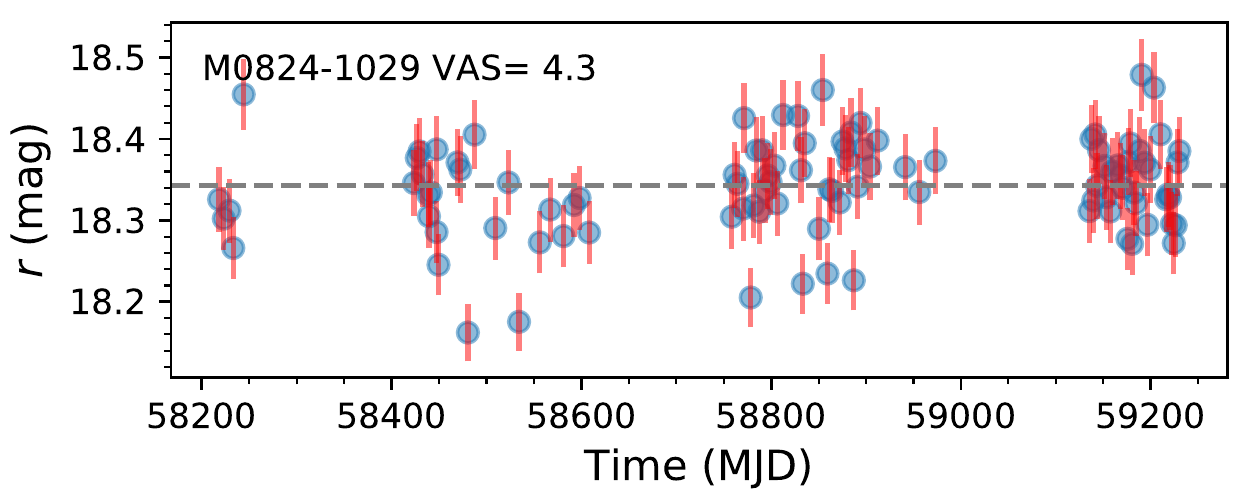} 
}}
\centerline{\hbox{ 
\includegraphics[trim = {0cm 0cm 0cm 0cm}, width=0.50\textwidth,angle=0]{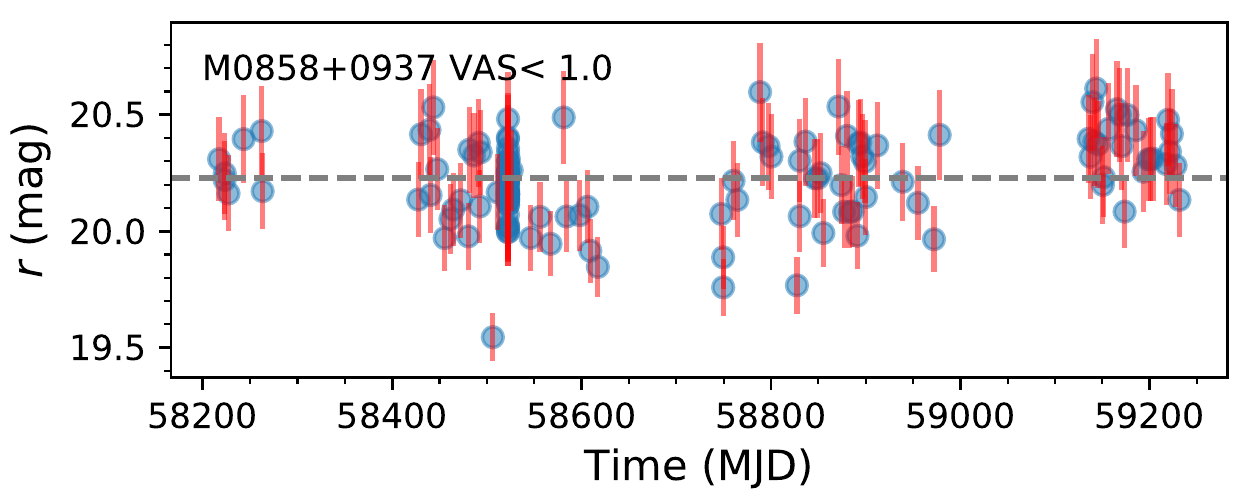} 
\includegraphics[trim = {0cm 0cm 0cm 0cm}, width=0.50\textwidth,angle=0]{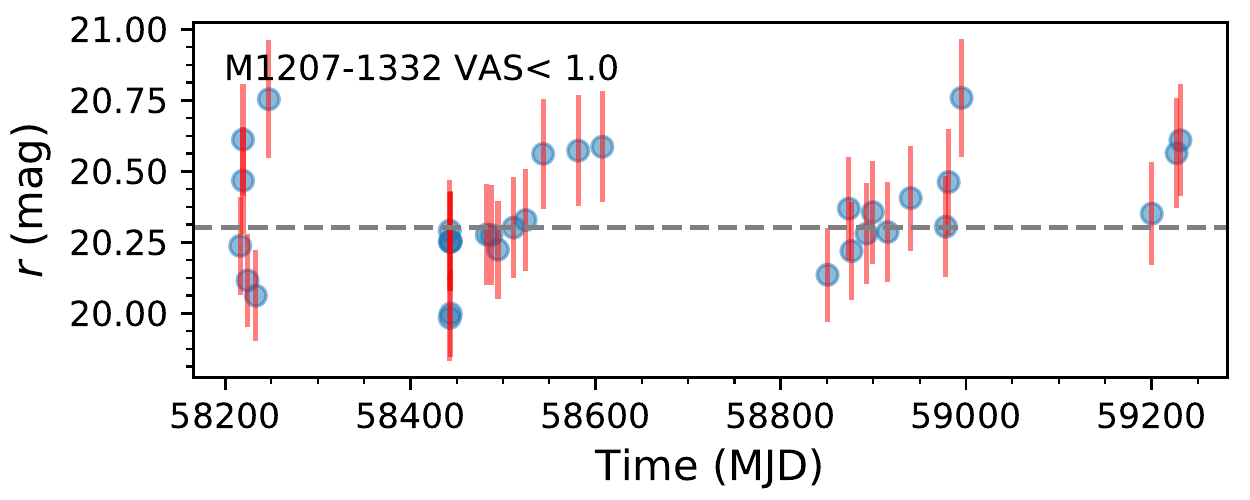} 
}}
\centerline{\hbox{ 
\includegraphics[trim = {0cm 0cm 0cm 0cm}, width=0.50\textwidth,angle=0]{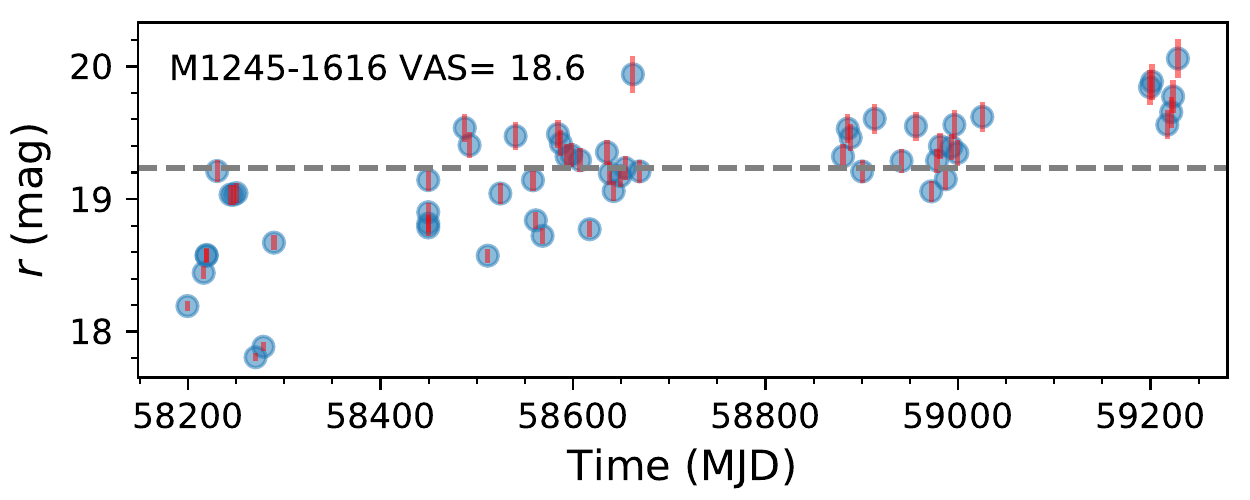} 
\includegraphics[trim = {0cm 0cm 0cm 0cm}, width=0.50\textwidth,angle=0]{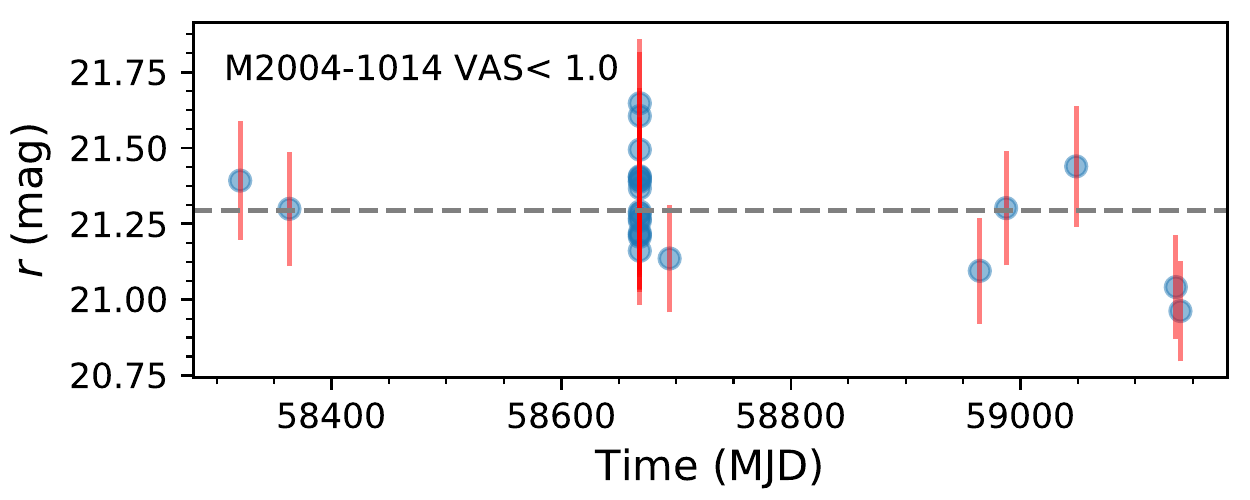} 
}}
\centerline{\hbox{ 
\includegraphics[trim = {0cm 0cm 0cm 0cm}, width=0.50\textwidth,angle=0]{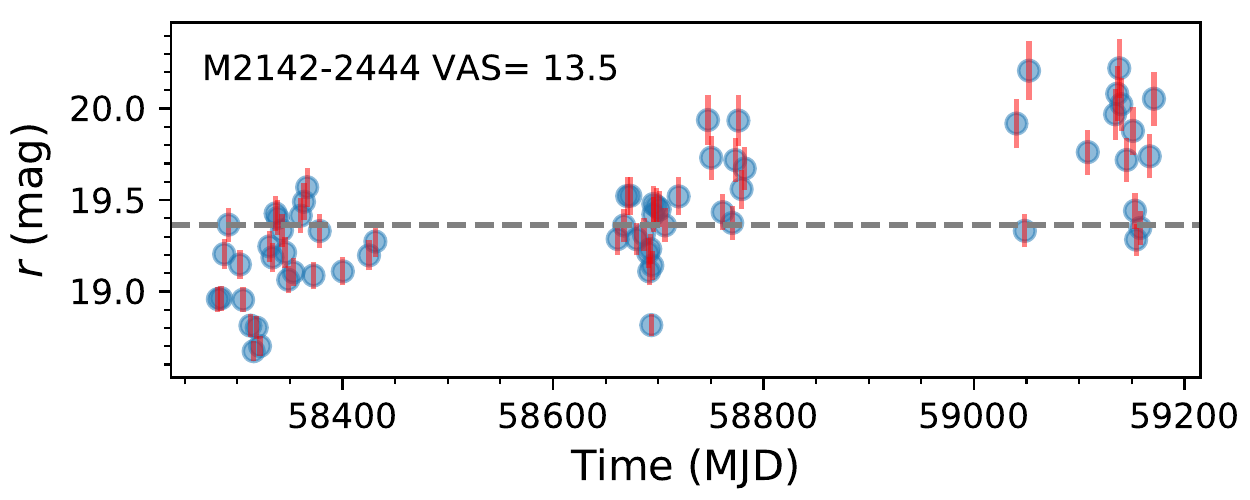} 
\includegraphics[trim = {0cm 0cm 0cm 0cm}, width=0.50\textwidth,angle=0]{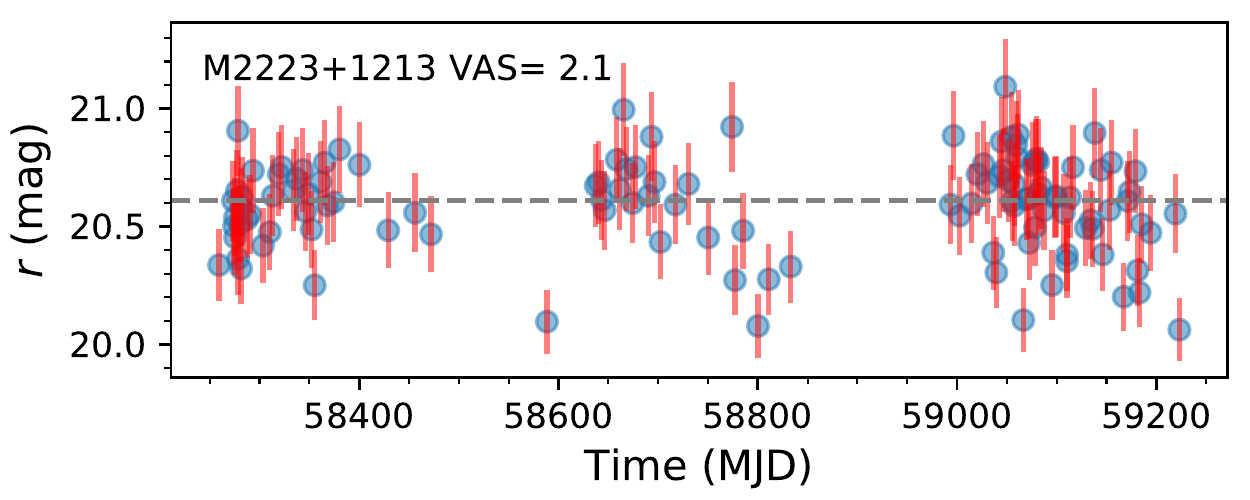}
}}
}}  
\vskip+0.0cm  
\caption{
ZTF light curves ($r$-band) and the variability amplitude strength (VAS) for targets with no emission lines.  The horizontal dashed line represents the median value. 
} 
\label{fig:ztf}   
\end{figure*} 

\begin{figure} 
\centerline{\hbox{
\centerline{\vbox{ 
\includegraphics[trim = {0cm 0cm 0cm 0cm}, width=0.50\textwidth,angle=0]{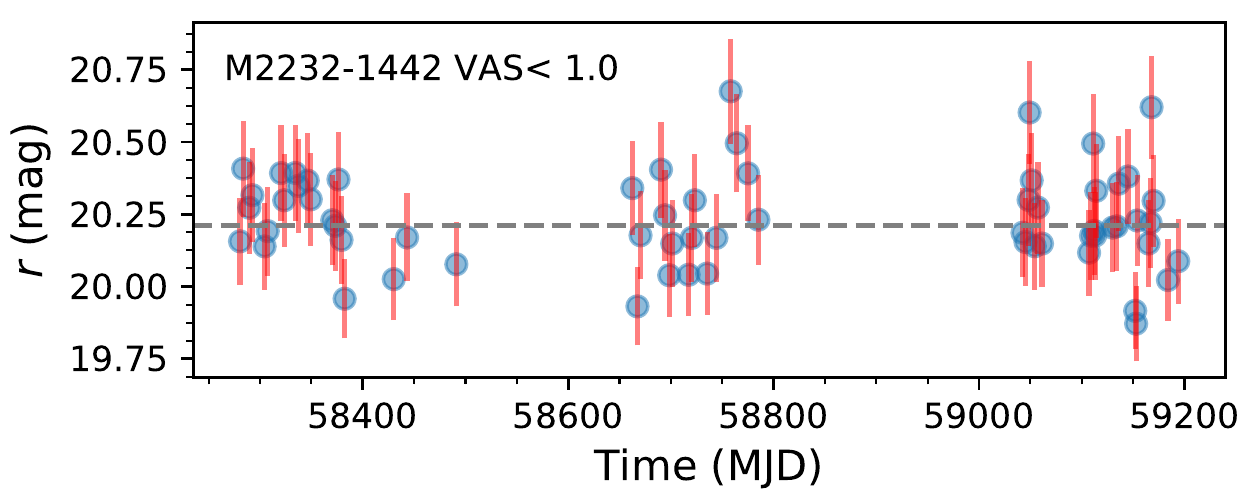} 
\includegraphics[trim = {0cm 0cm 0cm 0cm}, width=0.50\textwidth,angle=0]{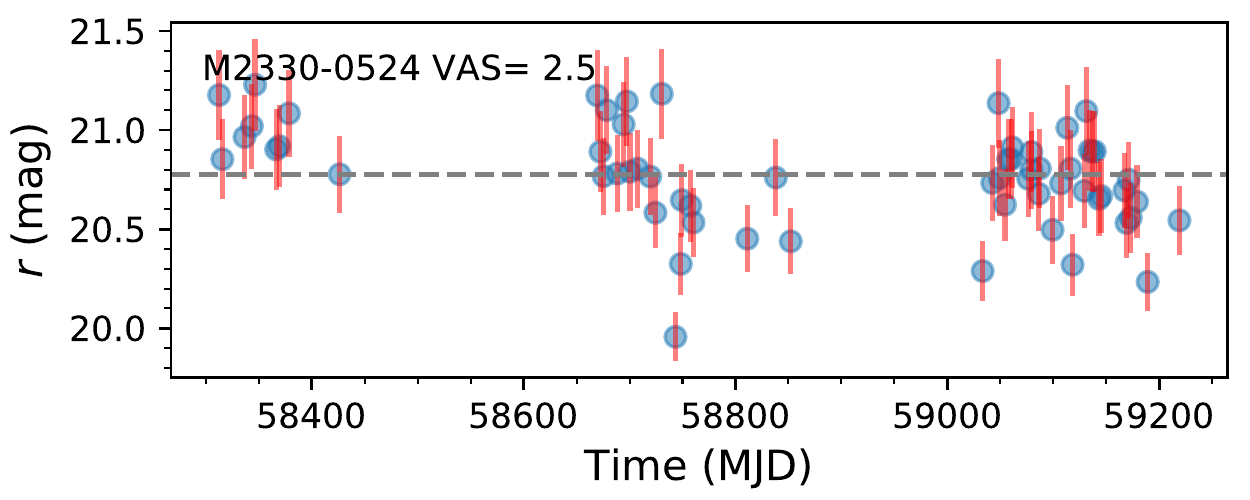} 
\includegraphics[trim = {0cm 0cm 0cm 0cm}, width=0.50\textwidth,angle=0]{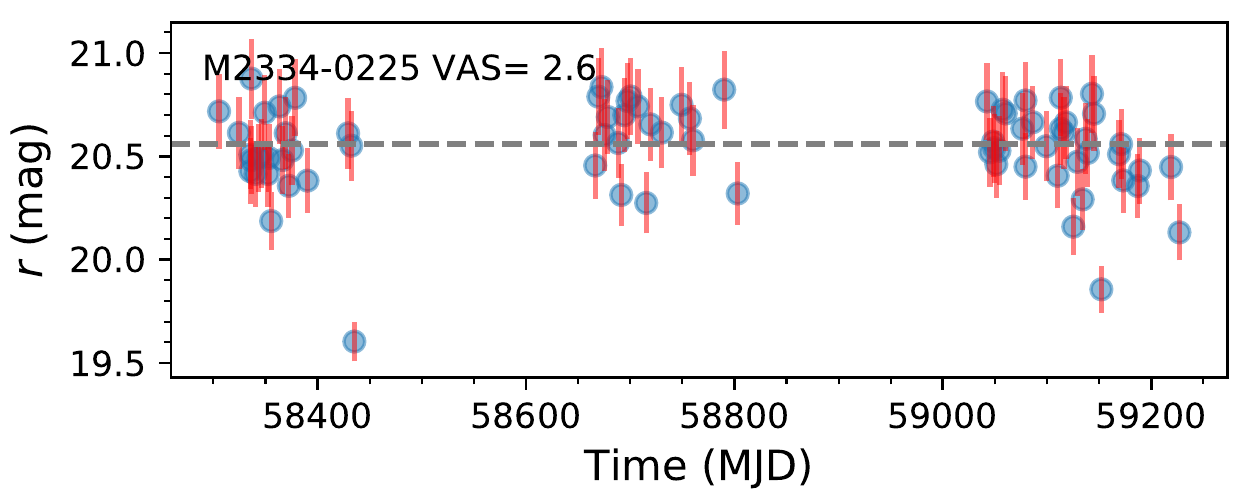} 
}}
}}  
\centering
	{\bf Figure \ref{fig:ztf}} {\it continued}.
\vskip+0.0cm  
\end{figure} 

\begin{figure*} 
\centerline{\vbox{
\centerline{\hbox{ 
\includegraphics[trim = {0cm 0cm 0cm 0cm}, width=0.30\textwidth,angle=0]{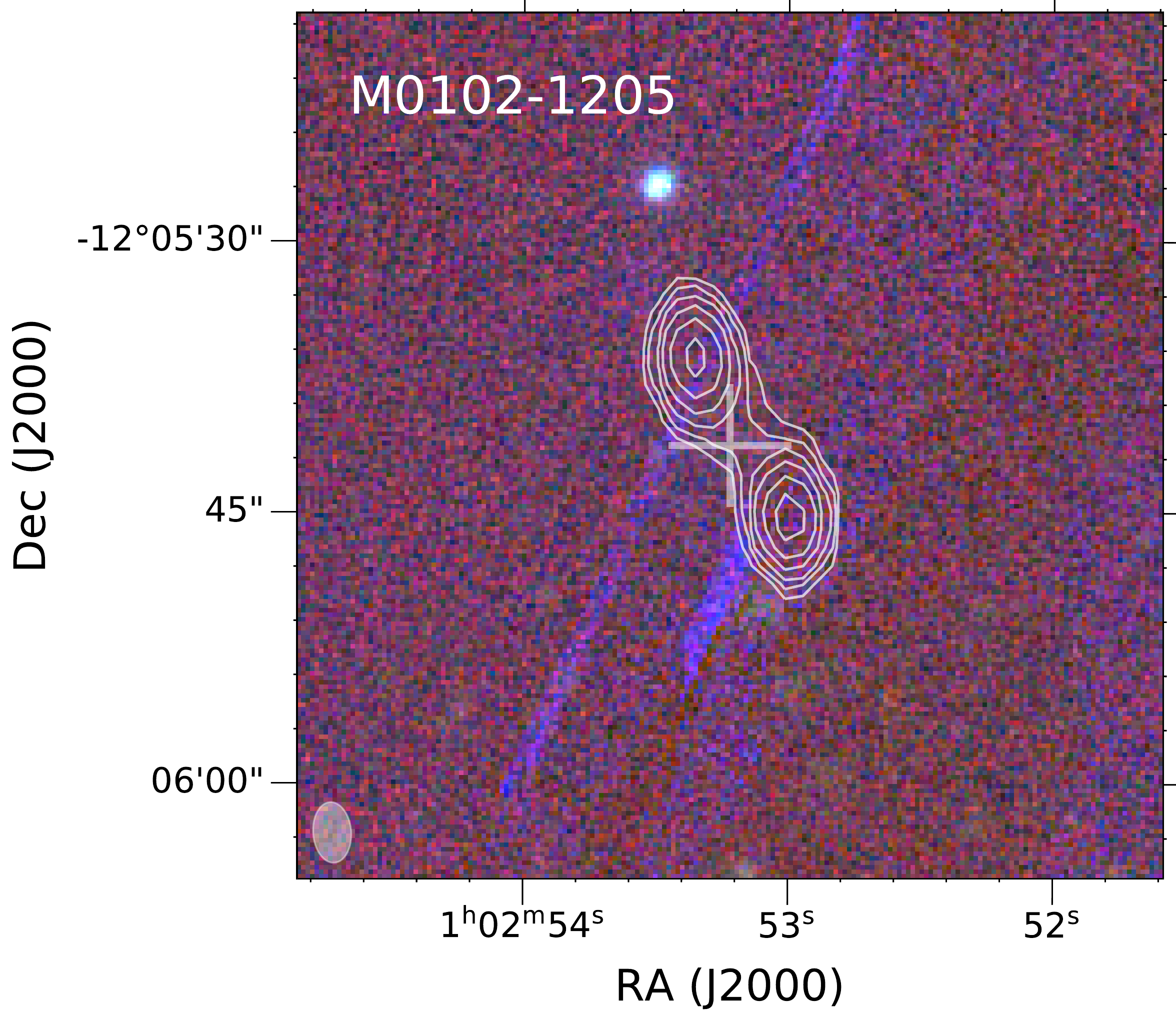} 
\includegraphics[trim = {0cm 0cm 0cm 0cm}, width=0.30\textwidth,angle=0]{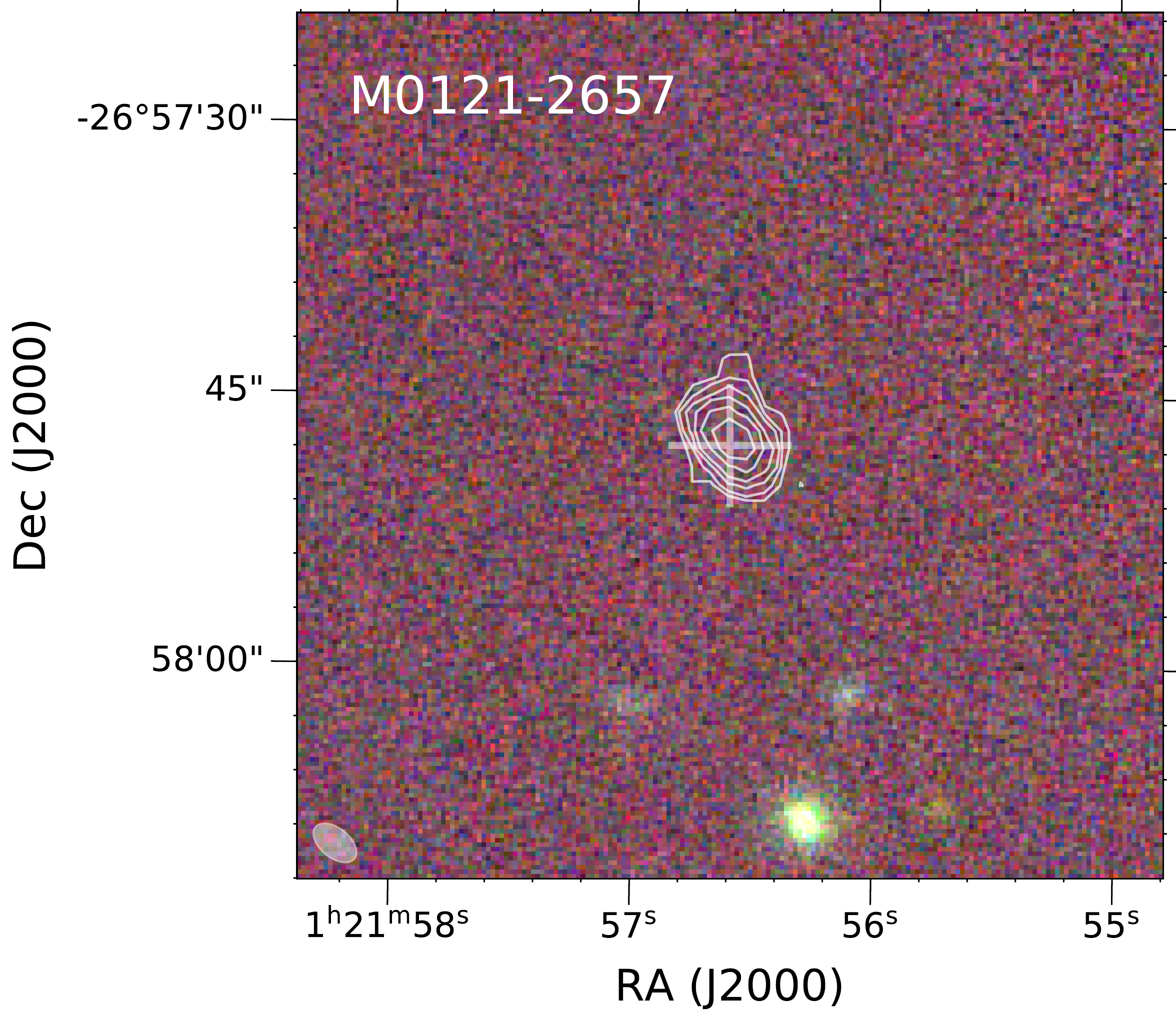} 
\includegraphics[trim = {0cm 0cm 0cm 0cm}, width=0.30\textwidth,angle=0]{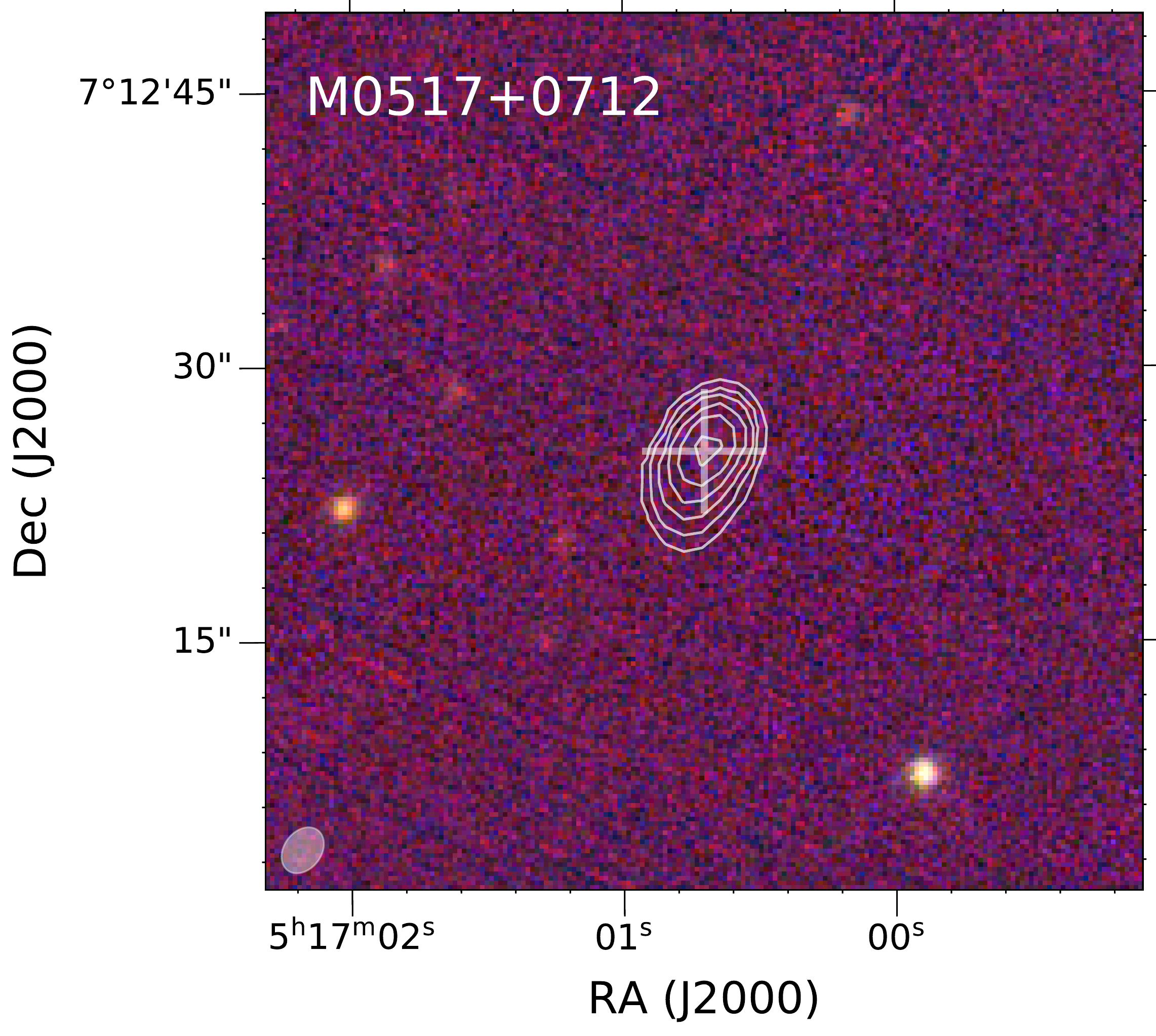} 
}}
\centerline{\hbox{ 
\includegraphics[trim = {0cm 0cm 0cm 0cm}, width=0.30\textwidth,angle=0]{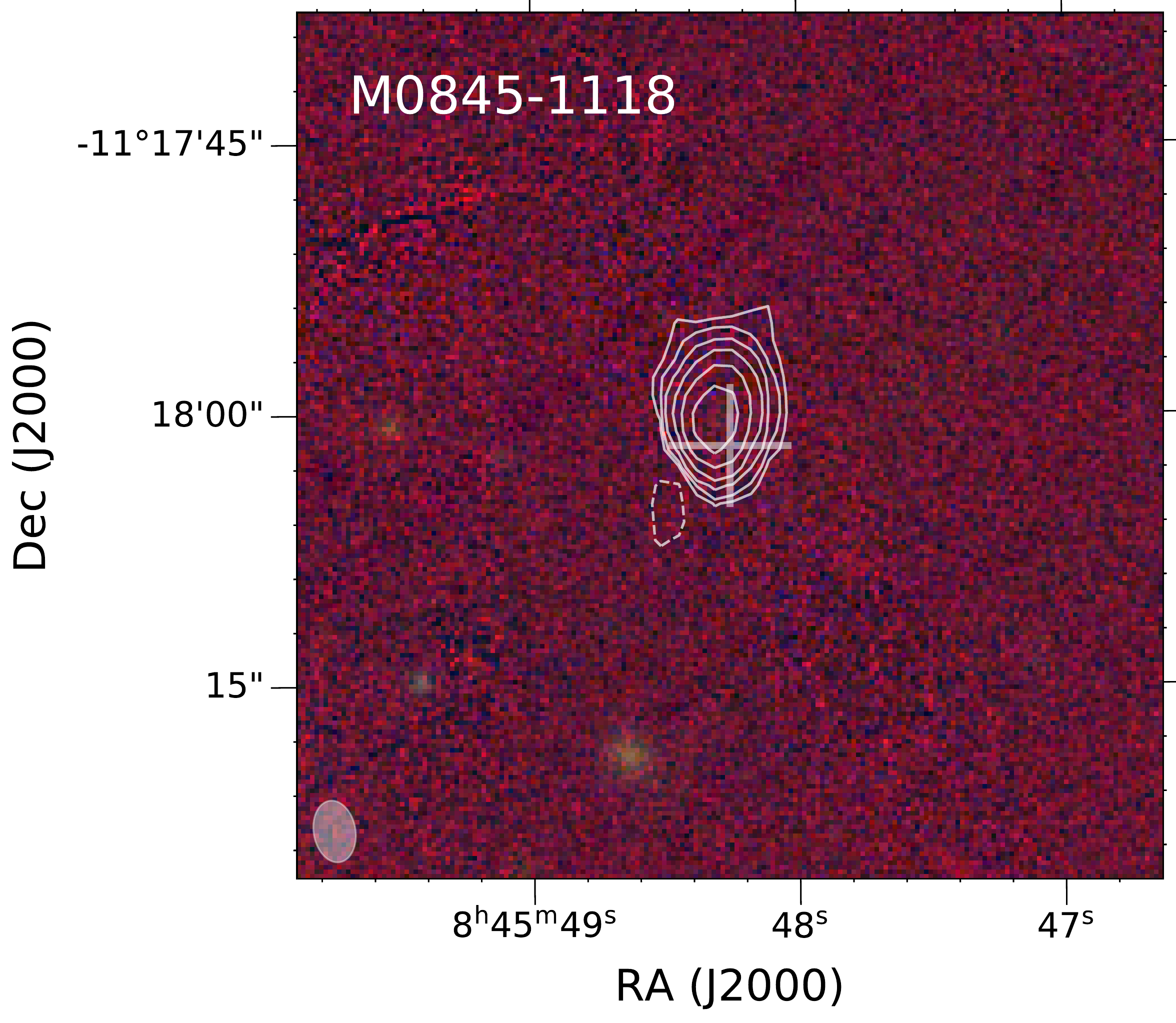}  
\includegraphics[trim = {0cm 0cm 0cm 0cm}, width=0.30\textwidth,angle=0]{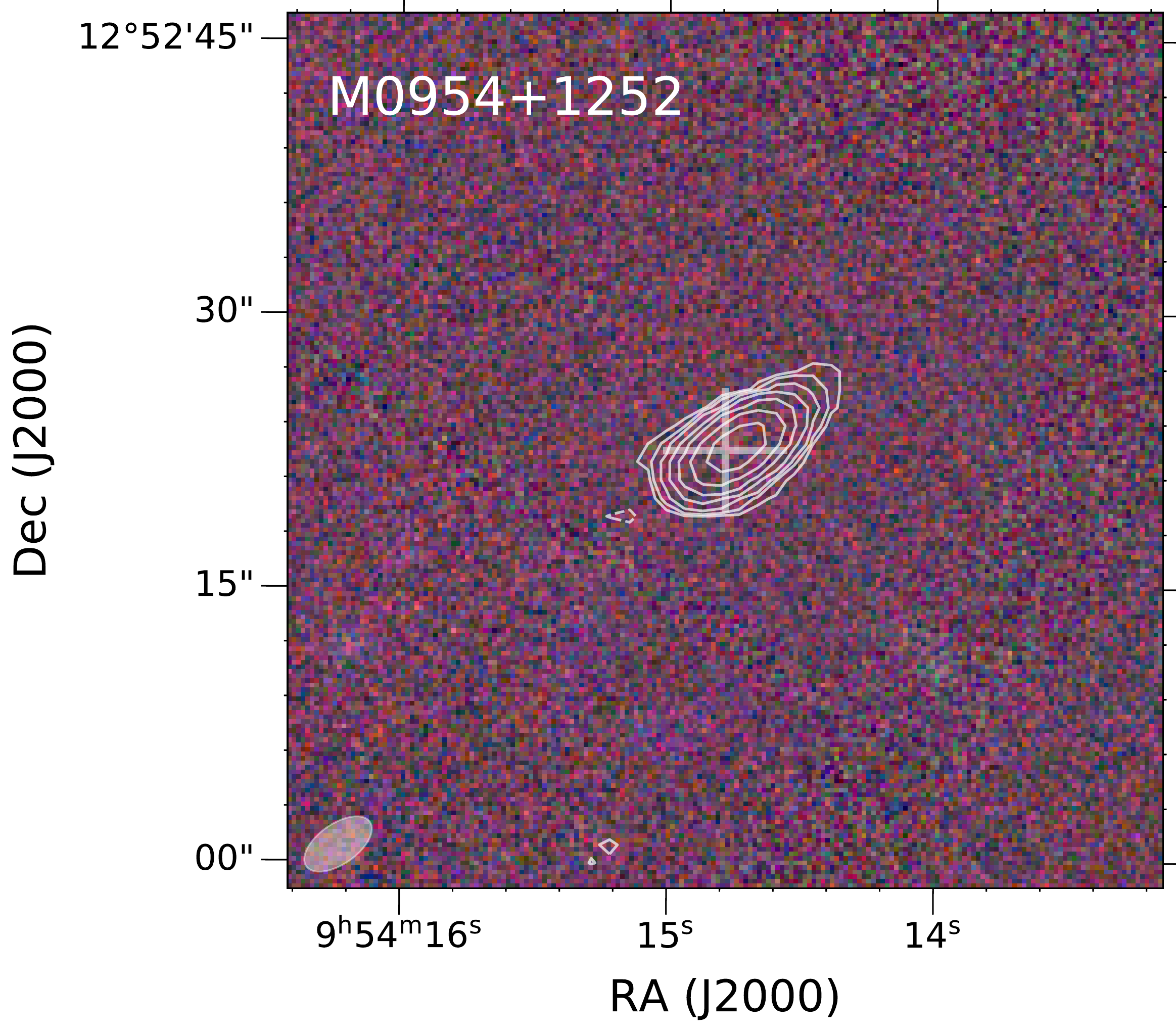}
\includegraphics[trim = {0cm 0cm 0cm 0cm}, width=0.30\textwidth,angle=0]{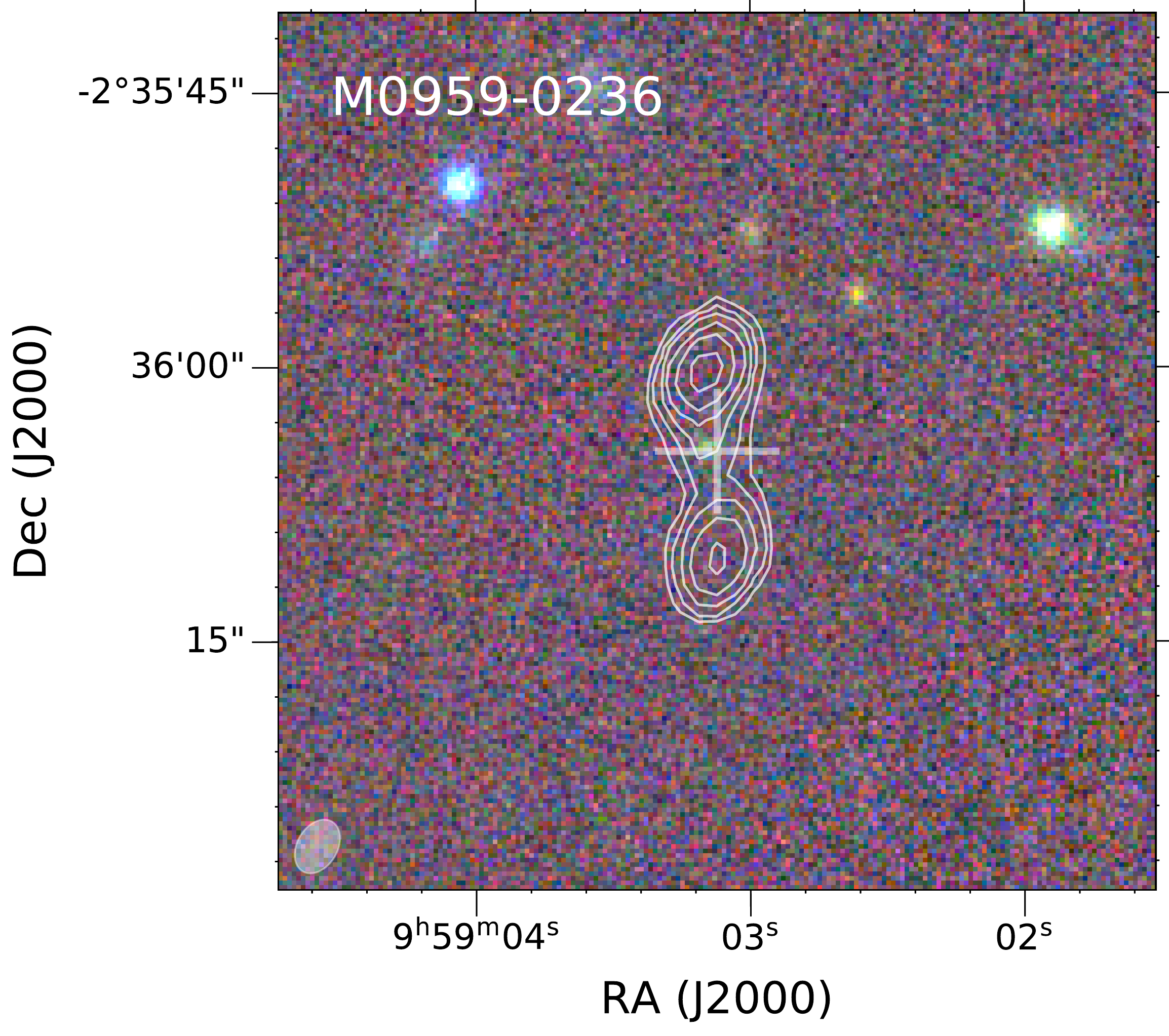} 
}}
\centerline{\hbox{ 
\includegraphics[trim = {0cm 0cm 0cm 0cm}, width=0.30\textwidth,angle=0]{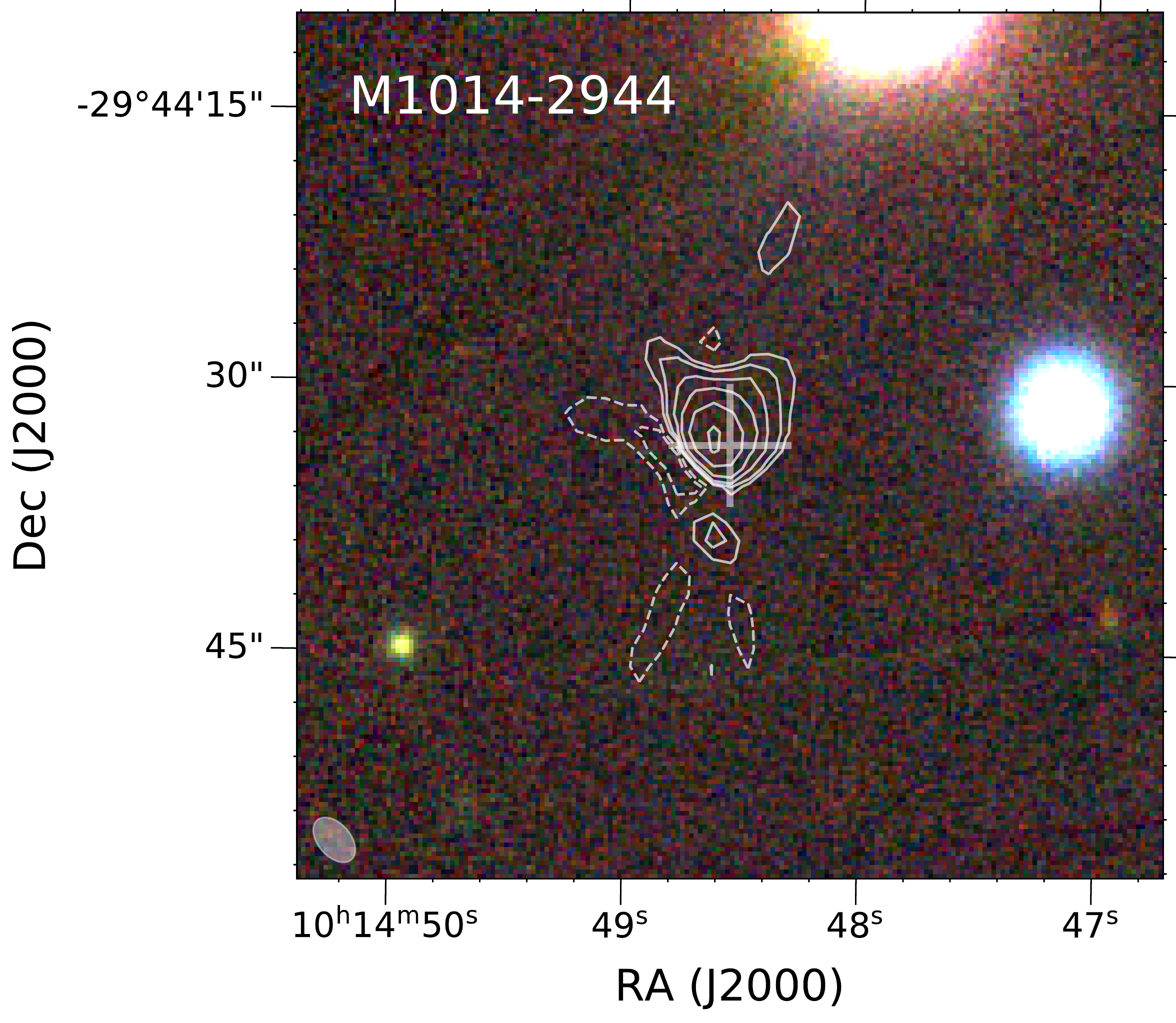} 
\includegraphics[trim = {0cm 0cm 0cm 0cm}, width=0.30\textwidth,angle=0]{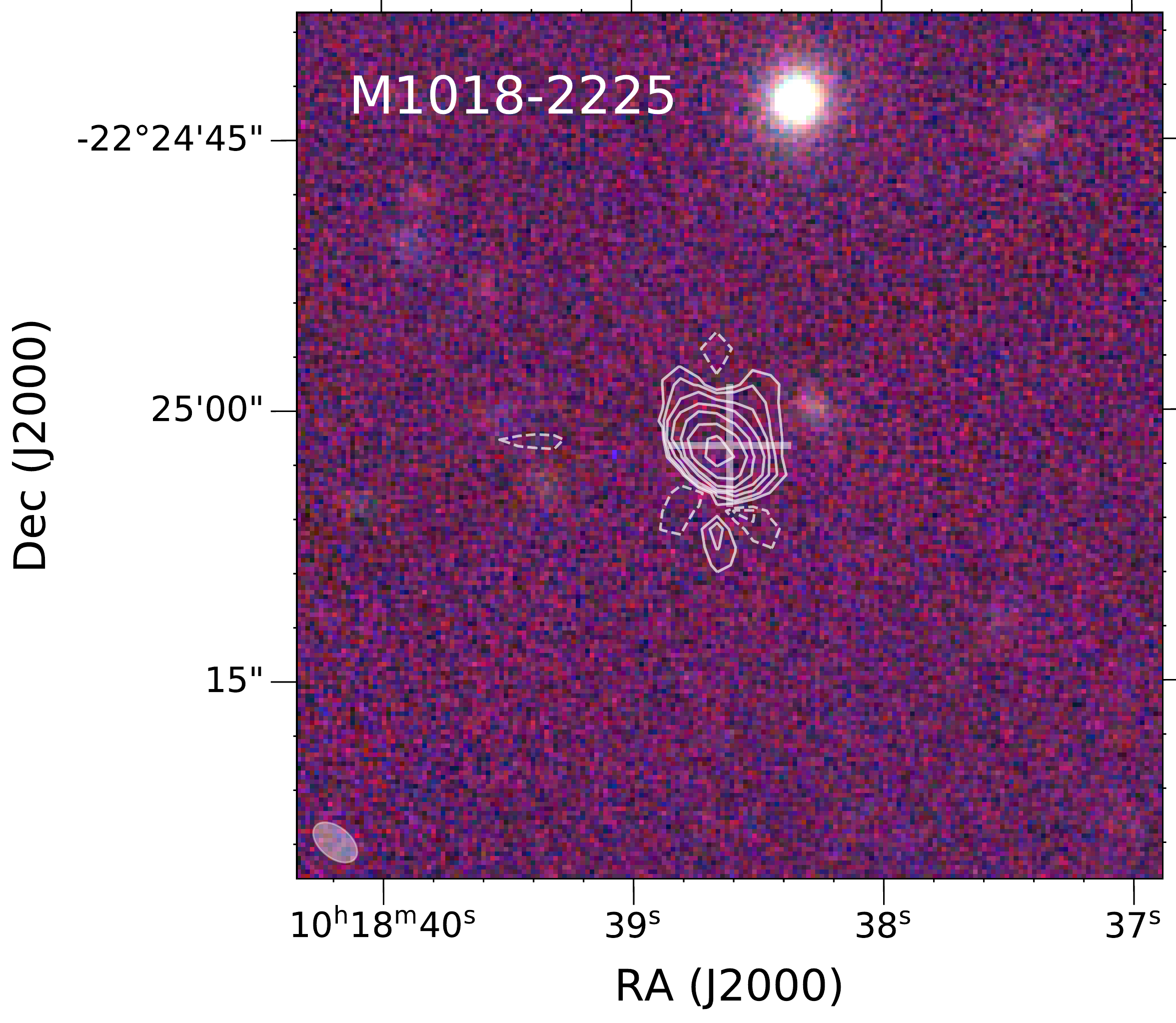}
\includegraphics[trim = {0cm 0cm 0cm 0cm}, width=0.30\textwidth,angle=0]{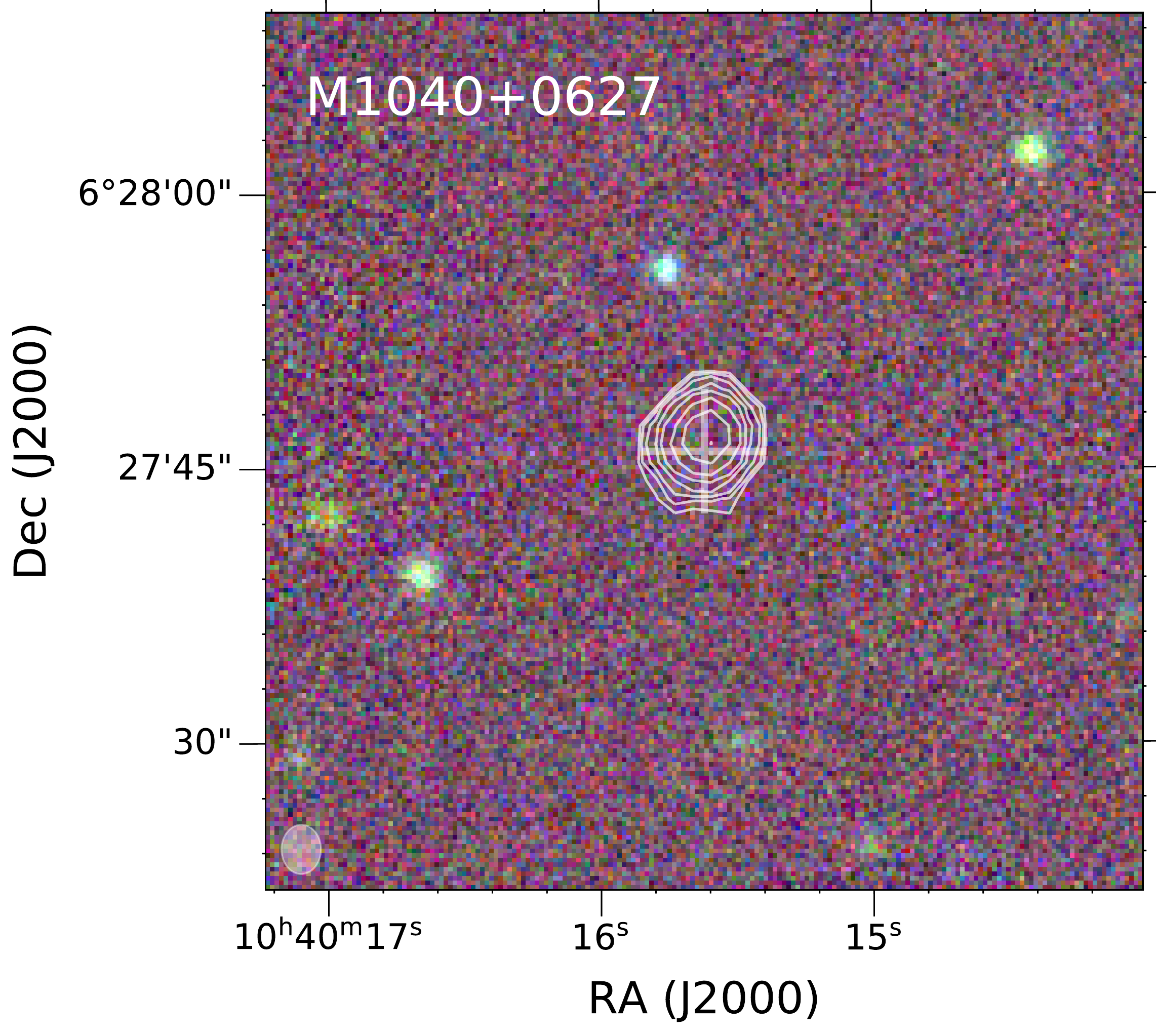}  
}}
\centerline{\hbox{ 
\includegraphics[trim = {0cm 0cm 0cm 0cm}, width=0.30\textwidth,angle=0]{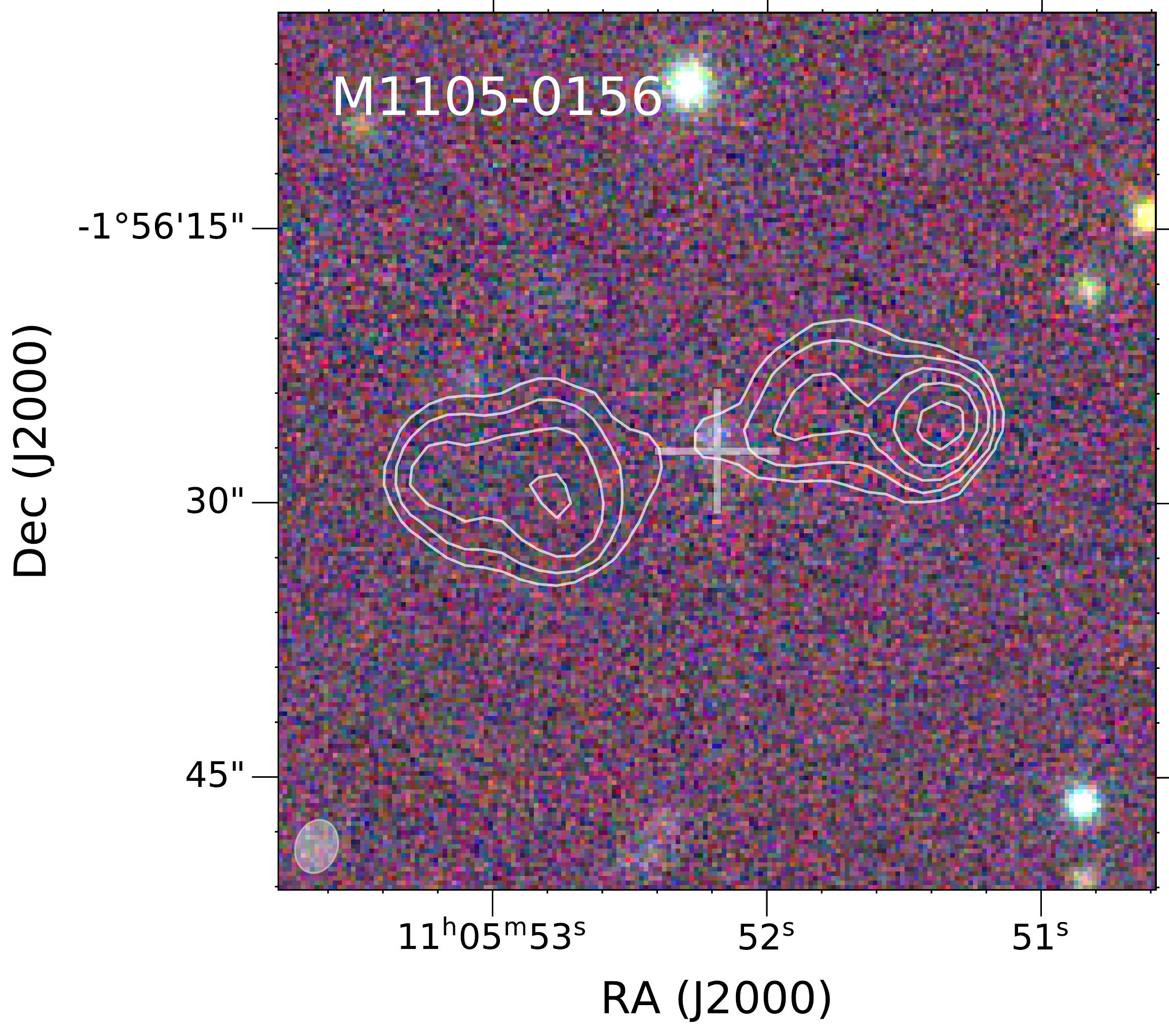} 
\includegraphics[trim = {0cm 0cm 0cm 0cm}, width=0.30\textwidth,angle=0]{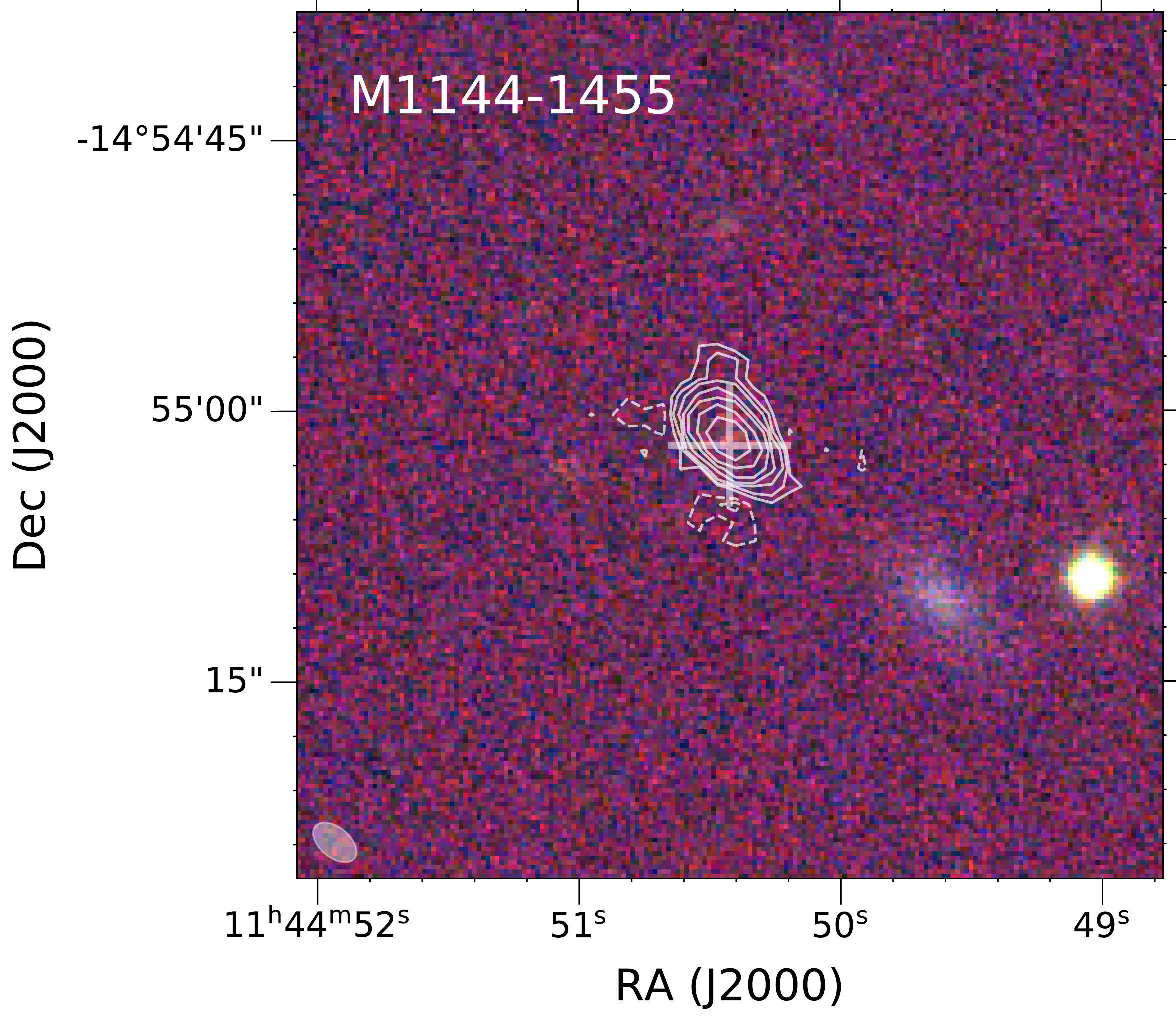}
\includegraphics[trim = {0cm 0cm 0cm 0cm}, width=0.30\textwidth,angle=0]{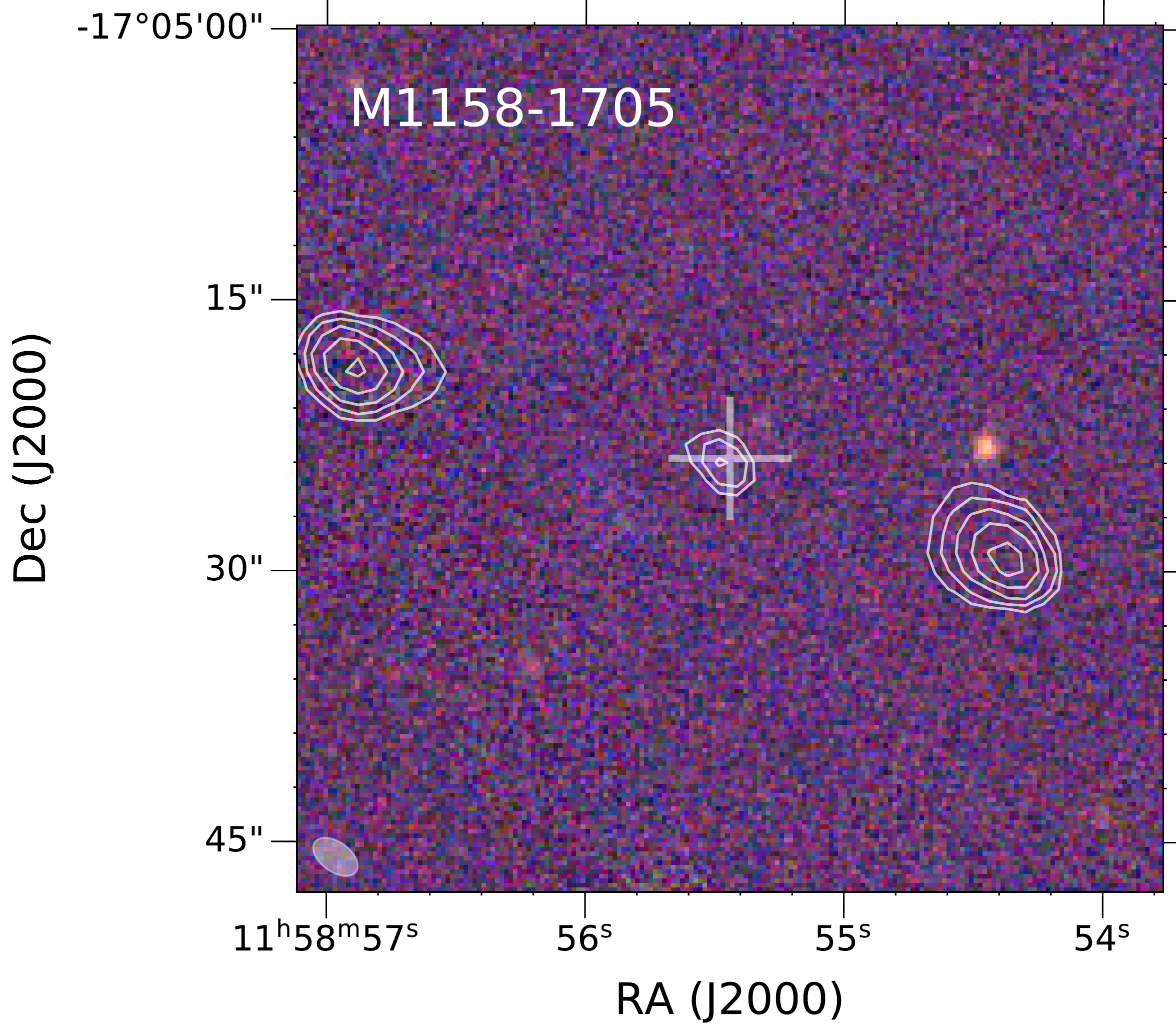}  
}}
}}  
\vskip+0.0cm  
\caption{
VLASS (3\,GHz) contours overlaid on PS1 $yig$ color composite images.   The contour levels are shown at 1.5$\times$(-1, 1, 2, 4, 8, ...)\,mJy\,beam$^{-1}$. The synthesized beam is shown at the bottom left corner.  The position of WISE source is marked with a cross. 
} 
\label{fig:maps}   
\end{figure*} 

\begin{figure*} 
\centerline{\vbox{
\centerline{\hbox{ 
\includegraphics[trim = {0cm 0cm 0cm 0cm}, width=0.30\textwidth,angle=0]{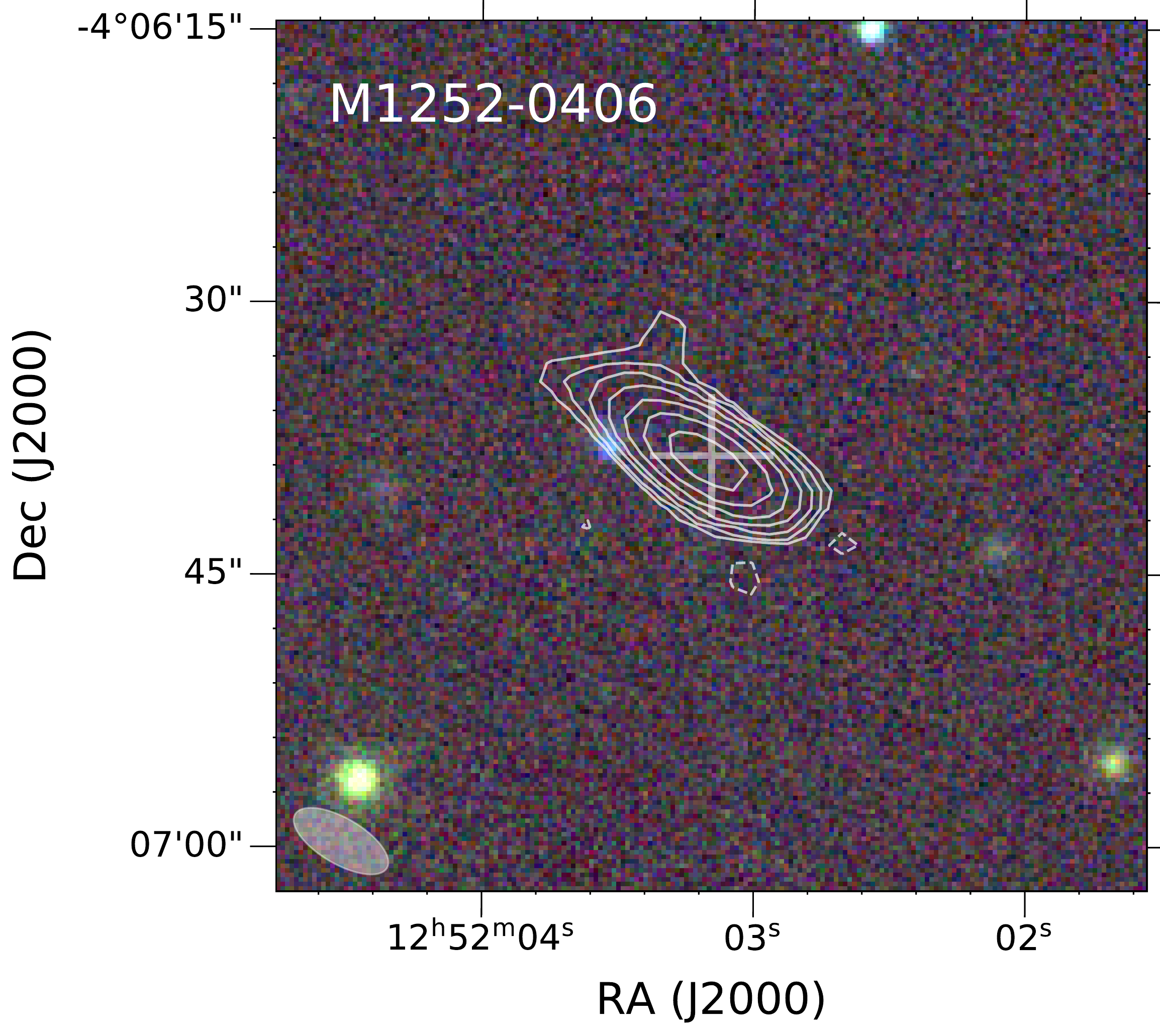} 
\includegraphics[trim = {0cm 0cm 0cm 0cm}, width=0.30\textwidth,angle=0]{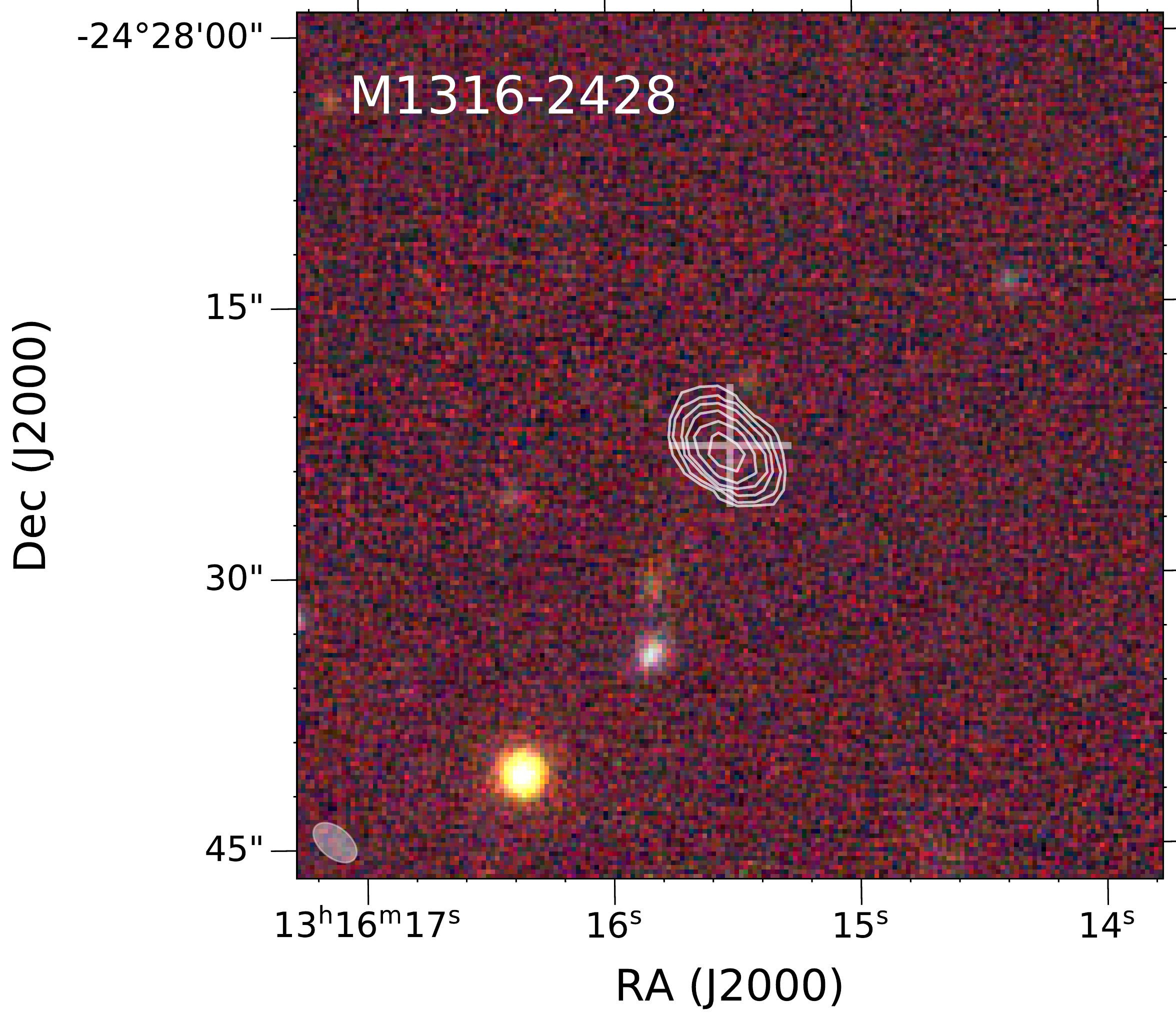} 
\includegraphics[trim = {0cm 0cm 0cm 0cm}, width=0.30\textwidth,angle=0]{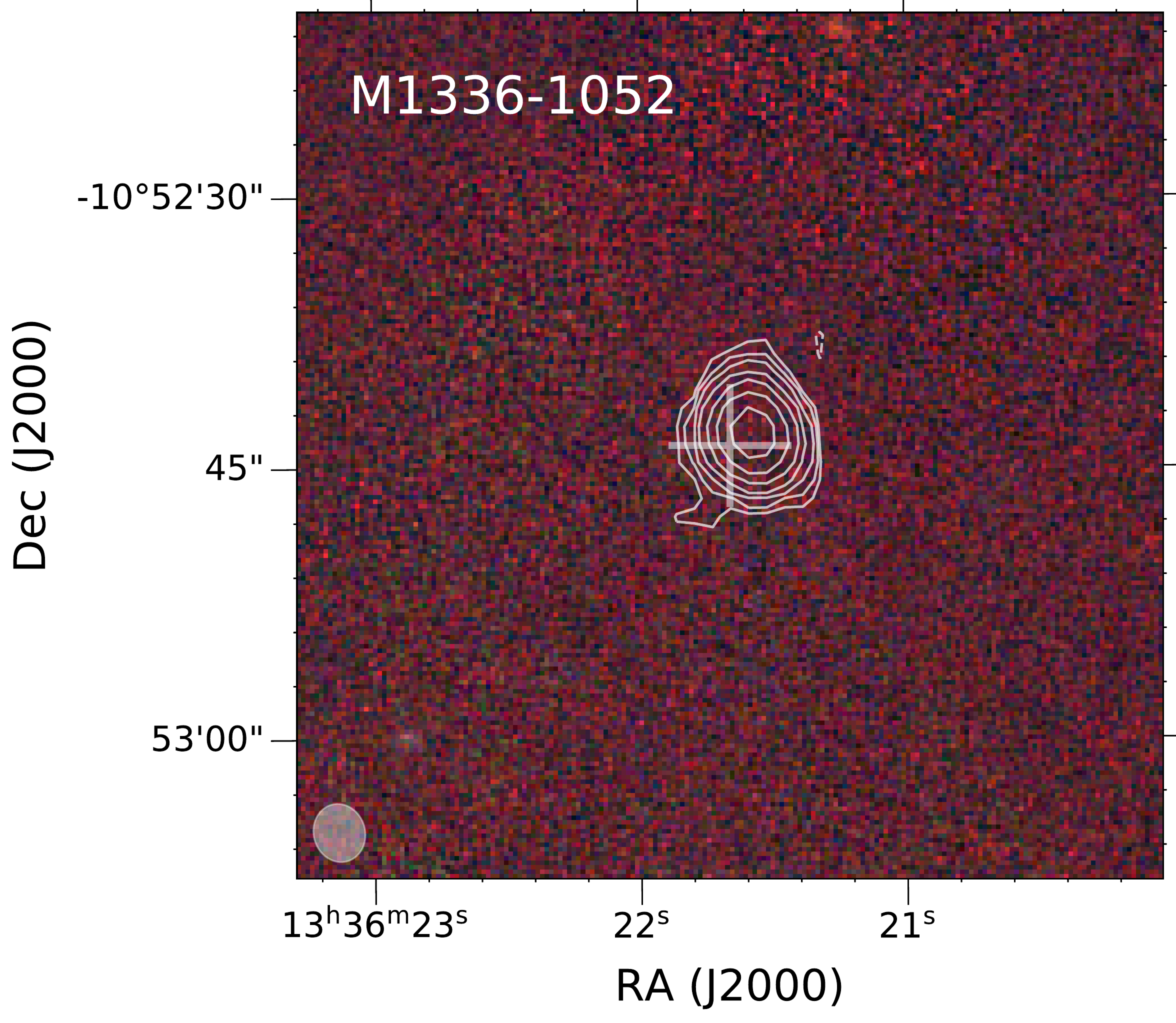} 
}}
\centerline{\hbox{ 
\includegraphics[trim = {0cm 0cm 0cm 0cm}, width=0.30\textwidth,angle=0]{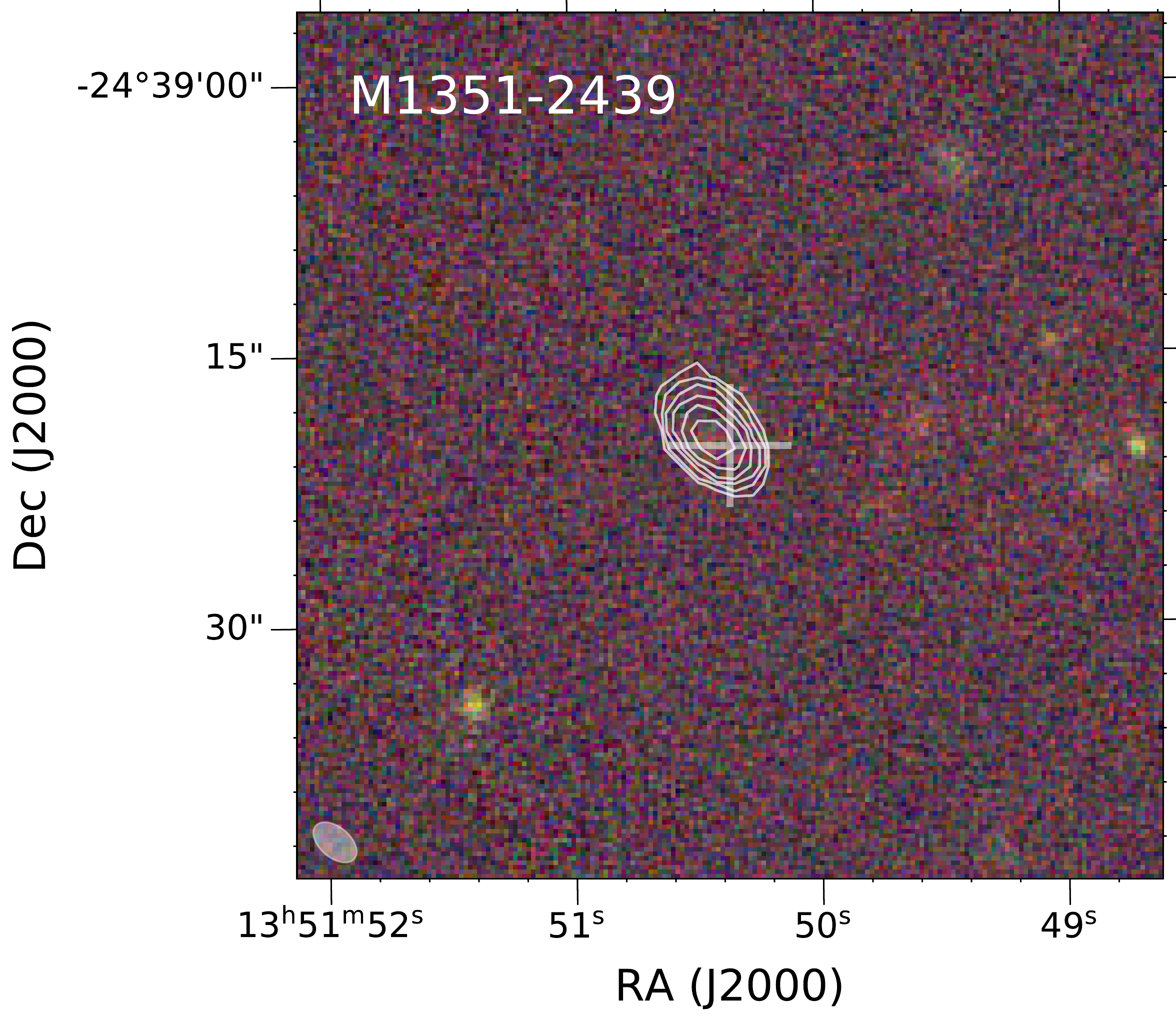}  
\includegraphics[trim = {0cm 0cm 0cm 0cm}, width=0.30\textwidth,angle=0]{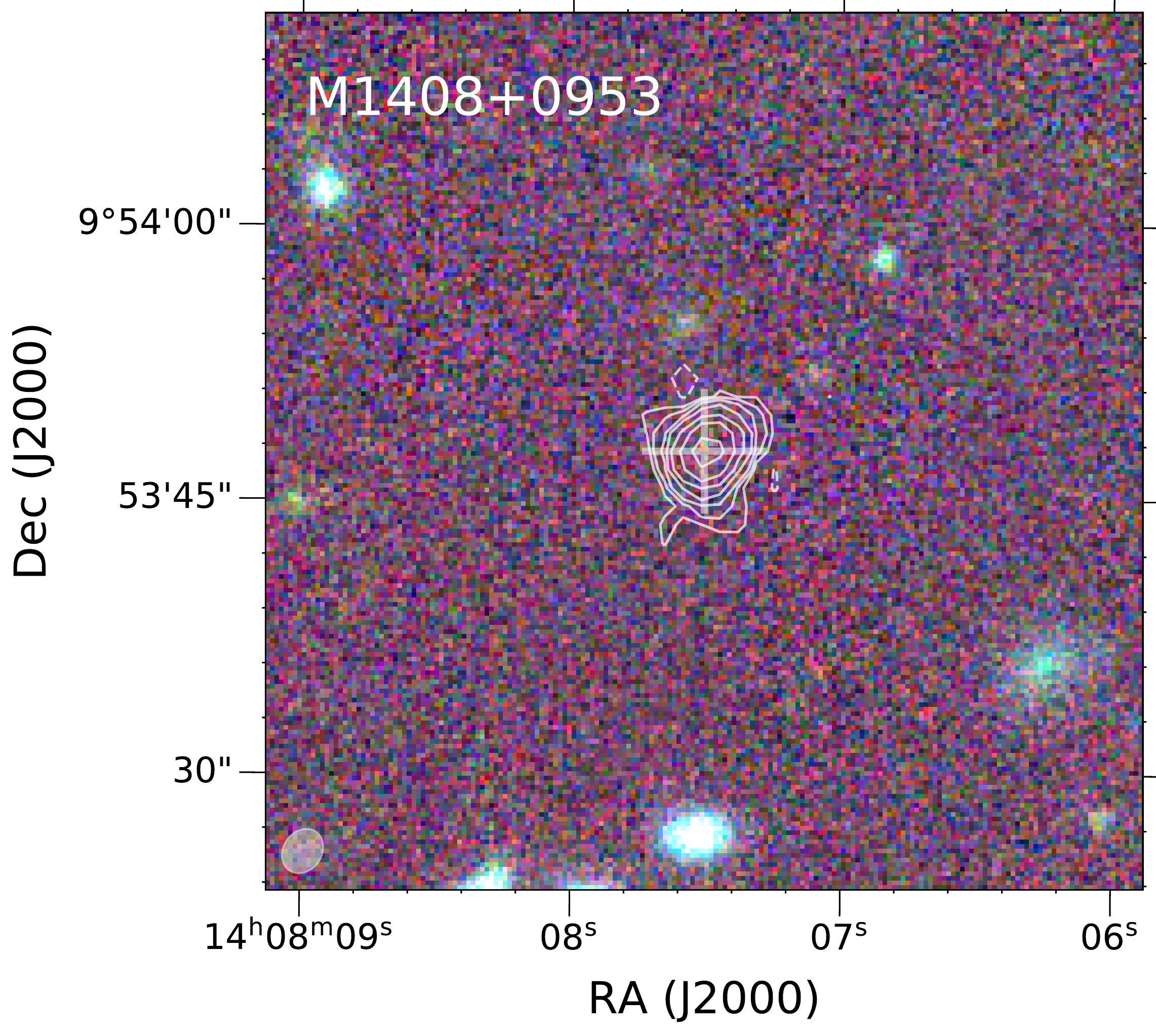}
\includegraphics[trim = {0cm 0cm 0cm 0cm}, width=0.30\textwidth,angle=0]{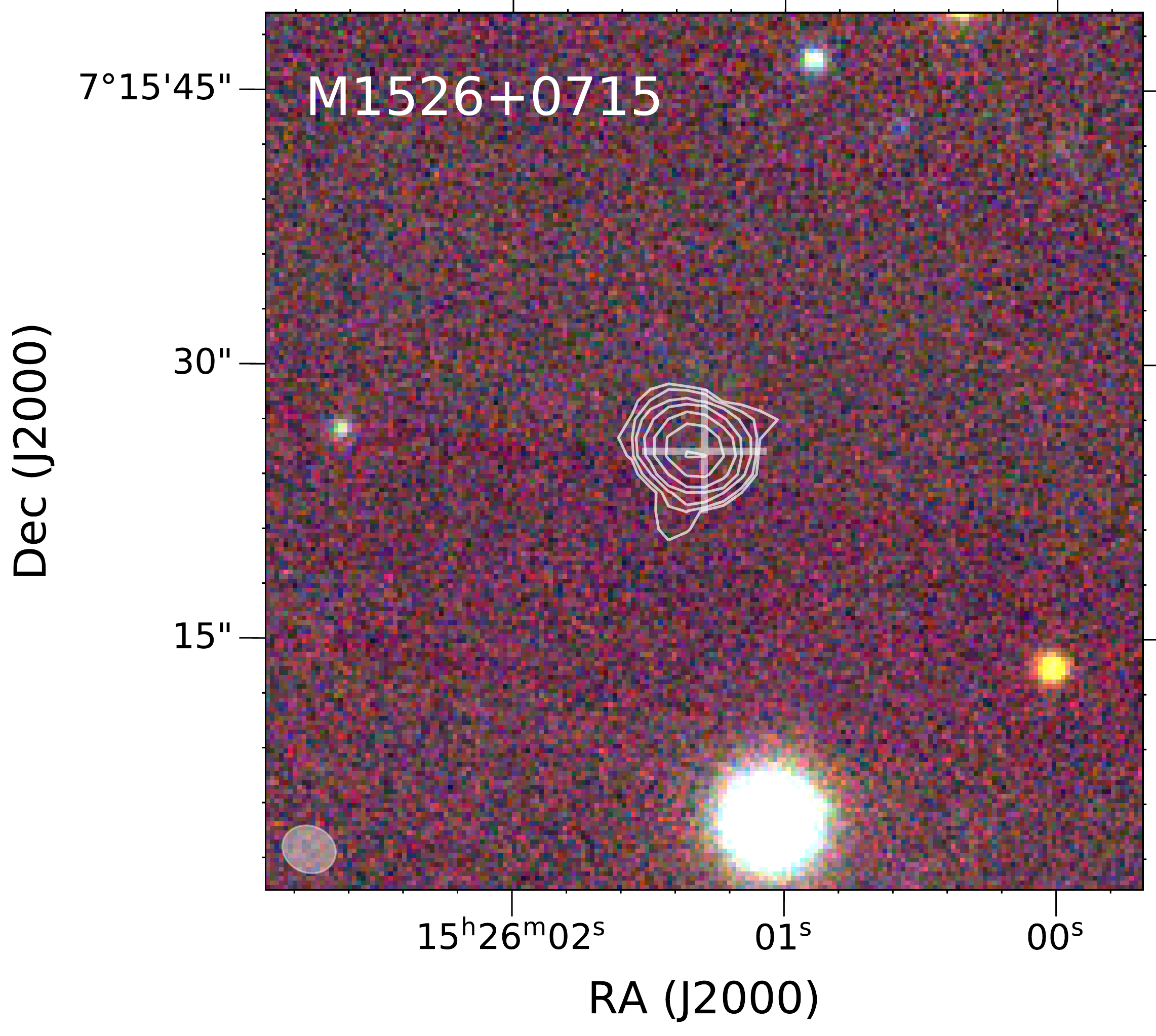} 
}}
\centerline{\hbox{ 
\includegraphics[trim = {0cm 0cm 0cm 0cm}, width=0.30\textwidth,angle=0]{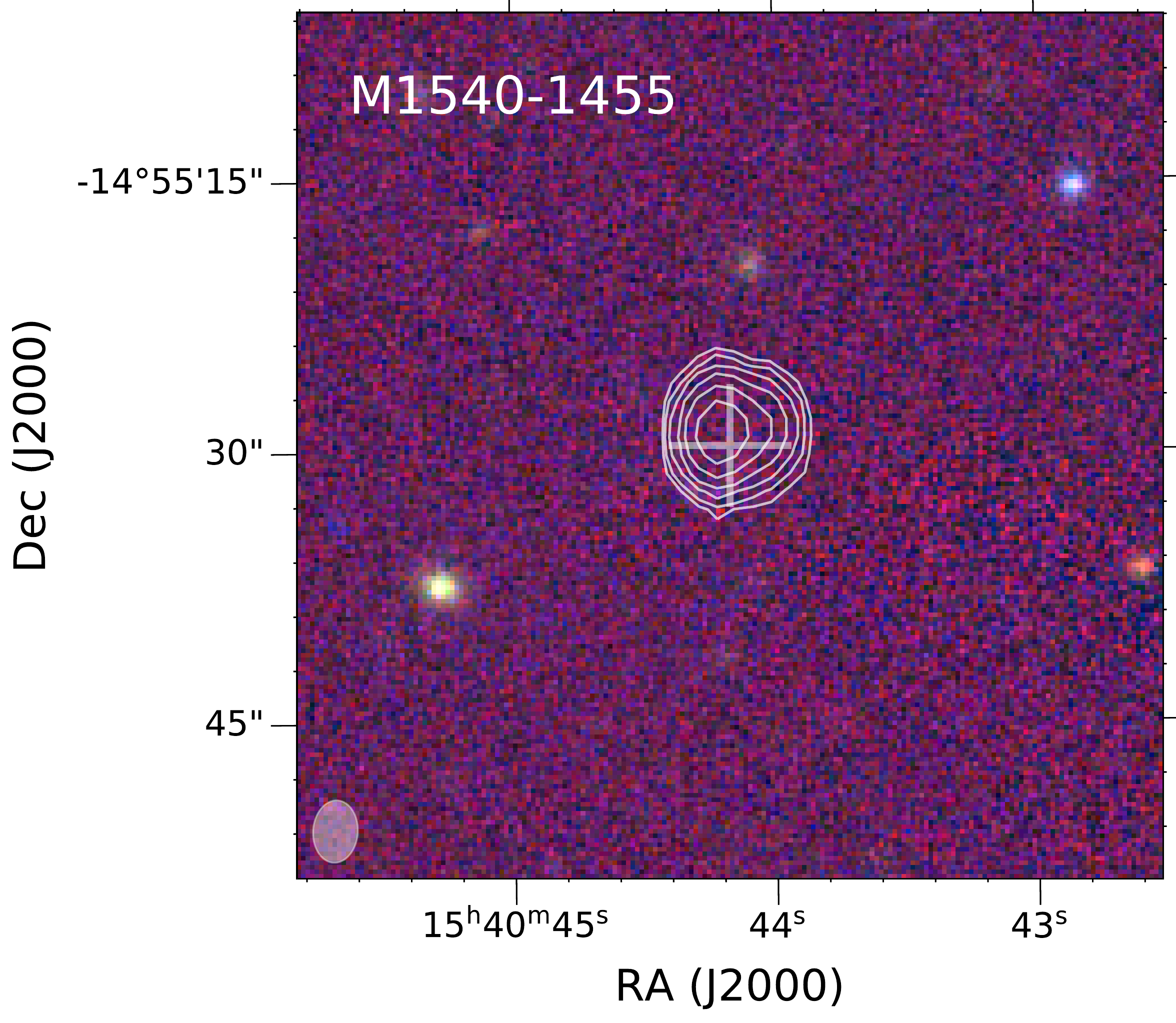} 
\includegraphics[trim = {0cm 0cm 0cm 0cm}, width=0.30\textwidth,angle=0]{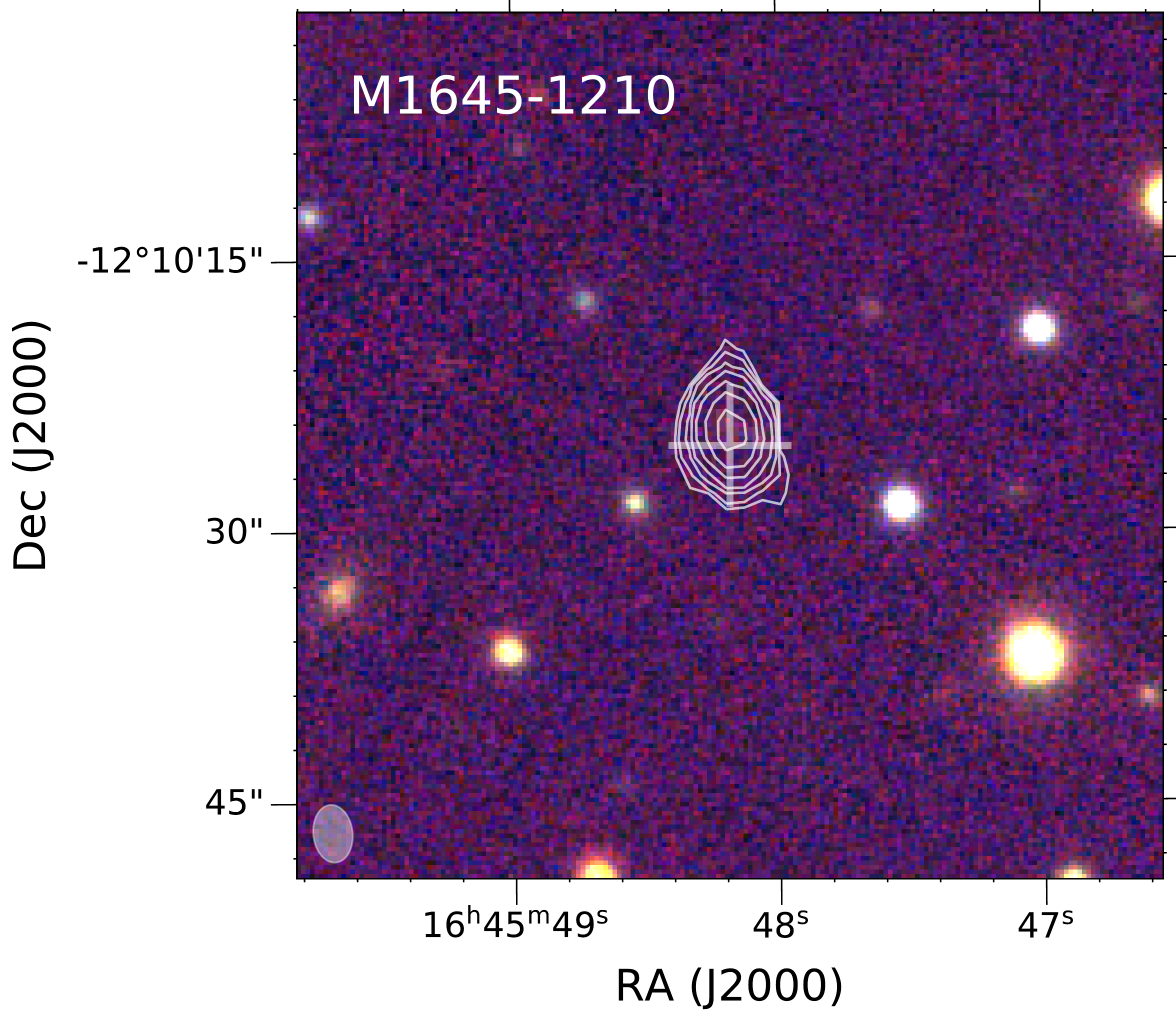}
\includegraphics[trim = {0cm 0cm 0cm 0cm}, width=0.30\textwidth,angle=0]{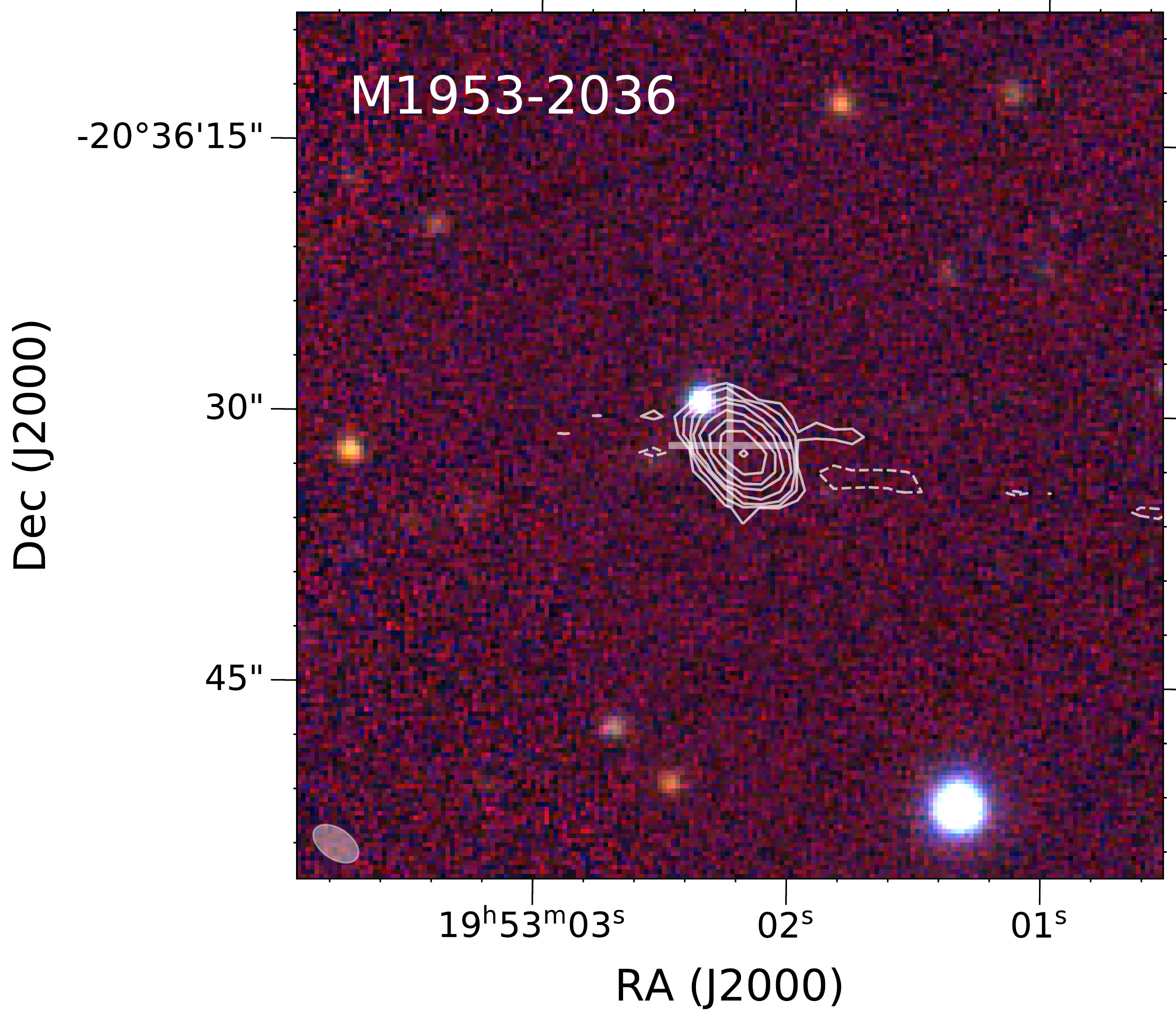}  
}}
\centerline{\hbox{ 
\includegraphics[trim = {0cm 0cm 0cm 0cm}, width=0.30\textwidth,angle=0]{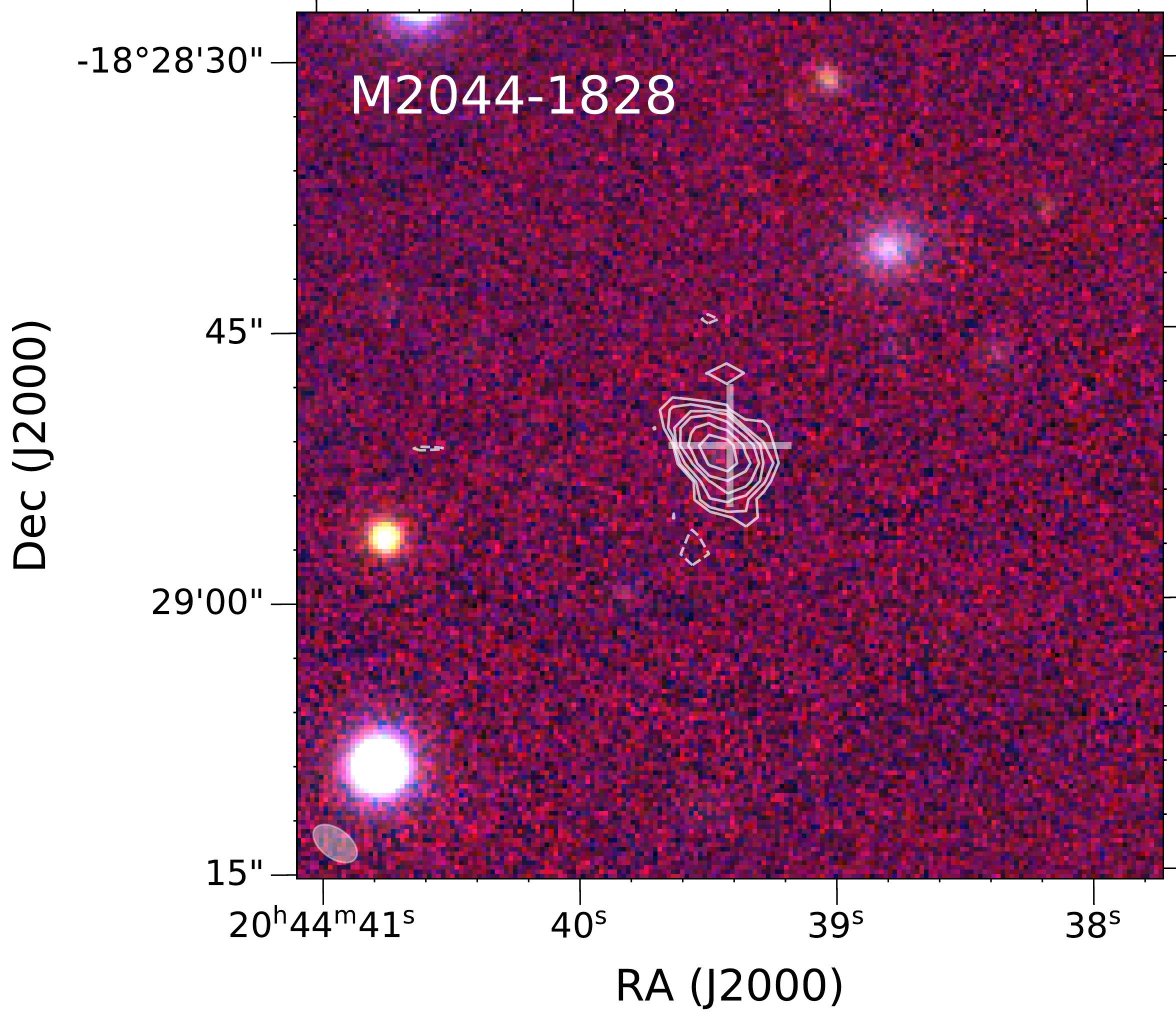} 
\includegraphics[trim = {0cm 0cm 0cm 0cm}, width=0.30\textwidth,angle=0]{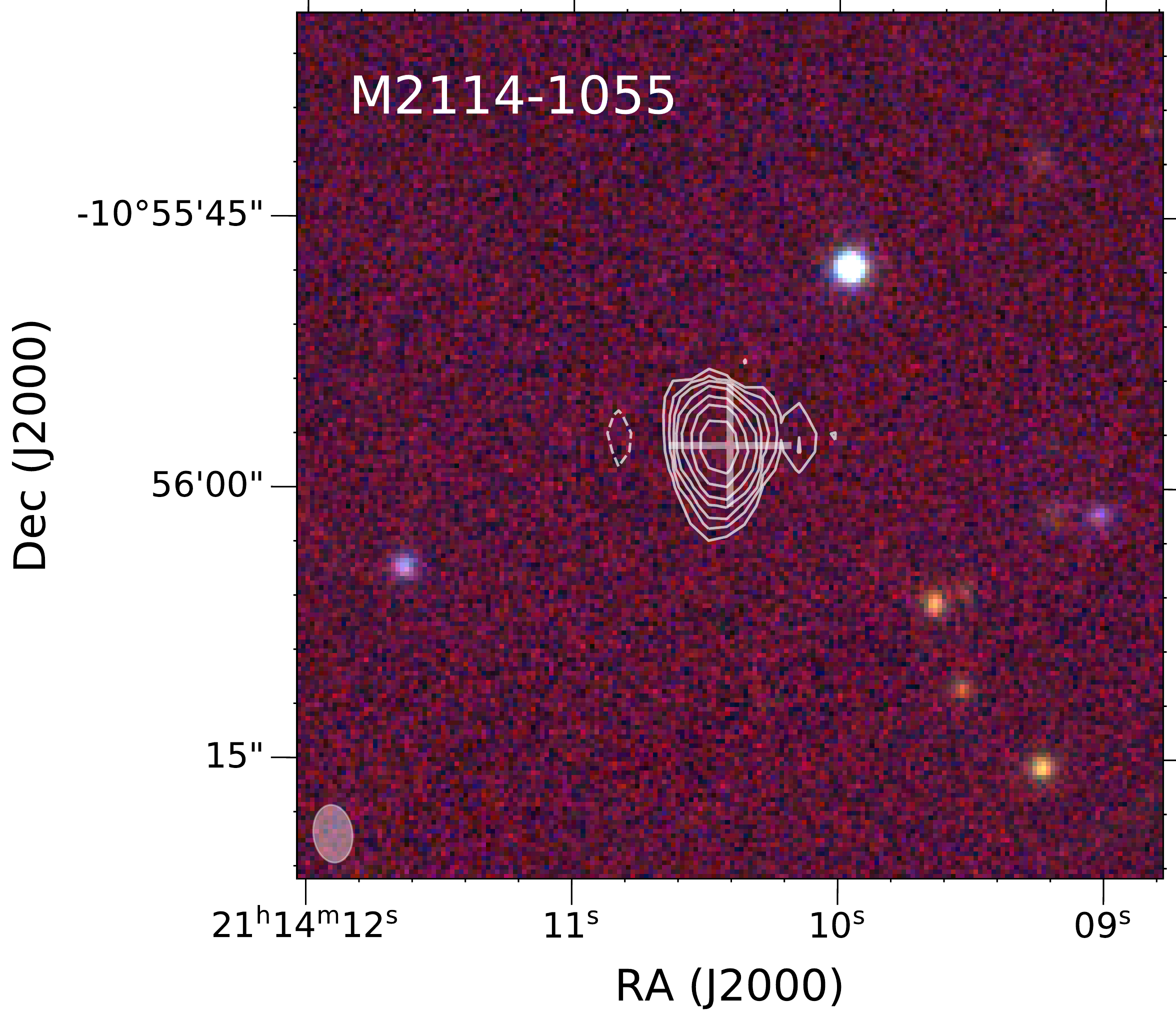}
\includegraphics[trim = {0cm 0cm 0cm 0cm}, width=0.30\textwidth,angle=0]{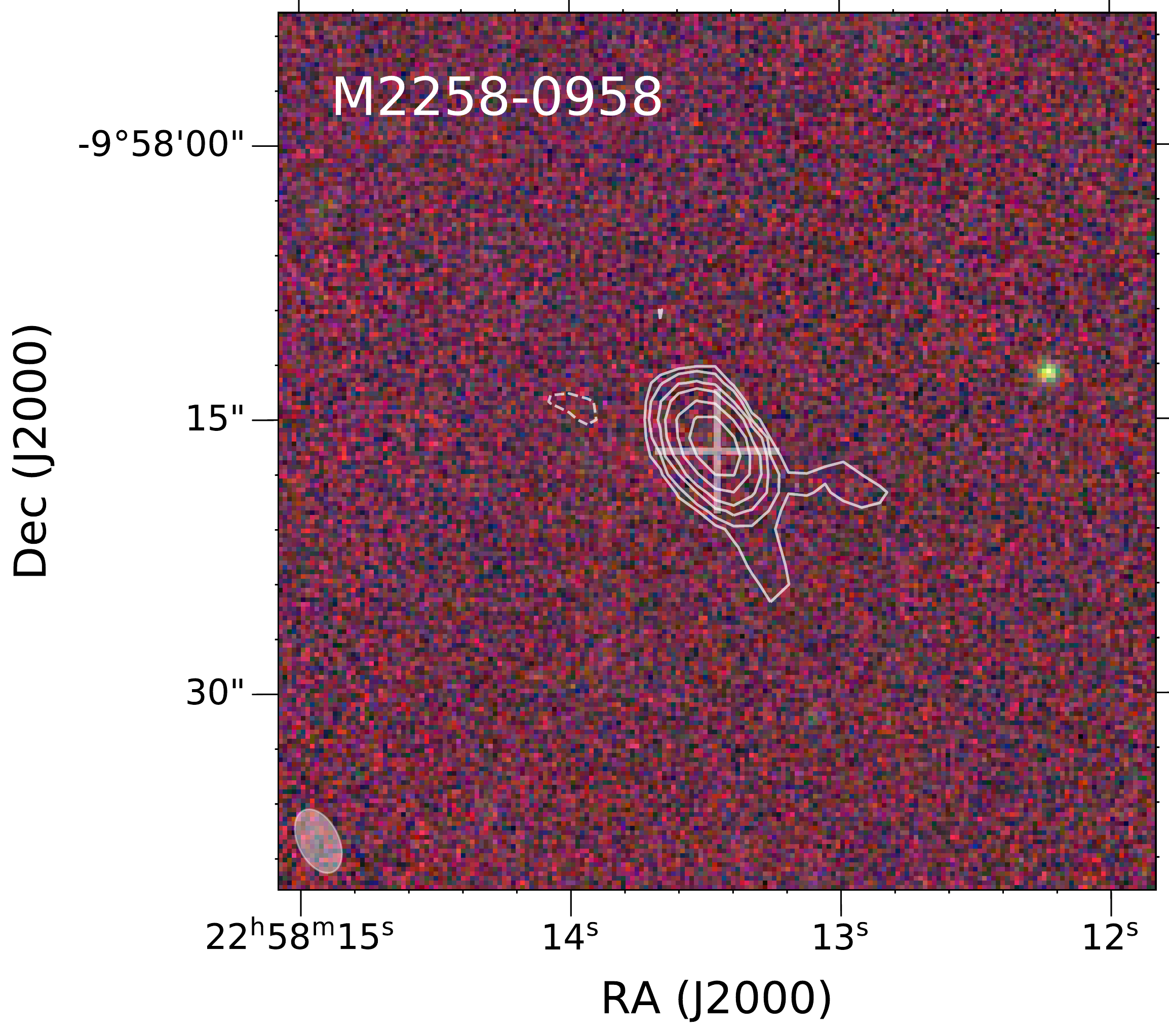}  
}}
}}  
\centering
	{\bf Figure \ref{fig:maps}} {\it continued}.
\vskip+0.0cm  
\end{figure*} 

\begin{longrotatetable}
\begin{deluxetable*}{ccccccccccccccccccccccccccccccc}
\tablecaption{Sample of radio sources for the uGMRT survey. }
\tabletypesize{\footnotesize}
\tablehead{
	\colhead{NVSS-ID} & \colhead{Flux} & \colhead{B$_{maj}$} & \colhead{B$_{min}$} & \colhead{B$_{pa}$} & \colhead{Redshift} & \colhead{Telescope} & \colhead{AGN-type}
	\\ & \colhead{WISE-ID} & \colhead{$RA_{\rm W}$} & \colhead{$Dec_{\rm W}$} & \colhead{W$1$}  & \colhead{$\delta W1$} & \colhead{$W2$} & \colhead{$\delta W2$} & \colhead{W$3$} & \colhead{$\delta W3$} 
	\\ & \colhead{PS1-ID}  & \colhead{nStackDets} & \colhead{nDetections} & \colhead{g} & \colhead{$\delta g$} & \colhead{r} & \colhead{$\delta r$} &  \colhead{i} & \colhead{$\delta i$} & \colhead{$z$} & \colhead{$\delta z$} & \colhead{$y$} & \colhead{$\delta y$} \\
	\vspace{-0.5cm} \\
	\colhead{(1) } & \colhead{ (2) } & \colhead{ (3)     } & \colhead{ (4)} & \colhead{ (5) } & \colhead{  (6)    } & \colhead{ (7)}  & \colhead{(8)} 
	\\  & \colhead{(9)} & \colhead{ (10)} & \colhead{ (11)} & \colhead{ (12)} & \colhead{(13)} & \colhead{(14)}  & \colhead{(15)} & \colhead{(16)} & \colhead{(17)} \\ 
	& \colhead{(18)} & \colhead{(19)} & \colhead{(20)} & \colhead{(21)} & \colhead{(22)} & \colhead{(23)} & \colhead{(24)} & \colhead{(25)} & \colhead{(26)} & \colhead{(27)} & \colhead{(28)} & \colhead{(29)}  & \colhead{(30)}\\
	\vspace{-0.7cm} \\
} 
\startdata
%
%
\vspace{-0.1cm} \\
\multicolumn{15}{c}{\small {\bf Spectra with emission lines (250)}}   \\  
\vspace{-0.3cm} \\
M000234.52-052324.5 & 235.4 & $<$16.7 & $<$15.9 & - & 0.306 & SALT & G \\ 
 & J000234.55-052324.0 & 0.6439997 & -5.3900227 & 15 & 0.04 & 14.18 & 0.05  & 10.96 & 0.14  \\
 & PSO J000.6440-05.3900 & 5 & 77 & 21.26 & 0.08 & 20.31 & 0.02 & 19.95 & 0.02 & 19.18 & 0.03 & 19.23 & 0.05 \\
M000920.42-074057.5 & 269.9 & $<$18.8 & $<$16.7 & - & 1.672 & SALT & Q \\
 & J000920.43-074057.3 & 2.3351346 & -7.6826025 & 16.26 & 0.07 & 15.07 & 0.09  & 11.65 & 0.25  \\
\enddata
\tablecomments{ 
\scriptsize{Column 1: NVSS ID based on Right Ascension and Declination (J2000); M stands for MALS SALT-NOT survey. Column 2: 20\,cm flux density from NVSS.  Columns 3 - 5: deconvolved radio sources size from NVSS (major and minor axes in arcseconds, and position angle in degrees). 
Column 6: spectroscopic redshift based on SALT-NOT spectra. Column 7: telescope used for spectroscopy. SUGAR = observed with both SALT and NOT. Column 8: AGN-type (G or Q corresponding to narrow or broad line AGN).
Columns 9 - 17 : WISE ID and $W1$, $W2$ and $W3$ magnitudes (mag) with errors. 
Column 18 - 30: PS1 ID, number of stack detections, number of single epoch detections in all filters, and g, r, i, z and y magnitudes with errors from PS1-mean for sources with redshifts and ELLs, and PS1-stack for DFs.   \\
List of targets excluded from the NOT sample presented in Table~1 of \citet[][]{Krogager18}: M024944.50+123706.3, M051938.34+174641.2, M163906.46+114409.2, M162303.03+123958.4, M060012.92-071037.9, M162539.34-041616.3 and M172423.07+032632.5 (see also Section~\ref{sec:cands}).
}
}
\end{deluxetable*}
\end{longrotatetable}

\section{Notes on individual targets}
\label{sec:sampnotes}
Here we briefly summarize the properties of interesting individual targets relevant to discussions in Sections~\ref{sec:zl0p5} to \ref{sec:zg2}. 

\subsection{AGN at $z<0.5$}

\begin{enumerate}
     
\item M0314-0909  (\zem = 0.312) and M1024-0852  (\zem = 0.455): The associated radio emission has projected linear sizes of $240-250$\,kpc.

\item M0133$-$0256 (\zem = 0.259) and M1703+0645 (\zem =0.399): These show double peaked emission lines.  For the former, the AGN host also shows signs of interaction with a nearby galaxy. 

\item J0609+0217 (\zem = 0.174) and M1219$-$1809 (\zem = 0.189): The host galaxies are part of an interacting group of galaxies.

\end{enumerate}

\subsection{AGN at $0.5<z<1.0$}

\begin{enumerate}
    
\item M0909$-$3133 (\zem = 0.884): The radio emission associated with this NLAGN has an extent of 330\,kpc.  The optical spectrum is shown in Fig.~\ref{fig:emission}). 

\item M2220+1307 (\zem = 0.760): The projected linear size of radio emission associated with this BLAGN is 180\,kpc. 

\item M1440+0531, M2058-1736 and M2059-1440: These have distorted radio morphologies suggesting interaction with the ambient medium.

\end{enumerate}

\subsection{AGN at $1.0<z<1.9$}

\begin{enumerate}


\item M0552$-$3459 (\zem = 1.776): This NLAGN has clear detections of \heii, C~{\sc iii}], C~{\sc ii}] and Ne~{\sc iv} (see Fig.~\ref{fig:emission}).  The FWHM measured based on the C~{\sc iii}] emission line is 835\,\kms. At 3\,GHz, the associated radio emission is resolved into two lobes separated by $8.3$\arcsec (71\,kpc).

\item M1033$-$1210 (\zem = 1.328): \citet{Krogager18} identified this as a NLAGN based on the presence of narrow [O~{\sc ii}] line and the absence of broad emission lines like Mg~{\sc ii} and C~{\sc iii}]. In this case too the radio emission is resolved into a double lobed morphology with a separation of  $3.2$\arcsec (28\,kpc).

\item M0956$-$1317 (\zem = 1.198): The associated radio emission has an extent of  $23.3$\arcsec (194\,kpc), which is the largest in our sample among the AGN at $1.0<z<1.9$.

\end{enumerate}

\subsection{AGN at $1.9<z<3.5$}

\begin{enumerate}

\item M0516+0732 (\zem = 2.594): The radio emission associated with this NLAGN is compact with a deconvolved size of $<0.5$\arcsec, i.e., $<4$\,kpc. It also shows strong continuum emission in the optical spectrum.  The \civ\ emission profile may be affected by the associated absorption.  We measure the \lya\ to \civ\ emission line ratio, f$_{\lya}$/f$_{\civ}$ = 2.40$\pm$0.10. This is similar to the ratios observed in high-$z$ radio galaxies.  

\item M1315$-$2745 (\zem = 1.911):  The radio emission associated with this NLAGN exhibits a double-lobed morphology with an extent of $32.8$\arcsec (280\,kpc).  We detect emission lines of \civ, \heii, C~{\sc iii}], C~{\sc ii}] and Ne~{\sc iv}. The optical continuum emission is very weak. While we do not clearly detect the AGN in PS1 images, it is clearly seen in the WISE image.  The deconvolved velocity width of the \civ\ emission line is 375\,\kms. The ratio $f_{\civ}$/f$_{\heii}$ is found to be 1.88$\pm$0.34.  This is typical of ratios observed in high-$z$ radio galaxies \citep[see for example figure 6 of][and references therein]{Shukla2021} but more than what is seen in typical quasars.

\item M1455$-$0957 (\zem = 1.995):  The double-lobed radio emission associated with this NLAGN has an extent of $8.1$\arcsec (i.e., 69\,kpc). The optical and IR characteristics are also very similar to M1315$-$2745.  The \civ, \heii, C~{\sc iii}], C~{\sc ii}] and Ne~{\sc iv} emission lines are clearly detected. The deconvolved velocity width of the \civ\ emission line is 430\,\kms. We estimate $f_{\civ}$/f$_{\heii}$ = 0.69$\pm$0.14, which is again typical of the ratios observed in high-$z$ radio galaxies. 

\item M1513-2524 (\zem = 3.132): This NLAGN is associated with a double-lobed radio emission with an extent of $23.7^{\prime\prime}$ (184\,kpc). It exhibits extended \lya, \civ\ and \heii\ emission. The detailed analysis of this object is presented by \citet[][]{Shukla2021}.

\end{enumerate}



\def\aj{AJ}%
\def\actaa{Acta Astron.}%
\def\araa{ARA\&A}%
\def\apj{ApJ}%
\def\apjl{ApJ}%
\def\apjs{ApJS}%
\def\ao{Appl.~Opt.}%
\def\apss{Ap\&SS}%
\def\aap{A\&A}%
\def\aapr{A\&A~Rev.}%
\def\aaps{A\&AS}%
\def\azh{AZh}%
\def\baas{BAAS}%
\def\bac{Bull. astr. Inst. Czechosl.}%
\def\caa{Chinese Astron. Astrophys.}%
\def\cjaa{Chinese J. Astron. Astrophys.}%
\def\icarus{Icarus}%
\def\jcap{J. Cosmology Astropart. Phys.}%
\def\jrasc{JRASC}%
\def\mnras{MNRAS}%
\def\memras{MmRAS}%
\def\na{New A}%
\def\nar{New A Rev.}%
\def\pasa{PASA}%
\def\pra{Phys.~Rev.~A}%
\def\prb{Phys.~Rev.~B}%
\def\prc{Phys.~Rev.~C}%
\def\prd{Phys.~Rev.~D}%
\def\pre{Phys.~Rev.~E}%
\def\prl{Phys.~Rev.~Lett.}%
\def\pasp{PASP}%
\def\pasj{PASJ}%
\def\qjras{QJRAS}%
\def\rmxaa{Rev. Mexicana Astron. Astrofis.}%
\def\skytel{S\&T}%
\def\solphys{Sol.~Phys.}%
\def\sovast{Soviet~Ast.}%
\def\ssr{Space~Sci.~Rev.}%
\def\zap{ZAp}%
\def\nat{Nature}%
\def\iaucirc{IAU~Circ.}%
\def\aplett{Astrophys.~Lett.}%
\def\apspr{Astrophys.~Space~Phys.~Res.}%
\def\bain{Bull.~Astron.~Inst.~Netherlands}%
\def\fcp{Fund.~Cosmic~Phys.}%
\def\gca{Geochim.~Cosmochim.~Acta}%
\def\grl{Geophys.~Res.~Lett.}%
\def\jcp{J.~Chem.~Phys.}%
\def\jgr{J.~Geophys.~Res.}%
\def\jqsrt{J.~Quant.~Spec.~Radiat.~Transf.}%
\def\memsai{Mem.~Soc.~Astron.~Italiana}%
\def\nphysa{Nucl.~Phys.~A}%
\def\physrep{Phys.~Rep.}%
\def\physscr{Phys.~Scr}%
\def\planss{Planet.~Space~Sci.}%
\def\procspie{Proc.~SPIE}%
\let\astap=\aap
\let\apjlett=\apjl
\let\apjsupp=\apjs
\let\applopt=\ao
\bibliographystyle{aasjournal}
\bibliography{mybib}

\begin{thebibliography}{}
\expandafter\ifx\csname natexlab\endcsname\relax\def\natexlab#1{#1}\fi
\providecommand{\url}[1]{\href{#1}{#1}}
\providecommand{\dodoi}[1]{doi:~\href{http://doi.org/#1}{\nolinkurl{#1}}}
\providecommand{\doeprint}[1]{\href{http://ascl.net/#1}{\nolinkurl{http://ascl.net/#1}}}
\providecommand{\doarXiv}[1]{\href{https://arxiv.org/abs/#1}{\nolinkurl{https://arxiv.org/abs/#1}}}

\bibitem[{{Acero} {et~al.}(2015){Acero}, {Ackermann}, {Ajello}, {Albert},
  {Atwood}, {Axelsson}, {Baldini}, {Ballet}, {Barbiellini}, {Bastieri},
  {Belfiore}, {Bellazzini}, {Bissaldi}, {Blandford}, {Bloom}, {Bogart},
  {Bonino}, {Bottacini}, {Bregeon}, {Britto}, {Bruel}, {Buehler}, {Burnett},
  {Buson}, {Caliandro}, {Cameron}, {Caputo}, {Caragiulo}, {Caraveo},
  {Casandjian}, {Cavazzuti}, {Charles}, {Chaves}, {Chekhtman}, {Cheung},
  {Chiang}, {Chiaro}, {Ciprini}, {Claus}, {Cohen-Tanugi}, {Cominsky}, {Conrad},
  {Cutini}, {D'Ammando}, {de Angelis}, {DeKlotz}, {de Palma}, {Desiante},
  {Digel}, {Di Venere}, {Drell}, {Dubois}, {Dumora}, {Favuzzi}, {Fegan},
  {Ferrara}, {Finke}, {Franckowiak}, {Fukazawa}, {Funk}, {Fusco}, {Gargano},
  {Gasparrini}, {Giebels}, {Giglietto}, {Giommi}, {Giordano}, {Giroletti},
  {Glanzman}, {Godfrey}, {Grenier}, {Grondin}, {Grove}, {Guillemot}, {Guiriec},
  {Hadasch}, {Harding}, {Hays}, {Hewitt}, {Hill}, {Horan}, {Iafrate}, {Jogler},
  {J{\'o}hannesson}, {Johnson}, {Johnson}, {Johnson}, {Johnson}, {Kamae},
  {Kataoka}, {Katsuta}, {Kuss}, {La Mura}, {Landriu}, {Larsson}, {Latronico},
  {Lemoine-Goumard}, {Li}, {Li}, {Longo}, {Loparco}, {Lott}, {Lovellette},
  {Lubrano}, {Madejski}, {Massaro}, {Mayer}, {Mazziotta}, {McEnery},
  {Michelson}, {Mirabal}, {Mizuno}, {Moiseev}, {Mongelli}, {Monzani},
  {Morselli}, {Moskalenko}, {Murgia}, {Nuss}, {Ohno}, {Ohsugi}, {Omodei},
  {Orienti}, {Orlando}, {Ormes}, {Paneque}, {Panetta}, {Perkins},
  {Pesce-Rollins}, {Piron}, {Pivato}, {Porter}, {Racusin}, {Rando}, {Razzano},
  {Razzaque}, {Reimer}, {Reimer}, {Reposeur}, {Rochester}, {Romani},
  {Salvetti}, {S{\'a}nchez-Conde}, {Saz Parkinson}, {Schulz}, {Siskind},
  {Smith}, {Spada}, {Spandre}, {Spinelli}, {Stephens}, {Strong}, {Suson},
  {Takahashi}, {Takahashi}, {Tanaka}, {Thayer}, {Thayer}, {Thompson},
  {Tibaldo}, {Tibolla}, {Torres}, {Torresi}, {Tosti}, {Troja}, {Van Klaveren},
  {Vianello}, {Winer}, {Wood}, {Wood}, {Zimmer}, \& {Fermi-LAT
  Collaboration}}]{Acero15}
{Acero}, F., {Ackermann}, M., {Ajello}, M., {et~al.} 2015, \apjs, 218, 23,
  \dodoi{10.1088/0067-0049/218/2/23}

\bibitem[{{Assef} {et~al.}(2018){Assef}, {Stern}, {Noirot}, {Jun}, {Cutri}, \&
  {Eisenhardt}}]{Assef18}
{Assef}, R.~J., {Stern}, D., {Noirot}, G., {et~al.} 2018, \apjs, 234, 23,
  \dodoi{10.3847/1538-4365/aaa00a}

\bibitem[{{Assef} {et~al.}(2010){Assef}, {Kochanek}, {Brodwin}, {Cool},
  {Forman}, {Gonzalez}, {Hickox}, {Jones}, {Le Floc'h}, {Moustakas}, {Murray},
  \& {Stern}}]{Assef10}
{Assef}, R.~J., {Kochanek}, C.~S., {Brodwin}, M., {et~al.} 2010, \apj, 713,
  970, \dodoi{10.1088/0004-637X/713/2/970}

\bibitem[{{Balashev} \& {Noterdaeme}(2018)}]{Balashev18}
{Balashev}, S.~A., \& {Noterdaeme}, P. 2018, \mnras, 478, L7,
  \dodoi{10.1093/mnrasl/sly067}

\bibitem[{{Ballet} {et~al.}(2020){Ballet}, {Burnett}, {Digel}, \&
  {Lott}}]{Ballet20}
{Ballet}, J., {Burnett}, T.~H., {Digel}, S.~W., \& {Lott}, B. 2020, arXiv
  e-prints, arXiv:2005.11208.
\newblock \doarXiv{2005.11208}

\bibitem[{{Barrows} {et~al.}(2021){Barrows}, {Comerford}, {Stern}, \&
  {Assef}}]{Barrows21}
{Barrows}, R.~S., {Comerford}, J.~M., {Stern}, D., \& {Assef}, R.~J. 2021,
  arXiv e-prints, arXiv:2107.02815.
\newblock \doarXiv{2107.02815}

\bibitem[{{Becker} {et~al.}(2000){Becker}, {White}, {Gregg}, {Brotherton},
  {Laurent-Muehleisen}, \& {Arav}}]{Becker00}
{Becker}, R.~H., {White}, R.~L., {Gregg}, M.~D., {et~al.} 2000, \apj, 538, 72,
  \dodoi{10.1086/309099}

\bibitem[{{Becker} {et~al.}(1995){Becker}, {White}, \& {Helfand}}]{Becker95}
{Becker}, R.~H., {White}, R.~L., \& {Helfand}, D.~J. 1995, \apj, 450, 559,
  \dodoi{10.1086/176166}

\bibitem[{{Boettcher} {et~al.}(2021){Boettcher}, {Chen}, {Zahedy}, {Cooper},
  {Johnson}, {Rudie}, {Chen}, {Petitjean}, {Cantalupo}, {Cooksey},
  {Faucher-Gigu{\`e}re}, {Greene}, {Lopez}, {Mulchaey}, {Penton}, {Putman},
  {Rafelski}, {Rauch}, {Schaye}, {Simcoe}, \& {Walth}}]{Boettcher21}
{Boettcher}, E., {Chen}, H.-W., {Zahedy}, F.~S., {et~al.} 2021, \apj, 913, 18,
  \dodoi{10.3847/1538-4357/abf0a0}

\bibitem[{{Burgh} {et~al.}(2003){Burgh}, {Nordsieck}, {Kobulnicky}, {Williams},
  {O'Donoghue}, {Smith}, \& {Percival}}]{burgh2003}
{Burgh}, E.~B., {Nordsieck}, K.~H., {Kobulnicky}, H.~A., {et~al.} 2003, in
  Society of Photo-Optical Instrumentation Engineers (SPIE) Conference Series,
  Vol. 4841, \procspie, ed. M.~{Iye} \& A.~F.~M. {Moorwood}, 1463--1471

\bibitem[{{Chambers} {et~al.}(2016){Chambers}, {Magnier}, {Metcalfe},
  {Flewelling}, {Huber}, {Waters}, {Denneau}, {Draper}, {Farrow}, {Finkbeiner},
  {Holmberg}, {Koppenhoefer}, {Price}, {Rest}, {Saglia}, {Schlafly}, {Smartt},
  {Sweeney}, {Wainscoat}, {Burgett}, {Chastel}, {Grav}, {Heasley}, {Hodapp},
  {Jedicke}, {Kaiser}, {Kudritzki}, {Luppino}, {Lupton}, {Monet}, {Morgan},
  {Onaka}, {Shiao}, {Stubbs}, {Tonry}, {White}, {Ba{\~n}ados}, {Bell},
  {Bender}, {Bernard}, {Boegner}, {Boffi}, {Botticella}, {Calamida},
  {Casertano}, {Chen}, {Chen}, {Cole}, {Deacon}, {Frenk}, {Fitzsimmons},
  {Gezari}, {Gibbs}, {Goessl}, {Goggia}, {Gourgue}, {Goldman}, {Grant},
  {Grebel}, {Hambly}, {Hasinger}, {Heavens}, {Heckman}, {Henderson}, {Henning},
  {Holman}, {Hopp}, {Ip}, {Isani}, {Jackson}, {Keyes}, {Koekemoer}, {Kotak},
  {Le}, {Liska}, {Long}, {Lucey}, {Liu}, {Martin}, {Masci}, {McLean}, {Mindel},
  {Misra}, {Morganson}, {Murphy}, {Obaika}, {Narayan}, {Nieto-Santisteban},
  {Norberg}, {Peacock}, {Pier}, {Postman}, {Primak}, {Rae}, {Rai}, {Riess},
  {Riffeser}, {Rix}, {R{\"o}ser}, {Russel}, {Rutz}, {Schilbach}, {Schultz},
  {Scolnic}, {Strolger}, {Szalay}, {Seitz}, {Small}, {Smith}, {Soderblom},
  {Taylor}, {Thomson}, {Taylor}, {Thakar}, {Thiel}, {Thilker}, {Unger},
  {Urata}, {Valenti}, {Wagner}, {Walder}, {Walter}, {Watters}, {Werner},
  {Wood-Vasey}, \& {Wyse}}]{Chambers16}
{Chambers}, K.~C., {Magnier}, E.~A., {Metcalfe}, N., {et~al.} 2016, arXiv
  e-prints, arXiv:1612.05560.
\newblock \doarXiv{1612.05560}

\bibitem[{{Chang} {et~al.}(2017){Chang}, {Arsioli}, {Giommi}, \&
  {Padovani}}]{Chang17}
{Chang}, Y.~L., {Arsioli}, B., {Giommi}, P., \& {Padovani}, P. 2017, \aap, 598,
  A17, \dodoi{10.1051/0004-6361/201629487}

\bibitem[{{Coatman} {et~al.}(2016){Coatman}, {Hewett}, {Banerji}, \&
  {Richards}}]{Coatman16}
{Coatman}, L., {Hewett}, P.~C., {Banerji}, M., \& {Richards}, G.~T. 2016,
  \mnras, 461, 647, \dodoi{10.1093/mnras/stw1360}

\bibitem[{{Combes} {et~al.}(2021){Combes}, {Gupta}, {Muller}, {Balashev},
  {Jozsa}, {Srianand}, {Momjian}, {Noterdaeme}, {Kloeckner}, {Baker},
  {Boettcher}, {Bosma}, {Chen}, {Dutta}, {Jagannathan}, {Jose}, {Knowles},
  {Krogager}, {Kulkarni}, {Moodley}, {Pandey}, {Petitjean}, \&
  {Sekhar}}]{Combes21}
{Combes}, F., {Gupta}, N., {Muller}, S., {et~al.} 2021, arXiv e-prints,
  arXiv:2101.00188.
\newblock \doarXiv{2101.00188}

\bibitem[{{Condon} {et~al.}(1998){Condon}, {Cotton}, {Greisen}, {Yin},
  {Perley}, {Taylor}, \& {Broderick}}]{Condon98}
{Condon}, J.~J., {Cotton}, W.~D., {Greisen}, E.~W., {et~al.} 1998, \aj, 115,
  1693, \dodoi{10.1086/300337}

\bibitem[{{Crawford} {et~al.}(2010){Crawford}, {Still}, {Schellart}, {Balona},
  {Buckley}, {Dugmore}, {Gulbis}, {Kniazev}, {Kotze}, {Loaring}, {Nordsieck},
  {Pickering}, {Potter}, {Romero Colmenero}, {Vaisanen}, {Williams}, \&
  {Zietsman}}]{Crawford2010}
{Crawford}, S.~M., {Still}, M., {Schellart}, P., {et~al.} 2010, in \procspie,
  Vol. 7737, Observatory Operations: Strategies, Processes, and Systems III,
  773725

\bibitem[{{Cutri} \& {et al.}(2014)}]{Cutri14}
{Cutri}, R.~M., \& {et al.} 2014, VizieR Online Data Catalog, II/328

\bibitem[{{D'Abrusco} {et~al.}(2019){D'Abrusco}, {{\'A}lvarez Crespo},
  {Massaro}, {Campana}, {Chavushyan}, {Landoni}, {La Franca}, {Masetti},
  {Milisavljevic}, {Paggi}, {Ricci}, \& {Smith}}]{Dabrusco19}
{D'Abrusco}, R., {{\'A}lvarez Crespo}, N., {Massaro}, F., {et~al.} 2019, \apjs,
  242, 4, \dodoi{10.3847/1538-4365/ab16f4}

\bibitem[{{Denney} {et~al.}(2010){Denney}, {Peterson}, {Pogge}, {Adair},
  {Atlee}, {Au-Yong}, {Bentz}, {Bird}, {Brokofsky}, {Chisholm}, {Comins},
  {Dietrich}, {Doroshenko}, {Eastman}, {Efimov}, {Ewald}, {Ferbey}, {Gaskell},
  {Hedrick}, {Jackson}, {Klimanov}, {Klimek}, {Kruse}, {Lad{\'e}route}, {Lamb},
  {Leighly}, {Minezaki}, {Nazarov}, {Onken}, {Petersen}, {Peterson},
  {Poindexter}, {Sakata}, {Schlesinger}, {Sergeev}, {Skolski}, {Stieglitz},
  {Tobin}, {Unterborn}, {Vestergaard}, {Watkins}, {Watson}, \&
  {Yoshii}}]{Denney10}
{Denney}, K.~D., {Peterson}, B.~M., {Pogge}, R.~W., {et~al.} 2010, \apj, 721,
  715, \dodoi{10.1088/0004-637X/721/1/715}

\bibitem[{{Ellison} {et~al.}(2001){Ellison}, {Yan}, {Hook}, {Pettini}, {Wall},
  \& {Shaver}}]{Ellison01}
{Ellison}, S.~L., {Yan}, L., {Hook}, I.~M., {et~al.} 2001, \aap, 379, 393,
  \dodoi{10.1051/0004-6361:20011281}

\bibitem[{{Ellison} {et~al.}(2008){Ellison}, {York}, {Pettini}, \&
  {Kanekar}}]{Ellison08}
{Ellison}, S.~L., {York}, B.~A., {Pettini}, M., \& {Kanekar}, N. 2008, \mnras,
  388, 1349, \dodoi{10.1111/j.1365-2966.2008.13482.x}

\bibitem[{{Fabian}(2012)}]{Fabian12}
{Fabian}, A.~C. 2012, \araa, 50, 455,
  \dodoi{10.1146/annurev-astro-081811-125521}

\bibitem[{{Frank} \& {P{\'e}roux}(2010)}]{Frank10}
{Frank}, S., \& {P{\'e}roux}, C. 2010, \mnras, 406, 2235,
  \dodoi{10.1111/j.1365-2966.2010.16848.x}

\bibitem[{{Ghisellini} {et~al.}(2014){Ghisellini}, {Celotti}, {Tavecchio},
  {Haardt}, \& {Sbarrato}}]{Ghisellini14}
{Ghisellini}, G., {Celotti}, A., {Tavecchio}, F., {Haardt}, F., \& {Sbarrato},
  T. 2014, \mnras, 438, 2694, \dodoi{10.1093/mnras/stt2394}

\bibitem[{{Gupta} {et~al.}(2005){Gupta}, {Srianand}, \& {Saikia}}]{Gupta05}
{Gupta}, N., {Srianand}, R., \& {Saikia}, D.~J. 2005, \mnras, 361, 451,
  \dodoi{10.1111/j.1365-2966.2005.09169.x}

\bibitem[{{Gupta} {et~al.}(2017){Gupta}, {Srianand}, {Baan}, {Baker},
  {Beswick}, {Bhatnagar}, {Bhattacharya}, {Bosma}, {Carilli}, {Cluver},
  {Combes}, {Cress}, {Dutta}, {Fynbo}, {Heald}, {Hilton}, {Hussain}, {Jarvis},
  {Jozsa}, {Kamphuis}, {Kembhavi}, {Kerp}, {Kl{\"o}ckner}, {Krogager},
  {Kulkarni}, {Ledoux}, {Mahabal}, {Mauch}, {Moodley}, {Momjian}, {Morganti},
  {Noterdaeme}, {Oosterloo}, {Petitjean}, {Schr{\"o}der}, {Serra}, {Sievers},
  {Spekkens}, {V{\"a}is{\"a}nen}, {van der Hulst}, {Vivek}, {Wang}, {Wong}, \&
  {Zungu}}]{Gupta17mals}
{Gupta}, N., {Srianand}, R., {Baan}, W., {et~al.} 2017, ArXiv e-prints.
\newblock \doarXiv{1708.07371}

\bibitem[{{Gupta} {et~al.}(2021{\natexlab{a}}){Gupta}, {Jagannathan},
  {Srianand}, {Bhatnagar}, {Noterdaeme}, {Combes}, {Petitjean}, {Jose},
  {Pandey}, {Kaski}, {Baker}, {Balashev}, {Boettcher}, {Chen}, {Cress},
  {Dutta}, {Goedhart}, {Heald}, {J{\'o}zsa}, {Kamau}, {Kamphuis}, {Kerp},
  {Kl{\"o}ckner}, {Knowles}, {Krishnan}, {Krogager}, {Kulkarni}, {Momjian},
  {Moodley}, {Passmoor}, {Schr{\"o}eder}, {Sekhar}, {Sikhosana}, {Wagenveld},
  \& {Wong}}]{Gupta21}
{Gupta}, N., {Jagannathan}, P., {Srianand}, R., {et~al.} 2021{\natexlab{a}},
  \apj, 907, 11, \dodoi{10.3847/1538-4357/abcb85}

\bibitem[{{Gupta} {et~al.}(2021{\natexlab{b}}){Gupta}, {Srianand}, {Shukla},
  {Krogager}, {Noterdaeme}, {Combes}, {Dutta}, {Fynbo}, {Hilton}, {Momjian},
  {Moodley}, \& {Petitjean}}]{Gupta21hz}
{Gupta}, N., {Srianand}, R., {Shukla}, G., {et~al.} 2021{\natexlab{b}}, arXiv
  e-prints, arXiv:2103.09437.
\newblock \doarXiv{2103.09437}

\bibitem[{{Heckman} \& {Best}(2014)}]{Heckman14}
{Heckman}, T.~M., \& {Best}, P.~N. 2014, \araa, 52, 589,
  \dodoi{10.1146/annurev-astro-081913-035722}

\bibitem[{{Hickox} \& {Alexander}(2018)}]{Hickox18}
{Hickox}, R.~C., \& {Alexander}, D.~M. 2018, \araa, 56, 625,
  \dodoi{10.1146/annurev-astro-081817-051803}

\bibitem[{{Jauncey} {et~al.}(1991){Jauncey}, {Reynolds}, {Tzioumis}, {Muxlow},
  {Perley}, {Murphy}, {Preston}, {King}, {Patnaik}, {Jones}, {Meier}, {Bird},
  {Blair}, {Bunton}, {Clay}, {Costa}, {Duncan}, {Ferris}, {Gough}, {Hamilton},
  {Hoard}, {Kemball}, {Kesteven}, {Lobdell}, {Luiten}, {Mcculloch}, {Murray},
  {Nicholson}, {Rao}, {Savage}, {Sinclair}, {Skjerve}, {Taaffe}, {Wark}, \&
  {White}}]{Jauncey1991}
{Jauncey}, D.~L., {Reynolds}, J.~E., {Tzioumis}, A.~K., {et~al.} 1991, \nat,
  352, 132, \dodoi{10.1038/352132a0}

\bibitem[{{Jonas} \& {MeerKAT Team}(2016)}]{Jonas16}
{Jonas}, J., \& {MeerKAT Team}. 2016, in Proceedings of MeerKAT Science: On the
  Pathway to the SKA. 25-27 May, 2016 Stellenbosch, South Africa (MeerKAT2016),
  1

\bibitem[{{Kapahi} {et~al.}(1998){Kapahi}, {Athreya}, {van Breugel},
  {McCarthy}, \& {Subrahmanya}}]{Kapahi98}
{Kapahi}, V.~K., {Athreya}, R.~M., {van Breugel}, W., {McCarthy}, P.~J., \&
  {Subrahmanya}, C.~R. 1998, \apjs, 118, 275, \dodoi{10.1086/313144}

\bibitem[{{Kaspi} {et~al.}(2000){Kaspi}, {Smith}, {Netzer}, {Maoz}, {Jannuzi},
  \& {Giveon}}]{Kaspi00}
{Kaspi}, S., {Smith}, P.~S., {Netzer}, H., {et~al.} 2000, \apj, 533, 631,
  \dodoi{10.1086/308704}

\bibitem[{{Kobulnicky} {et~al.}(2003){Kobulnicky}, {Nordsieck}, {Burgh},
  {Smith}, {Percival}, {Williams}, \& {O'Donoghue}}]{kobul2003}
{Kobulnicky}, H.~A., {Nordsieck}, K.~H., {Burgh}, E.~B., {et~al.} 2003, in
  Society of Photo-Optical Instrumentation Engineers (SPIE) Conference Series,
  Vol. 4841, \procspie, ed. M.~{Iye} \& A.~F.~M. {Moorwood}, 1634--1644

\bibitem[{{Krogager} {et~al.}(2019){Krogager}, {Fynbo}, {M{\o}ller},
  {Noterdaeme}, {Heintz}, \& {Pettini}}]{Krogager19}
{Krogager}, J.-K., {Fynbo}, J. P.~U., {M{\o}ller}, P., {et~al.} 2019, \mnras,
  486, 4377, \dodoi{10.1093/mnras/stz1120}

\bibitem[{{Krogager} {et~al.}(2018){Krogager}, {Gupta}, {Noterdaeme}, {Ranjan},
  {Fynbo}, {Srianand}, {Petitjean}, {Combes}, \& {Mahabal}}]{Krogager18}
{Krogager}, J.~K., {Gupta}, N., {Noterdaeme}, P., {et~al.} 2018, \apjs, 235,
  10, \dodoi{10.3847/1538-4365/aaab51}

\bibitem[{{Lacy} {et~al.}(2004){Lacy}, {Storrie-Lombardi}, {Sajina},
  {Appleton}, {Armus}, {Chapman}, {Choi}, {Fadda}, {Fang}, {Frayer},
  {Heinrichsen}, {Helou}, {Im}, {Marleau}, {Masci}, {Shupe}, {Soifer},
  {Surace}, {Teplitz}, {Wilson}, \& {Yan}}]{Lacy04}
{Lacy}, M., {Storrie-Lombardi}, L.~J., {Sajina}, A., {et~al.} 2004, \apjs, 154,
  166, \dodoi{10.1086/422816}

\bibitem[{{Lacy} {et~al.}(2020){Lacy}, {Baum}, {Chandler}, {Chatterjee},
  {Clarke}, {Deustua}, {English}, {Farnes}, {Gaensler}, {Gugliucci},
  {Hallinan}, {Kent}, {Kimball}, {Law}, {Lazio}, {Marvil}, {Mao}, {Medlin},
  {Mooley}, {Murphy}, {Myers}, {Osten}, {Richards}, {Rosolowsky}, {Rudnick},
  {Schinzel}, {Sivakoff}, {Sjouwerman}, {Taylor}, {White}, {Wrobel},
  {Andernach}, {Beasley}, {Berger}, {Bhatnager}, {Birkinshaw}, {Bower},
  {Brandt}, {Brown}, {Burke-Spolaor}, {Butler}, {Comerford}, {Demorest}, {Fu},
  {Giacintucci}, {Golap}, {G{\"u}th}, {Hales}, {Hiriart}, {Hodge}, {Horesh},
  {Ivezi{\'c}}, {Jarvis}, {Kamble}, {Kassim}, {Liu}, {Loinard}, {Lyons},
  {Masters}, {Mezcua}, {Moellenbrock}, {Mroczkowski}, {Nyland}, {O'Dea},
  {O'Sullivan}, {Peters}, {Radford}, {Rao}, {Robnett}, {Salcido}, {Shen},
  {Sobotka}, {Witz}, {Vaccari}, {van Weeren}, {Vargas}, {Williams}, \&
  {Yoon}}]{Lacy20}
{Lacy}, M., {Baum}, S.~A., {Chandler}, C.~J., {et~al.} 2020, \pasp, 132,
  035001, \dodoi{10.1088/1538-3873/ab63eb}

\bibitem[{{Laor}(2000)}]{Laor00}
{Laor}, A. 2000, \apjl, 543, L111, \dodoi{10.1086/317280}

\bibitem[{{Ledoux} {et~al.}(2003){Ledoux}, {Petitjean}, \&
  {Srianand}}]{Ledoux03}
{Ledoux}, C., {Petitjean}, P., \& {Srianand}, R. 2003, \mnras, 346, 209,
  \dodoi{10.1046/j.1365-2966.2003.07082.x}

\bibitem[{{Liu} {et~al.}(2010){Liu}, {Shen}, {Strauss}, \& {Greene}}]{Liu10}
{Liu}, X., {Shen}, Y., {Strauss}, M.~A., \& {Greene}, J.~E. 2010, \apj, 708,
  427, \dodoi{10.1088/0004-637X/708/1/427}

\bibitem[{{Lu} {et~al.}(2007){Lu}, {Wang}, {Zhou}, \& {Wu}}]{Lu07}
{Lu}, Y., {Wang}, T., {Zhou}, H., \& {Wu}, J. 2007, \aj, 133, 1615,
  \dodoi{10.1086/512034}

\bibitem[{{Masci} {et~al.}(2019){Masci}, {Laher}, {Rusholme}, {Shupe}, {Groom},
  {Surace}, {Jackson}, {Monkewitz}, {Beck}, {Flynn}, {Terek}, {Landry},
  {Hacopians}, {Desai}, {Howell}, {Brooke}, {Imel}, {Wachter}, {Ye}, {Lin},
  {Cenko}, {Cunningham}, {Rebbapragada}, {Bue}, {Miller}, {Mahabal}, {Bellm},
  {Patterson}, {Juri{\'c}}, {Golkhou}, {Ofek}, {Walters}, {Graham}, {Kasliwal},
  {Dekany}, {Kupfer}, {Burdge}, {Cannella}, {Barlow}, {Van Sistine}, {Giomi},
  {Fremling}, {Blagorodnova}, {Levitan}, {Riddle}, {Smith}, {Helou}, {Prince},
  \& {Kulkarni}}]{Masci19}
{Masci}, F.~J., {Laher}, R.~R., {Rusholme}, B., {et~al.} 2019, \pasp, 131,
  018003, \dodoi{10.1088/1538-3873/aae8ac}

\bibitem[{{Massaro} {et~al.}(2009){Massaro}, {Giommi}, {Leto}, {Marchegiani},
  {Maselli}, {Perri}, {Piranomonte}, \& {Sclavi}}]{Massaro09}
{Massaro}, E., {Giommi}, P., {Leto}, C., {et~al.} 2009, \aap, 495, 691,
  \dodoi{10.1051/0004-6361:200810161}

\bibitem[{{Mateos} {et~al.}(2012){Mateos}, {Alonso-Herrero}, {Carrera},
  {Blain}, {Watson}, {Barcons}, {Braito}, {Severgnini}, {Donley}, \&
  {Stern}}]{Mateos12}
{Mateos}, S., {Alonso-Herrero}, A., {Carrera}, F.~J., {et~al.} 2012, \mnras,
  426, 3271, \dodoi{10.1111/j.1365-2966.2012.21843.x}

\bibitem[{{Mathur} \& {Nair}(1997)}]{Mathur97}
{Mathur}, S., \& {Nair}, S. 1997, \apj, 484, 140, \dodoi{10.1086/304327}

\bibitem[{{Mauch} {et~al.}(2003){Mauch}, {Murphy}, {Buttery}, {Curran},
  {Hunstead}, {Piestrzynski}, {Robertson}, \& {Sadler}}]{Mauch03}
{Mauch}, T., {Murphy}, T., {Buttery}, H.~J., {et~al.} 2003, \mnras, 342, 1117,
  \dodoi{10.1046/j.1365-8711.2003.06605.x}

\bibitem[{{McConnell} {et~al.}(2020){McConnell}, {Hale}, {Lenc}, {Banfield},
  {Heald}, {Hotan}, {Leung}, {Moss}, {Murphy}, {O'Brien}, {Pritchard}, {Raja},
  {Sadler}, {Stewart}, {Thomson}, {Whiting}, {Allison}, {Amy}, {Anderson},
  {Ball}, {Bannister}, {Bell}, {Bock}, {Bolton}, {Bunton}, {Chippendale},
  {Collier}, {Cooray}, {Cornwell}, {Diamond}, {Edwards}, {Gupta}, {Hayman},
  {Heywood}, {Jackson}, {Koribalski}, {Lee-Waddell}, {McClure-Griffiths}, {Ng},
  {Norris}, {Phillips}, {Reynolds}, {Roxby}, {Schinckel}, {Shields},
  {Tremblay}, {Tzioumis}, {Voronkov}, \& {Westmeier}}]{Mcconnell20}
{McConnell}, D., {Hale}, C.~L., {Lenc}, E., {et~al.} 2020, \pasa, 37, e048,
  \dodoi{10.1017/pasa.2020.41}

\bibitem[{{Mullaney} {et~al.}(2013){Mullaney}, {Alexander}, {Fine}, {Goulding},
  {Harrison}, \& {Hickox}}]{Mullaney13}
{Mullaney}, J.~R., {Alexander}, D.~M., {Fine}, S., {et~al.} 2013, \mnras, 433,
  622, \dodoi{10.1093/mnras/stt751}

\bibitem[{{Murphy} \& {Liske}(2004)}]{Murphy04}
{Murphy}, M.~T., \& {Liske}, J. 2004, \mnras, 354, L31,
  \dodoi{10.1111/j.1365-2966.2004.08374.x}

\bibitem[{{Muzahid} {et~al.}(2015){Muzahid}, {Srianand}, \&
  {Charlton}}]{Muzahid15}
{Muzahid}, S., {Srianand}, R., \& {Charlton}, J. 2015, \mnras, 448, 2840,
  \dodoi{10.1093/mnras/stv133}

\bibitem[{{Nandra} {et~al.}(1997){Nandra}, {George}, {Mushotzky}, {Turner}, \&
  {Yaqoob}}]{Nandra97}
{Nandra}, K., {George}, I.~M., {Mushotzky}, R.~F., {Turner}, T.~J., \&
  {Yaqoob}, T. 1997, \apj, 476, 70, \dodoi{10.1086/303600}

\bibitem[{{Netzer} {et~al.}(2007){Netzer}, {Lutz}, {Schweitzer}, {Contursi},
  {Sturm}, {Tacconi}, {Veilleux}, {Kim}, {Rupke}, {Baker}, {Dasyra},
  {Mazzarella}, \& {Lord}}]{Netzer07}
{Netzer}, H., {Lutz}, D., {Schweitzer}, M., {et~al.} 2007, \apj, 666, 806,
  \dodoi{10.1086/520716}

\bibitem[{{Noterdaeme} {et~al.}(2008){Noterdaeme}, {Ledoux}, {Petitjean}, \&
  {Srianand}}]{Noterdaeme08}
{Noterdaeme}, P., {Ledoux}, C., {Petitjean}, P., \& {Srianand}, R. 2008, \aap,
  481, 327, \dodoi{10.1051/0004-6361:20078780}

\bibitem[{{Noterdaeme} {et~al.}(2009){Noterdaeme}, {Petitjean}, {Ledoux}, \&
  {Srianand}}]{Noterdaeme09dla}
{Noterdaeme}, P., {Petitjean}, P., {Ledoux}, C., \& {Srianand}, R. 2009, \aap,
  505, 1087, \dodoi{10.1051/0004-6361/200912768}

\bibitem[{{Noterdaeme} {et~al.}(2015){Noterdaeme}, {Srianand}, {Rahmani},
  {Petitjean}, {P{\^a}ris}, {Ledoux}, {Gupta}, \& {L{\'o}pez}}]{Noterdaeme15}
{Noterdaeme}, P., {Srianand}, R., {Rahmani}, H., {et~al.} 2015, \aap, 577, A24,
  \dodoi{10.1051/0004-6361/201425376}

\bibitem[{{Onken} {et~al.}(2004){Onken}, {Ferrarese}, {Merritt}, {Peterson},
  {Pogge}, {Vestergaard}, \& {Wandel}}]{Onken04}
{Onken}, C.~A., {Ferrarese}, L., {Merritt}, D., {et~al.} 2004, \apj, 615, 645,
  \dodoi{10.1086/424655}

\bibitem[{{Padovani} {et~al.}(2017){Padovani}, {Alexander}, {Assef}, {De
  Marco}, {Giommi}, {Hickox}, {Richards}, {Smol{\v{c}}i{\'c}},
  {Hatziminaoglou}, {Mainieri}, \& {Salvato}}]{Padovani17}
{Padovani}, P., {Alexander}, D.~M., {Assef}, R.~J., {et~al.} 2017, \aapr, 25,
  2, \dodoi{10.1007/s00159-017-0102-9}

\bibitem[{{Parks} {et~al.}(2018){Parks}, {Prochaska}, {Dong}, \&
  {Cai}}]{Parks18}
{Parks}, D., {Prochaska}, J.~X., {Dong}, S., \& {Cai}, Z. 2018, \mnras, 476,
  1151, \dodoi{10.1093/mnras/sty196}

\bibitem[{{Pei} {et~al.}(1991){Pei}, {Fall}, \& {Bechtold}}]{Pei91}
{Pei}, Y.~C., {Fall}, S.~M., \& {Bechtold}, J. 1991, \apj, 378, 6,
  \dodoi{10.1086/170401}

\bibitem[{{Petitjean} {et~al.}(2000){Petitjean}, {Srianand}, \&
  {Ledoux}}]{Petitjean00}
{Petitjean}, P., {Srianand}, R., \& {Ledoux}, C. 2000, \aap, 364, L26

\bibitem[{{Planck Collaboration} {et~al.}(2020){Planck Collaboration},
  {Aghanim}, {Akrami}, {Ashdown}, {Aumont}, {Baccigalupi}, {Ballardini},
  {Banday}, {Barreiro}, {Bartolo}, {Basak}, {Battye}, {Benabed}, {Bernard},
  {Bersanelli}, {Bielewicz}, {Bock}, {Bond}, {Borrill}, {Bouchet}, {Boulanger},
  {Bucher}, {Burigana}, {Butler}, {Calabrese}, {Cardoso}, {Carron},
  {Challinor}, {Chiang}, {Chluba}, {Colombo}, {Combet}, {Contreras}, {Crill},
  {Cuttaia}, {de Bernardis}, {de Zotti}, {Delabrouille}, {Delouis}, {Di
  Valentino}, {Diego}, {Dor{\'e}}, {Douspis}, {Ducout}, {Dupac}, {Dusini},
  {Efstathiou}, {Elsner}, {En{\ss}lin}, {Eriksen}, {Fantaye}, {Farhang},
  {Fergusson}, {Fernandez-Cobos}, {Finelli}, {Forastieri}, {Frailis},
  {Fraisse}, {Franceschi}, {Frolov}, {Galeotta}, {Galli}, {Ganga},
  {G{\'e}nova-Santos}, {Gerbino}, {Ghosh}, {Gonz{\'a}lez-Nuevo}, {G{\'o}rski},
  {Gratton}, {Gruppuso}, {Gudmundsson}, {Hamann}, {Handley}, {Hansen},
  {Herranz}, {Hildebrandt}, {Hivon}, {Huang}, {Jaffe}, {Jones}, {Karakci},
  {Keih{\"a}nen}, {Keskitalo}, {Kiiveri}, {Kim}, {Kisner}, {Knox},
  {Krachmalnicoff}, {Kunz}, {Kurki-Suonio}, {Lagache}, {Lamarre}, {Lasenby},
  {Lattanzi}, {Lawrence}, {Le Jeune}, {Lemos}, {Lesgourgues}, {Levrier},
  {Lewis}, {Liguori}, {Lilje}, {Lilley}, {Lindholm}, {L{\'o}pez-Caniego},
  {Lubin}, {Ma}, {Mac{\'\i}as-P{\'e}rez}, {Maggio}, {Maino}, {Mandolesi},
  {Mangilli}, {Marcos-Caballero}, {Maris}, {Martin}, {Martinelli},
  {Mart{\'\i}nez-Gonz{\'a}lez}, {Matarrese}, {Mauri}, {McEwen}, {Meinhold},
  {Melchiorri}, {Mennella}, {Migliaccio}, {Millea}, {Mitra},
  {Miville-Desch{\^e}nes}, {Molinari}, {Montier}, {Morgante}, {Moss}, {Natoli},
  {N{\o}rgaard-Nielsen}, {Pagano}, {Paoletti}, {Partridge}, {Patanchon},
  {Peiris}, {Perrotta}, {Pettorino}, {Piacentini}, {Polastri}, {Polenta},
  {Puget}, {Rachen}, {Reinecke}, {Remazeilles}, {Renzi}, {Rocha}, {Rosset},
  {Roudier}, {Rubi{\~n}o-Mart{\'\i}n}, {Ruiz-Granados}, {Salvati}, {Sandri},
  {Savelainen}, {Scott}, {Shellard}, {Sirignano}, {Sirri}, {Spencer},
  {Sunyaev}, {Suur-Uski}, {Tauber}, {Tavagnacco}, {Tenti}, {Toffolatti},
  {Tomasi}, {Trombetti}, {Valenziano}, {Valiviita}, {Van Tent}, {Vibert},
  {Vielva}, {Villa}, {Vittorio}, {Wandelt}, {Wehus}, {White}, {White},
  {Zacchei}, \& {Zonca}}]{Planck20}
{Planck Collaboration}, {Aghanim}, N., {Akrami}, Y., {et~al.} 2020, \aap, 641,
  A6, \dodoi{10.1051/0004-6361/201833910}

\bibitem[{{Pontzen} \& {Pettini}(2009)}]{Pontzen09}
{Pontzen}, A., \& {Pettini}, M. 2009, \mnras, 393, 557,
  \dodoi{10.1111/j.1365-2966.2008.14193.x}

\bibitem[{{Pramesh Rao} \& {Subrahmanyan}(1988)}]{Rao1988}
{Pramesh Rao}, A., \& {Subrahmanyan}, R. 1988, \mnras, 231, 229,
  \dodoi{10.1093/mnras/231.2.229}

\bibitem[{{Rector} \& {Stocke}(2001{\natexlab{a}})}]{Stocke01}
{Rector}, T.~A., \& {Stocke}, J.~T. 2001{\natexlab{a}}, \aj, 122, 565,
  \dodoi{10.1086/321179}

\bibitem[{{Rector} \& {Stocke}(2001{\natexlab{b}})}]{Rector01}
---. 2001{\natexlab{b}}, \aj, 122, 565, \dodoi{10.1086/321179}

\bibitem[{{Richards} {et~al.}(2002){Richards}, {Fan}, {Newberg}, {Strauss},
  {Vanden Berk}, {Schneider}, {Yanny}, {Boucher}, {Burles}, {Frieman}, {Gunn},
  {Hall}, {Ivezi{\'c}}, {Kent}, {Loveday}, {Lupton}, {Rockosi}, {Schlegel},
  {Stoughton}, {SubbaRao}, \& {York}}]{Richards02}
{Richards}, G.~T., {Fan}, X., {Newberg}, H.~J., {et~al.} 2002, \aj, 123, 2945,
  \dodoi{10.1086/340187}

\bibitem[{{Richards} {et~al.}(2006){Richards}, {Lacy}, {Storrie-Lombardi},
  {Hall}, {Gallagher}, {Hines}, {Fan}, {Papovich}, {Vanden Berk}, {Trammell},
  {Schneider}, {Vestergaard}, {York}, {Jester}, {Anderson}, {Budav{\'a}ri}, \&
  {Szalay}}]{Richards06}
{Richards}, G.~T., {Lacy}, M., {Storrie-Lombardi}, L.~J., {et~al.} 2006, \apjs,
  166, 470, \dodoi{10.1086/506525}

\bibitem[{{Sanders} \& {Mirabel}(1996)}]{Sanders96}
{Sanders}, D.~B., \& {Mirabel}, I.~F. 1996, \araa, 34, 749,
  \dodoi{10.1146/annurev.astro.34.1.749}

\bibitem[{{Satyapal} {et~al.}(2017){Satyapal}, {Secrest}, {Ricci}, {Ellison},
  {Rothberg}, {Blecha}, {Constantin}, {Gliozzi}, {McNulty}, \&
  {Ferguson}}]{Satyapal17}
{Satyapal}, S., {Secrest}, N.~J., {Ricci}, C., {et~al.} 2017, \apj, 848, 126,
  \dodoi{10.3847/1538-4357/aa88ca}

\bibitem[{{Shankar} {et~al.}(2008){Shankar}, {Dai}, \& {Sivakoff}}]{Shankar08}
{Shankar}, F., {Dai}, X., \& {Sivakoff}, G.~R. 2008, \apj, 687, 859,
  \dodoi{10.1086/591488}

\bibitem[{{Shen} {et~al.}(2011){Shen}, {Richards}, {Strauss}, {Hall},
  {Schneider}, {Snedden}, {Bizyaev}, {Brewington}, {Malanushenko},
  {Malanushenko}, {Oravetz}, {Pan}, \& {Simmons}}]{Shen11}
{Shen}, Y., {Richards}, G.~T., {Strauss}, M.~A., {et~al.} 2011, \apjs, 194, 45,
  \dodoi{10.1088/0067-0049/194/2/45}

\bibitem[{{Shukla} {et~al.}(2021{\natexlab{a}}){Shukla}, {Srianand}, {Gupta},
  {Petitjean}, {Baker}, {Krogager}, \& {Noterdaeme}}]{Shukla2021}
{Shukla}, G., {Srianand}, R., {Gupta}, N., {et~al.} 2021{\natexlab{a}}, \mnras,
  501, 5362, \dodoi{10.1093/mnras/staa3977}

\bibitem[{{Shukla} {et~al.}(2021{\natexlab{b}}){Shukla}, {Srianand}, {Gupta},
  {Petitjean}, {Baker}, {Krogager}, \& {Noterdaeme}}]{Shukla21samp}
---. 2021{\natexlab{b}}, arXiv e-prints, arXiv:2109.00576.
\newblock \doarXiv{2109.00576}

\bibitem[{{Sikora} \& {Begelman}(2013)}]{Sikora13}
{Sikora}, M., \& {Begelman}, M.~C. 2013, \apjl, 764, L24,
  \dodoi{10.1088/2041-8205/764/2/L24}

\bibitem[{{Sikora} {et~al.}(2007){Sikora}, {Stawarz}, \& {Lasota}}]{Sikora07}
{Sikora}, M., {Stawarz}, {\L}., \& {Lasota}, J.-P. 2007, \apj, 658, 815,
  \dodoi{10.1086/511972}

\bibitem[{{Srianand} {et~al.}(2012){Srianand}, {Gupta}, {Petitjean},
  {Noterdaeme}, {Ledoux}, {Salter}, \& {Saikia}}]{Srianand12dla}
{Srianand}, R., {Gupta}, N., {Petitjean}, P., {et~al.} 2012, \mnras, 421, 651,
  \dodoi{10.1111/j.1365-2966.2011.20342.x}

\bibitem[{{Srianand} \& {Kembhavi}(1997)}]{Srianand97}
{Srianand}, R., \& {Kembhavi}, A. 1997, \apj, 478, 70, \dodoi{10.1086/303764}

\bibitem[{{Stern} {et~al.}(2005){Stern}, {Eisenhardt}, {Gorjian}, {Kochanek},
  {Caldwell}, {Eisenstein}, {Brodwin}, {Brown}, {Cool}, {Dey}, {Green},
  {Jannuzi}, {Murray}, {Pahre}, \& {Willner}}]{Stern05}
{Stern}, D., {Eisenhardt}, P., {Gorjian}, V., {et~al.} 2005, \apj, 631, 163,
  \dodoi{10.1086/432523}

\bibitem[{{Stern} {et~al.}(2012){Stern}, {Assef}, {Benford}, {Blain}, {Cutri},
  {Dey}, {Eisenhardt}, {Griffith}, {Jarrett}, {Lake}, {Masci}, {Petty},
  {Stanford}, {Tsai}, {Wright}, {Yan}, {Harrison}, \& {Madsen}}]{Stern12}
{Stern}, D., {Assef}, R.~J., {Benford}, D.~J., {et~al.} 2012, \apj, 753, 30,
  \dodoi{10.1088/0004-637X/753/1/30}

\bibitem[{{Stickel} {et~al.}(1991){Stickel}, {Padovani}, {Urry}, {Fried}, \&
  {Kuehr}}]{Stickel91}
{Stickel}, M., {Padovani}, P., {Urry}, C.~M., {Fried}, J.~W., \& {Kuehr}, H.
  1991, \apj, 374, 431, \dodoi{10.1086/170133}

\bibitem[{{Taylor} {et~al.}(2009){Taylor}, {Stil}, \& {Sunstrum}}]{Taylor09}
{Taylor}, A.~R., {Stil}, J.~M., \& {Sunstrum}, C. 2009, \apj, 702, 1230,
  \dodoi{10.1088/0004-637X/702/2/1230}

\bibitem[{{Urry} \& {Padovani}(1995)}]{Urry95}
{Urry}, C.~M., \& {Padovani}, P. 1995, \pasp, 107, 803, \dodoi{10.1086/133630}

\bibitem[{{Vaughan} {et~al.}(2003){Vaughan}, {Edelson}, {Warwick}, \&
  {Uttley}}]{Vaughan03}
{Vaughan}, S., {Edelson}, R., {Warwick}, R.~S., \& {Uttley}, P. 2003, \mnras,
  345, 1271, \dodoi{10.1046/j.1365-2966.2003.07042.x}

\bibitem[{{Vestergaard} \& {Peterson}(2006)}]{Vestergaard06}
{Vestergaard}, M., \& {Peterson}, B.~M. 2006, \apj, 641, 689,
  \dodoi{10.1086/500572}

\bibitem[{{Waters} {et~al.}(2020){Waters}, {Magnier}, {Price}, {Chambers},
  {Burgett}, {Draper}, {Flewelling}, {Hodapp}, {Huber}, {Jedicke}, {Kaiser},
  {Kudritzki}, {Lupton}, {Metcalfe}, {Rest}, {Sweeney}, {Tonry}, {Wainscoat},
  \& {Wood-Vasey}}]{Waters20}
{Waters}, C.~Z., {Magnier}, E.~A., {Price}, P.~A., {et~al.} 2020, \apjs, 251,
  4, \dodoi{10.3847/1538-4365/abb82b}

\bibitem[{{Wiklind} \& {Combes}(1996)}]{wiklind1996}
{Wiklind}, T., \& {Combes}, F. 1996, \nat, 379, 139, \dodoi{10.1038/379139a0}

\bibitem[{{Wolfe} {et~al.}(2005){Wolfe}, {Gawiser}, \& {Prochaska}}]{Wolfe05}
{Wolfe}, A.~M., {Gawiser}, E., \& {Prochaska}, J.~X. 2005, \araa, 43, 861

\bibitem[{{Wolfe} \& {Prochaska}(2000)}]{Wolfe00}
{Wolfe}, A.~M., \& {Prochaska}, J.~X. 2000, \apj, 545, 591,
  \dodoi{10.1086/317860}

\bibitem[{{Wright} {et~al.}(2010){Wright}, {Eisenhardt}, {Mainzer}, {Ressler},
  {Cutri}, {Jarrett}, {Kirkpatrick}, {Padgett}, {McMillan}, {Skrutskie},
  {Stanford}, {Cohen}, {Walker}, {Mather}, {Leisawitz}, {Gautier}, {McLean},
  {Benford}, {Lonsdale}, {Blain}, {Mendez}, {Irace}, {Duval}, {Liu}, {Royer},
  {Heinrichsen}, {Howard}, {Shannon}, {Kendall}, {Walsh}, {Larsen}, {Cardon},
  {Schick}, {Schwalm}, {Abid}, {Fabinsky}, {Naes}, \& {Tsai}}]{Wright10}
{Wright}, E.~L., {Eisenhardt}, P. R.~M., {Mainzer}, A.~K., {et~al.} 2010, \aj,
  140, 1868, \dodoi{10.1088/0004-6256/140/6/1868}

\bibitem[{{Zahedy} {et~al.}(2020){Zahedy}, {Chen}, {Boettcher}, {Rauch},
  {French}, \& {Zabludoff}}]{Zahedy20}
{Zahedy}, F.~S., {Chen}, H.-W., {Boettcher}, E., {et~al.} 2020, \apjl, 904,
  L10, \dodoi{10.3847/2041-8213/abc48d}

\end{thebibliography}

\end{document}